\documentclass[fleqn,usenatbib]{mnras} 

\usepackage{newtxtext,newtxmath}

\usepackage[T1]{fontenc}
\usepackage{ae,aecompl}

\usepackage{graphicx}	
\usepackage{amsmath}	
\usepackage{pdflscape}	
\usepackage{color,soul} 
\usepackage{appendix}
\usepackage{verbatim}

\newcommand{\lymana}{Lyman-$\alpha$}		
\newcommand{\lya}{Ly$\alpha$}			
\newcommand{\lyb}{Ly$\beta$}				
\newcommand{\lye}{Ly$\epsilon$}
\newcommand{\Lymana}{Lyman-$\alpha$}		
\newcommand{\Lya}{Ly$\alpha$}			
\newcommand{\Lyb}{Ly$\beta$}				

\newcommand{\kms}{\,\,{\rm km}\,{\rm s}^{-1}}	

\def\HI{\ion{H}{i}}
\def\HII{\ion{H}{ii}}
\def\CII{\ion{C}{ii}}
\def\CIII{\ion{C}{iii}}
\def\CIV{\ion{C}{iv}}
\def\SiII{\ion{Si}{ii}}
\def\SiIII{\ion{Si}{iii}}
\def\SiIV{\ion{Si}{iv}}

\def\NIII{\ion{N}{iii}}

\def\NV{\ion{N}{v}}
\def\OI{\ion{O}{i}}
\def\OVI{\ion{O}{vi}}
\def\AlII{\ion{Al}{ii}}
\def\AlIII{\ion{Al}{iii}}
\def\FeII{\ion{Fe}{ii}}
\def\MgII{\ion{Mg}{ii}}

\def\HeII{\ion{He}{ii}}
\newcommand{\hMpc}{h^{-1}{\rm Mpc}}
\newcommand{\ullim}[2]{\raisebox{0.5ex}{\tiny$\substack{+$#2$ \\ -$#1$}$}}

\defcitealias{Pieri2014}{P14}
\defcitealias{Pieri2010Stacking}{P10}
\defcitealias{Lyke2020}{DR16Q}
\defcitealias{Perez-Rafols2022}{PR22}

\makeatletter
\graphicspath{{figures/}}
\makeatother

\title[Studying the Circumgalactic Medium with SBLAs]{
A Strong Blend in the Morning: \\
Studying the Circumgalactic Medium Before Cosmic Noon with Strong, Blended \lymana\ Forest Systems}

\author[S. Morrison et al.]{Sean Morrison,$^{1,2}$
Debopam Som,$^{1,3,4}$
Matthew M. Pieri,$^{1}$\thanks{E-mail: matthew.pieri@lam.fr}
Ignasi P\'{e}rez-R\`{a}fols,$^{1,5}$ \newauthor
and Michael Blomqvist$^{1}$
\\
$^{1}$ Aix Marseille University, CNRS, LAM, Laboratoire d'Astrophysique de Marseille, Marseille, France\\
$^{2}$ Department of Astronomy, University of Illinois at Urbana-Champaign, Urbana, IL 61801, USA,\\
$^{3}$ Ohio State University\\
$^{4}$ Space Telescope Science Institute\\
$^{5}$ Sorbonne Universit\'{e}, Universit\'{e} Paris Diderot, CNRS/IN2P3, Laboratoire de Physique Nucl\'{e}aire et de Hautes Energies,\\ LPNHE, 4 Place Jussieu, F-75252 Paris, France
}

\date{Accepted XXX. Received YYY; in original form ZZZ}

\pubyear{2023}

\begin{document}
\label{firstpage}
\pagerange{\pageref{firstpage}--\pageref{lastpage}}
\maketitle

\begin{abstract}

We study of the properties of a new class of circumgalactic medium absorbers identified in the \lymana\ forest: "Strong, Blended \Lymana" (or SBLA) absorption systems. We study SBLAs at $2.4<z<3.1$ in SDSS-IV/eBOSS spectra by their strong extended \Lymana\ absorption complexes covering 138 $\kms$ with an integrated $\log (N_{HI}/$cm$^{-2}) =16.04$\ullim{0.06}{0.05} and Doppler parameter $b=18.1$\ullim{0.4}{0.7}$\kms$. 
 Clustering with the \Lymana\ forest provides a large-scale structure bias of $b = 2.34\pm0.06$ and halo mass estimate of $M_h \approx 10^{12}{\rm h^{-1}M_{\sun}}$ for our SBLA sample. 
We measure the ensemble mean column densities of 22 metal features in the SBLA composite spectrum and find that no single-population multiphase model for them is viable. We therefore explore the underlying SBLA population by forward modelling the SBLA absorption distribution.
Based on covariance measurements and favoured populations we find that $\approx 25$\% of our SBLAs have stronger metals.
Using silicon only we find that our strong metal SBLAs trace gas with a $\log(n_H / $cm$^{-3}) > -2.40$ for $T=10^{3.5}$K and show gas clumping on $<210$ parsec scales.
We fit multiphase models to this strong sub-population and find a low ionization phase with $n_H=1$cm$^{-3}$,  $T=10^{3.5}$K  and $[X/H]=0.8$,  an intermediate ionization phase with $\log(n_H / $cm$^{-3}) = -3.05$, $T=10^{3.5}$K and $[X/H]=-0.8$, and a poorly constrained higher ionization phase. We find that the low ionization phase favours cold, dense super-solar metallicity gas with a clumping scale of just 0.009 parsecs.

\end{abstract}




\section{Introduction}
\label{sec:intro}

The history of the universe can be thought of as an evolution through a series of distinct epochs; the hot Big Bang, the dark ages, the epoch of the first stars, hydrogen reionization,  the galaxy formative era reaching a crescendo when the star formation rate peaks at  $z \approx 2$ \citep{Madau2014}, and finally a gradual decline in star formation activity (driven in-part by dark energy driving the expansion of the universe) leading to the mature galaxies we see today. The star formation rate peak is often known as `cosmic noon'. The period leading up to that epoch (which we might call the `cosmic morning') is one of the most active periods in the history of the universe. This is the epoch where gas is actively accreting onto galaxies and fuelling imminent star formation. It is also the epoch where galaxies increasingly respond to star formation and eject outflows into their surrounding environments. The zone where accretion and outflows occur is known as the `circumgalactic medium' (or CGM), in general regions outside galaxies are known as the `intergalactic medium' (or IGM).

The cosmic morning is also notable as the epoch where key UV atomic transitions are redshifted into the optical window allowing us to study them from the ground-based observatories in great detail. In particular, the richly informative \lymana\ (\lya) forest is well-studied at  $z>2$, typically towards background quasars \citep{GunnPeterson1965,Lynds1971}.
This leads to samples of \lymana\ forest spectra going from a few hundred at $z<2$ to a few hundred thousand at $z>2$.

This combination of high-observability and high-information-content is encouraging for developing an understanding of galaxy formation, however, progress has been held back by the fact that at these high redshifts galaxies are faint and so have been observed in much smaller numbers than the active galactic nuclei hosting quasars. Wide-area surveys of galaxies at $z>2$ are on their way (e.g. HETDEX, \citealt{hill2008} and PFS, \citealt{Takada2014}) 
but in the meantime and in complement to them, we can study galaxies in absorption. 

The most widely known and accepted way of doing this is to study damped \Lymana\ systems (or DLAs; \citealt{Wolfe2005}), which are systems with a column density $N_{HI}>10^{20.3}$cm$^{-2}$ such that ionizing photons cannot penetrate them. These systems are typically easy to identify in absorption through their wide damping wings. A wider category of systems (including DLAs) that do not allow the passage of ionizing photons (i.e. self-shielded) are named Lyman limit systems (or LLSs), which have column densities $N_{HI}>10^{17.2}$cm$^{-2}$. Partial Lyman limit systems  with $10^{16.2}$cm$^{-2}< N_{HI}<10^{17.2}$cm$^{-2}$ absorb a sufficient fraction of ionizing photons and modify ionization fractions of species they host (though the lower boundary  of this group is somewhat ill-defined).  DLAs are thought to have a particularly strong connection to galaxies since the densities inferred are approximately sufficient to provoke star formation (e.g. \citealt{Rafelski2011}). LLSs are less clear, sometimes they are thought to be closely associated with galaxies and in other cases they are thought to trace cold streams of inflowing gas (e.g. \citealt{Fumagalli2011}). 

Self-shielded systems cover a small covering fraction of the CGM (typically defined as regions within the virial radius of a galaxy hosting dark matter halo).
The overwhelming majority of CGM regions are not detectable as Lyman limit systems but are optically thin with
$10^{14}$cm$^{-2} \lesssim N_{HI} \lesssim 10^{16}$cm$^{-2} $
(e.g. \citealt{Fumagalli2011} and  \citealt{Hummels2019}). Conversely, many of these strong optically thin systems are not CGM systems but instead probe diffuse IGM gas. In other words, these systems are critically important tracers of the CGM but their CGM/IGM classification  is challenging. Furthermore given that lines with $N_{HI} \gtrsim 10^{14}$cm$^{-2} $ are on the flat part of the curve of growth (e.g. \citealt{Charlton2000}) and therefore suffer from degeneracy between column density and line broadening even the column density itself is a non-trivial measurement. 

We explore a wider sample of CGM systems that are not optically thick to ionizing photons, nor do they require a challenging estimation of column density for confirmation. The sample requires only that the absorption in the \lymana\ forest be strong and blended. This population has already been studied in \cite{Pieri2010Stacking} and followed up in \cite{Pieri2014} through spectral stacking. We return to this sample with a refined error analysis of the stacked spectra, a formalised measurement of column densities, halo mass constraints and more extensive interpretation, in particular modelling of the underlying metal populations in the stack of spectra.

There are various observational studies of the CGM that provide gas details such as thermal properties, density, metallicity, sometimes with respect to a galactic disk, often as a function of impact parameter to the likely closest galaxy (e.g. \citealt{Steidel2010, Bouche2013, Werk2014, Augustin2018, Qu2022}). SINFONI and MUSE integral field unit data have provided a particular boost to the detail and statistics of CGM observations (e.g. \citealt{Peroux2016, Fumagalli2014, Fossati2021}). 

Despite the exquisite detail offered by these datasets, an unbiased, large statistical sample of spectra is needed in order to develop a global picture of the CGM. Obtaining such samples with this level of detail remains a distant prospect. Hence, we take a brute force approach. We identify CGM regions as a general population with diverse gas properties studied simultaneously. These samples number in the 10s or even 100s of thousands recovered from SDSS spectra of $z>2$ quasars. The selection function is simple (see \autoref{sec:sampleselection}) but the challenge resides in the interpretation of this rich but mixed sample. Complexity exists not only in the unresolved phases, but also in the diversity of systems selected.

In previous work \citep{Pieri2010Stacking, Pieri2014,Morrison2021, Yang2022} the focus has been to interpret the multi-phase properties of a hypothetical mean system that is measured with high precision in the composite spectrum of the ensemble. We  revisit these measurements, and go further to study the underlying populations of metals features: both their individual expected populations and the degree of covariance between them. We focus in particular on a strong population of metals that we infer and find signal of metal rich, high-density, cold gas clumping on remarkably small-scales. Much remains to be explored but we offer a framework for studying the CGM in the largest \Lymana\ forest samples.
In light of the improved understanding outlined here, we define these absorption systems (initially discovered in \citealt{Pieri2010Stacking}) as a new class
of CGM absorption systems defined by the both absorption strength and clustering on $\sim 100\kms$ scales, and we name them `Strong, Blended \Lymana' or SBLA absorption systems. Over the coming decade quasar surveys at $z>2$ will grow and will increasingly be accompanied by galaxy surveys at the same redshifts, making this statistical population analysis an increasingly powerful tool.

This publication is structured as follows. We begin by describing the dataset (including quasar continuum fitting to isolate the foreground transmission spectrum). In \autoref{sec:sampleselection} we describe various ways of selecting SBLAs for different purity and absorption strength before settling on a analysis sample in subsequent sections. We then review the stacking methodology in \autoref{sec:stackprocess} and follow this in  \autoref{sec:errore2e} with a comprehensive end-to-end error analysis of the resulting composite spectrum of SBLAs. In \autoref{sec:sblamass} we present large-scale structure biases  for SBLAs and inferences regarding their halo masses. In \autoref{sec:interpret_comp} we begin to explore our results with measurements in the composite  \HI\ and metal column densities, the sensitivity to physical conditions and the covariance between metal lines. We then go on to model and constrain the underlying absorber populations and explore the properties of the strong metal population in \autoref{sec:abspop}. We follow up with an extensive discussion \autoref{sec:discussion} and conclusions.  

We also provide appendices on the correlation function methodology used to measure structure bias (Appendix~\ref{extra:corrmethod}),  details on the error analysis (Appendix~\ref{extra:errore2e}), SBLAs studied in high-resolution spectra (Appendix~\ref{extra:HR_SBLAs}), and finally measurements of the covariance between our metal features (Appendix~\ref{extra:covariance}).

\section{Data}
\label{sec:data}
SDSS-IV \citep{Blanton2017} carried out three spectroscopic surveys using the 2.5-meter Sloan telescope \citep{Sloan} in New Mexico. These surveys included APOGEE-2 (an infrared survey of the Milky Way Stars), Extended Baryon Oscillation Spectroscopic Survey (eBOSS; a optical cosmological survey of quasars and galaxies) and MaNGA (an optical IFU survey of $\sim$10,000 nearby galaxies). eBOSS, an extension of the SDSS-III \citep{Eisenstein2011, Dawson2013} BOSS survey, utilizes the BOSS spectrograph.

The BOSS instrument \citep{Smee2013} employs a twin spectrograph design with each spectrograph separating incoming light into a blue and a red camera. The resulting spectral coverage is over 3450\AA\ -- 10,400\AA\ with a resolving power ($\lambda$/$\Delta\lambda$) ranging between $\sim$ 1650 (near the blue end) to $\sim$ 2500 (near the red end). 
We discard regions with a 100 pixel boxcar smoothed signal-to-noise ratio (S/N)~$<1$, in order to exclude from our analysis regions of sharply falling S/N at the blue extreme of SDSS-IV quasar spectra. 
Pixels flagged by the pipeline, pixels around bright sky lines and the observed Galactic \ion{Ca}{II} H\&K absorption lines were also masked throughout our stacking analysis.  

We use a high redshift quasar sample derived from the final data release of eBOSS quasars (\citealt{Lyke2020}, hereafter \citetalias{Lyke2020}) from SDSS-IV Data Release 16 \citep{SDSS_DR16}. The spectra of objects targeted as quasars are reduced and calibrated by the SDSS spectroscopic pipeline \citep{Bolton2012} which also classifies and determines the redshifts of sources automatically. Unlike the quasar catalogues from DR12Q \citep{Paris2017} and earlier, the additional quasars in DR16Q are primarily selected via the automated pipeline, with a small visually inspected sample for validation.

Ensuring the availability of enough \lya\ forest pixels required for an accurate continuum estimate restricts the minimum redshift of our quasar sample to $z_{em} \ge$ 2.15. We also discard DR16Q quasars with median \lya\ forest S/N $<0.2$ pixel$^{-1}$ or median S/N $<0.5$ pixel$^{-1}$ over 1268 \AA\ -- 1380 \AA\ given the difficulty in the accurate estimation of continua for very noisy spectra. Since the presence of BAL troughs contaminate the \lya\ forest with intrinsic quasar absorption and likely affects continuum estimation, we discard quasars flagged as BAL quasars in DR16Q. Pixels which were flagged by the pipeline as problematic during the extraction, flux calibration or sky subtraction process were excluded from our analysis. Spectra of DR16Q quasars with more than 20$\%$ pixels within 1216 $< \lambda_{rest} <$ 1600 \AA\ or in the \lya\ forest region flagged to be unreliable by the pipeline were discarded.

DLAs and their wings (where the estimated flux decrement is $>5\%$) in our data were masked using the DLA catalogue internal to the \citetalias{Lyke2020} catalogue, presented in \cite{Chabanier2022} and based on the \citet{Parks2018} convolutional neural network deep learning algorithm designed to identify and characterise DLAs. Spectra with more than one DLA are entirely discarded throughout our analysis.

Further steps are taken to prepare the data for the selection of \lya\ systems to be stacked and the spectra to be stacked themselves. Steps taken for the calculation of forest correlation functions are explained separately in \autoref{sec:sblamass})

\subsection{Preparation for \texorpdfstring{\Lymana}{Lyman-alpha} absorber selection}

We take two approaches for the normalisation of the quasar continua in our stacking analysis.
For SBLA detection we follow the method described in \cite{Lee2013}
 over 1040 -- 1185 \AA\ in the rest frame.
 The modified version of the MF-PCA technique presented in \citet{Lee2012} fits the 1216 $< \lambda_{rest} <$ 1600 \AA\ region of a quasar spectrum with PCA templates providing a prediction for the continuum shape in the \lya\ forest. The predicted continuum is then re-scaled to match the expected evolution of the \lya\ forest mean flux from \citet{Faucher-Giguere2008}. The above definition of the forest region avoids contamination from higher order Lyman series lines and conservatively excludes the quasar proximity zone.

\begin{figure}
	\begin{center}
		\includegraphics[angle=0,width=.99\linewidth]{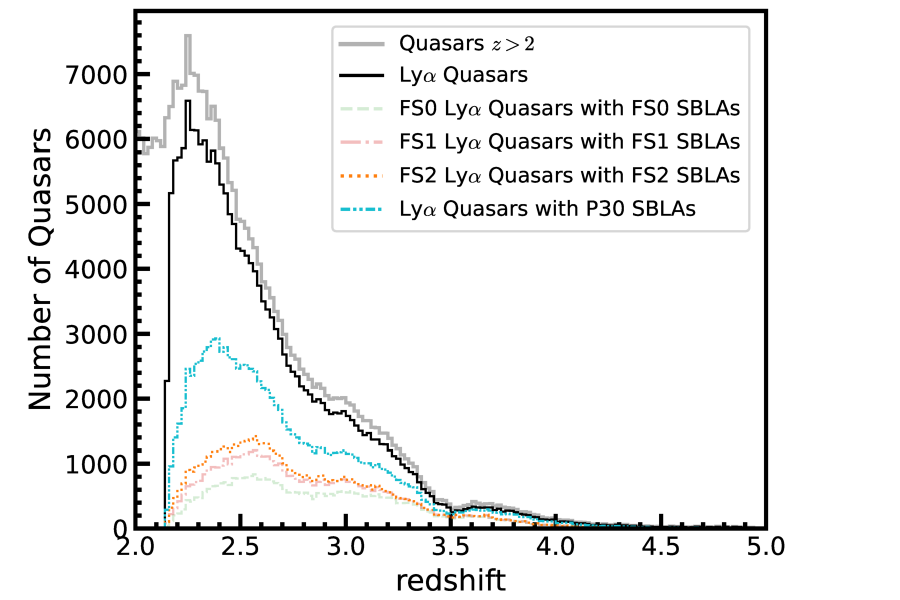}
		\caption{Redshift distribution of the 198,405 quasars in our initial sample is shown in black. The thick grey solid curve represents the distribution of all $z\ge2$ quasars from DR16Q. Also shown are the 4 samples of SBLAs FS0 (light green dashed line), FS1 (light red dashed dotted line), FS2 (orange dotted line), and {P30} (dashed double dotted line) as discussed in  \autoref{sec:sampleselection}. 
  }
		\label{fig:qsampledistrib}
	\end{center}
\end{figure}

We  discard any spectrum for which the estimated continuum turns out to be negative. 
Metal absorption lines are masked using a 3$\sigma$ iterative flagging of outlier pixels redward of the \lya\ forest from a spline fit of the continua. With all the cuts mentioned above, we are left with an analysis sample of 198,405 quasars
with a redshift distribution shown in \autoref{fig:qsampledistrib} along with the distribution of all $z\ge2$ quasars from DR16Q.

\subsection{Preparation of spectra to be stacked}

The mean-flux regulated PCA continua described above provide robust estimates of the \lya\ forest absorption and are therefore well-suited for the search and selection of SBLAs for stacking. However, these continua are limited to $<$1600 \AA\ in the quasar rest frame and present discontinuities due to the mean-flux regulation process. For spectra to be stacked, we required wide wavelength coverage without discontinuities and so we use spline fitting.

We split each spectrum into 25\AA\ chunks over the entire observed spectral range and calculate the median flux in each spectral chunk before fitting a cubic spline to these flux nodes. Pixels falling 1$\sigma$ below the fit within the \lya\ forest or 2$\sigma$ below outside the forest are then rejected and the flux nodes are recalculated followed by a re-evaluation of the spline fit. This absorption-rejection is iterated until convergence to estimate the quasar continuum.

The cubic spline fitting breaks down in regions with large gradients in the continuum, usually near the centres of strong quasar emission lines. We, therefore, mask data around the peaks of emission features commonly seen in quasar spectra before the continuum fitting is performed. In addition, as sharp edges (caused by missing data as a result of masking the emission peaks) can induce instability in the fits using the smooth cubic spline function, we discard 
a buffer region around the emission line masks. The extents of the masked region ($\lambda_{mask}$) and the corresponding buffer ($\pm\lambda_{buffer}$), in the quasar rest frame, depend on the typical strength of the emission line concerned and are listed in \autoref{tab:emimasks} along with the rest frame locations of the emission line centres.

\begin{table}
	\caption{Emission line masks and buffer regions used in cubic-spline continuum estimation. All wavelengths listed are in quasar rest frame.}
	\label{tab:emimasks}
	\begin{center}
	\begin{tabular}{@{}lccc}
		\hline
		Emission Line	& $\lambda_{rest}$  &	$\lambda_{mask}$	&$\pm\lambda_{buffer}$	\\
					    &	(\AA)			    &	(\AA)			        &	(\AA)				    \\
		\hline
		\hline
		\lyb			&	1033.03			&	1023 -- 1041		&		5				\\
		\lya			&	1215.67			&	1204 -- 1240		&		10				\\
		\OI			    &	1305.42			&	1298 -- 1312		&		5				\\
		\SiIV			&	1396.76			&	1387 -- 1407		&		10				\\
		\CIV			&	1549.06			&	1533 -- 1558		&		10				\\
		\HeII			&	1637.84			&	1630 -- 1645		&		5				\\
		\CIII			&	1908.73			&	1890 -- 1919		&		10				\\
		\MgII			&	2798.75			&	2788 -- 2811		&		5				\\
		\hline
	\end{tabular}
	\end{center}
\end{table}

\section{\texorpdfstring{Selection of Strong, Blended Lyman $\alpha$ forest absorption systems}{Selection of Strong, Blended Lyman $\alpha$ forest absorption systems}}
\label{sec:sampleselection}

When analysing the absorption in the \lya\ forest, typically two approaches are taken. One may treat the forest as a series of discrete identifiable systems such can be fit as Voigt profiles and therefore derive their column densities and thermal and/or turbulent broadening. Alternatively one may treat the forest as a continually fluctuating gas density field and therefore take each pixel in the spectrum and infer a measurement of gas opacity (the so-called `fluctuating Gunn-Peterson approximation'). For the former, the assumption is that the gas can be resolved into a discrete set of clouds, which is physically incorrect for the \lya\ forest as a whole but a useful approximation in some conditions. For the latter, it is assumed that line broadening effects are subdominant to the complex density structure as a function of redshift in the \lya\ forest.

In this work, we take the second approach, selecting absorption systems based on the measured flux transmission
in a spectral bin in the forest, $F_{\textrm{\lya}}$.
The absorbers in our sample are selected from wavelengths of 1040 \AA\ $<\lambda<$1185 \AA\ in the quasar rest frame. This range was chosen to eliminate the selection of 
\Lyb\ absorbers and exclude regions of elevated continuum fitting noise from \Lyb\ and \OVI\ emission lines at the blue limit, and absorbers proximate to the quasar (within $7563\kms$) at the red limit.

We follow the method of \cite{Pieri2014} \citepalias[hereafter][]{Pieri2014} to generate their three strongest absorber samples, which they argued select CGM systems with varying purity. We limit ourselves to $2.4<z_{abs}<3.1$ to retain sample homogeneity with varying wavelength. Without this limit there would be limited sample overlap across the composite spectrum (the blue end of the composite  would measure exclusively higher redshift SBLAs and the red end would measure exclusively lower redshift SBLAs). Specifically, \citepalias{Pieri2014} chose this redshift range to allow simultaneous measurement of both the full Lyman series and \MgII\ absorption.
We take main samples explored in \citetalias{Pieri2014} using a  signal-to-noise per pixel $>3$ over a 100 pixel boxcar. Of the 198405 quasars available,  68525 quasars had forest regions of sufficient quality necessary for the recovery of Strong, Blended \lymana\ absorbers.

These samples are: FS0 with $-0.05\leq F_{\textrm{\lya}} <0.05$, FS1 with $0.05\leq F_{\textrm{\lya}} <0.15$, and FS2 with $0.15\leq F_{\textrm{\lya}} <0.25$. 
The numbers of systems identified are given in \autoref{tab:SBLA_Samples}. This is approximately quadruple the number of SBLAs with respect to \citetalias{Pieri2014} (though they were not given this name at the time). We also consider samples defined by their purity as discussed below.

All remaining absorbers (after the flagging discussed in the previous section) are assumed to arise due to the \lya\ transition with  $2.4<z<3.1$, and are available for selection. Given the strength of absorbers selected here this is a fair assumption and in cases this is not true, the effect is easily controlled for (e.g. `the shadow \SiIII' features discussed in \citetalias{Pieri2014}). 
The spectral resolution of the BOSS spectrograph varies from $R = 1560$ at 3700\AA\ to $R= 2270$ at 6000\AA. For chosen redshift range the resolution at the wavelength of the selected \lya\ absorption is $R\approx 1800$ and this is therefore our effective spectral resolution throughout this work. This equates to $167\kms$ or 2.4 pixels in the native SDSS wavelength solution. This allows us to rebin the spectra by a factor of 2 before selection of our \lya\ absorbers to reduce noise and improve our selection of absorbers. It has the added benefit of precluding double-counting of absorbers within a single resolution element. This results in the selection of absorbers on velocity scales of $\sim 138\kms$. Given that \Lymana\ absorbers have a median Doppler parameter of $b\approx 30\kms$ (and $\sigma =10\kms$; \citealt{Hu1995,Rudie2012}) our absorber selection is both a function of absorber strength and absorber blending. More detail is provided on the meaning of this selection function in \citetalias{Pieri2014}. 

One of the key results of \citetalias{Pieri2014} was that regions of the \lya\ forest with transmission less than 25\% in bins of $138\kms$ are typically associated with the CGM of Lyman break galaxies (using Keck HIRES and VLT UVES spectra with nearby Lyman break galaxies).
The metal properties in the composite spectra were strongly supportive of this picture. We further reinforce this picture with improved metal measurements, constraints on their halo mass both from large-scale clustering, and arguments regarding halo circular velocities (\autoref{sec:sblamass}). 
Given the weight of evidence that these systems represent a previously unclassified sample of galaxies in absorption, we chose to explicitly define them as a new class and name them "Strong, Blended \Lymana"  (SBLAs) forest absorption systems. The preferred definition here is a noiseless transmitted \lya\ flux $F_{\textrm{\lya}} <0.25$ over bins of $138\kms$ for consistency with this Lyman break galaxy comparison and comparison with \citetalias{Pieri2014}. 
Refinement of this class of SBLAs and/or alternative classes of SBLAs are possible with modifications of transmission or velocity scale. In the arguments that follow, statements regarding purity refer specifically to the successful recovery of this idealised SBLA class.

As pointed out in \autoref{sec:data}, DLAs from \citetalias{Lyke2020} (presented in \citealt{Chabanier2022})
are masked in our selection of SBLAs, however, no catalogue of Lyman limit systems (LLS) are available and are therefore potentially among the SBLA sample. As \citetalias{Pieri2014} discussed at length, even if one assumes that all LLS are selected (which is not a given) no more than 3.7\% of SBLAs should be a LLS. SBLAs are much more numerous and this is not surprising in light of simulations (e.g. \citealt{Hummels2019}) showing that the covering fraction of LLS (including DLA) is small compared to regions of integrated column density $\approx 10^{16}$cm$^{-2}$ we find here. The presence of even small numbers of Lyman limit systems can be impactful for our ionization corrections, however, and we return to this topic in \autoref{subsec:modmet} and \autoref{sec:abspop}.

\subsection{\texorpdfstring{Using \lya}{Using Lyman alpha} Mocks to Characterise Sample Purity}
The FS0 sample provides higher purity SBLA selection than FS1 or FS2 \citepalias{Pieri2014}. However, we note that there exists sets of systems that do not meet these requirements but have equivalent or better purity compared to subsets of FS0 systems with limiting S/N or flux.
For example, systems with $F_{\textrm{\lya}} = 0.06$ and S/N/\AA\ $= 10$ will have a higher SBLA purity than systems with $F_{\textrm{\lya}} = 0.04$ and  S/N/\AA\ $\approx$ 3, even though the latter meets the requirements for sample FS0 and the former does not. 

We have therefore explored the optimal combination of interdependent S/N and flux transmission thresholds to obtain a desired limiting purity. We used the official SDSS BOSS \lymana\ forest mock data-set produced for DR11 \citep{Bautista2015} without the addition of 
DLAs (which are masked in our analysis) and metal absorption lines (which are rare, particularly for the strong, blended absorption studied here).
The signal-to-noise was calculated using a 100-pixel boxcar smoothing of the unbinned data (replicating the selection function in the data), and then was rebinned to match the resolution used in our selection function. We then compared the observed (noise-in) \lya\ flux transmission in the mocks with the true (noiseless) flux transmission of these system in various ranges of observed flux transmission and S/N. The purity is the fraction of systems selected which meets the SBLA definition of true (noiseless) flux transmission $F_{\textrm{\lya}} <0.25$. We then accept ranges that meet a given purity requirement.

We estimated the purity for a grid of S/N/\AA\ $>0.4$ (and step size of 0.2) and $-0.05\leq$F$<0.35$ (and step size of 0.05). The flux and S/N/\AA\ of the selected lines in the real data are compared to this grid to give an estimate of the purity of the selection.
By building samples in this way we are not limited to the high signal-to-noise data used in \citetalias{Pieri2014}. Though we focus on FS0 for consistency with \citepalias{Pieri2014}, we demonstrate here how expanded samples can be prepared.

Using this approach, we propose three additional samples defined by their limiting SBLA purities.
Noting that the mean purity of the FS0 sample of $\approx 90\%$, we produce a sample of 90\% minimum purity, which we label P90. We do indeed obtain a more optimal sample with both higher mean purity and nearly double the number of SBLAs with sample P90 compared to FS0.  We further produce samples with minimum purity  of 75\% and 30\%, labelled P75 and P30 respectively. The numbers and resulting mean purity taken from these mock tests are showing in \autoref{tab:SBLA_Samples}. These tests indicate that there are around 200,000 SBLAs between $2.4<z<3.1$ are present in data. 

Our companion paper, \citet{Perez-Rafols2022}, uses a version of our P30 sample without a redshift limit to measure large-scale structure clustering. This provided us with 742,832 SBLAs. Assuming that our inferred purity for P30 is correct for this sample also, we obtain around half a million true SBLAs in our most inclusive sample. This is more than an order of magnitude more CGM systems than our DLA sample.

\begin{table}
    \caption{Possible $2.4<z<3.1$ SBLA samples, their flux transmission boundaries (in 138$\kms$ bins and their purity to true (noiseless) flux transmission of $ F_{\textrm{\lya}}<0.25$}
    \label{tab:SBLA_Samples}
    \begin{center}
    \begin{tabular}{@{}lcccc}
       \hline
       Sample & F\textsubscript{lower} & F\textsubscript{upper} & \textless{Purity(\%)}\textgreater &
       Number of systems\\
       \hline
       \hline
       FS0 & -0.05 & 0.05 &  89 & 42,210 \\
       FS1 &  0.05 & 0.15 &  81 &  86,938 \\
       FS2 &  0.15 & 0.25 &  55 & 141,544 \\ 
       P30 & -0.05 & 0.25$^a$$^b$ & 63 & 335,259 \\
       P75 & -0.05 & 0.25$^a$ & 90 & 124,955 \\
       P90 & -0.05 & 0.25$^a$ & 97 &  74,660 \\
        \hline
        \multicolumn{5}{l}{$^a$ Hard limit. True maximum is a function }\\
        \multicolumn{5}{l}{\hspace{1em}of S/N tuned for desired minimum purity.}\\    
        \multicolumn{5}{l}{$^b$ Redshift limited version of  sample used in \citet{Perez-Rafols2022}.}\\    
   \end{tabular}
    \end{center}

\end{table}

\section{Stacking procedure}
\label{sec:stackprocess}

We follow the method originally set out in \citet{Pieri2010Stacking} \citepalias[hereafter][]{Pieri2010Stacking} and further elaborated in \citetalias{Pieri2014} for building composite spectra of \lya\ forest absorbers through the process of stacking SDSS spectra.  For every selected \lya\ absorber with redshift $z_\alpha$ the entire continuum fitted quasar spectrum is treated initially as if it were the spectrum of that system alone. In practise, one produces a rest frame spectrum of that absorber by dividing the wavelength array by $(1+z_\alpha)$. This is done many times for many different selected absorbers (sometimes using the quasar spectra more than once). This ensemble of SBLA rest frame spectra constitutes the stack of spectra to be analysed. Typically one collapses this stack to a single value at every wavelength 
using some statistic. In \citetalias{Pieri2010Stacking} and \citetalias{Pieri2014} two statistics were applied; the median and the arithmetic mean (though in some circumstances the geometric mean may be the more suitable choice). In \autoref{sec:abspop} below we will explore what we can learn from the full population of absorbers and relax the implicit assumption that all systems in a given sample are the same. In this work we will focus on the arithmetic mean with no S/N weighting for reasons which will become clear in \autoref{sec:abspop}. Stating this explicitly, we calculate the mean of the stack of spectra (or ‘mean stacked spectrum’) as
\begin{equation}
F_S(\lambda_r) = \sum_{i=1}^{n} F_i(\lambda_r) /  n
\end{equation}
where $\lambda_r$ indicates the wavelength in the rest frame of the SBLA system selected and the set of $i=1,n$ indicates SBLAs that contribute a measurement at the specified rest frame wavelength.

\begin{figure*}
\begin{center}
\includegraphics[angle=0, width=.99\textwidth]{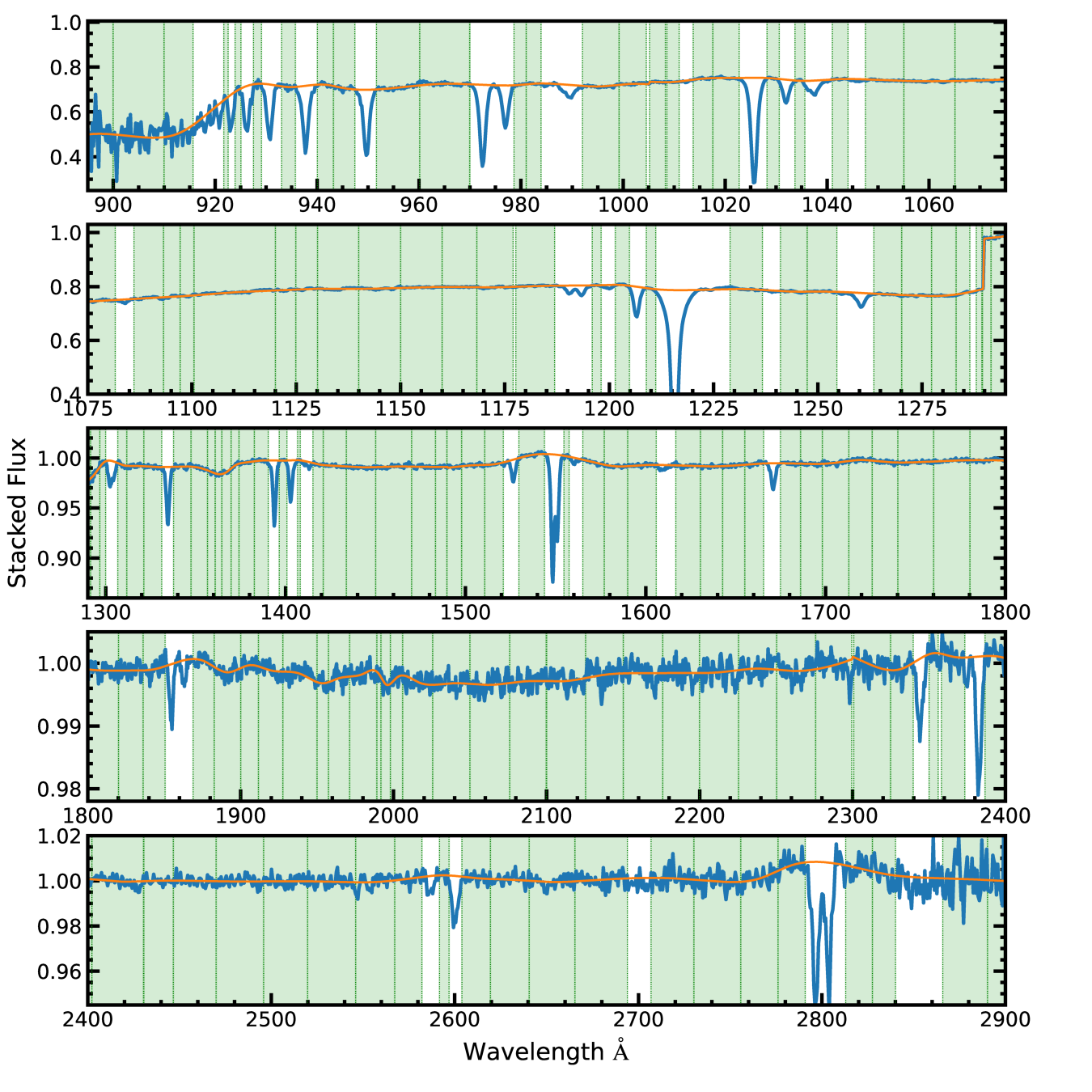}
\end{center}
\caption{The stacked spectrum of the SBLA system sample FS0 (systems selected with flux in the range $-0.05 \le F<0.05$ (FS0) is plotted with solid blue curve. The stacked spectrum show broad continuum variations resulting from uncorrelated absorption. The overlaid orange curve represents this pseudo-continuum. The regions used to estimate the pseudo-continuum are shown as green shaded regions within vertical green dashed lines.}
\label{fig:stack}
\end{figure*}

Following the method of \citetalias{Pieri2010Stacking} and \citetalias{Pieri2014}, in order to calculate the arithmetic mean, we sigma clip the high and low 3\% of the stack of spectra to reduce our sensitivity to outliers. We also allow that the overwhelming majority of the absorption in the spectra are not associated with the selected SBLAs. These unassociated absorbers do not generate any absorption features correlated with our selected \lya, but they do have an impact on  the mean stacked spectrum.  When a mean stacked spectrum is calculated, a broad suppression of transmitted flux is seen (see \autoref{fig:stack}). Since this absorption is not associated with the selected systems, it is therefore undesirable in the pursuit of a composite absorption spectrum of the selected systems. 

The stacked flux departs from unity even in regions where Lyman-series and metals 
features are not expected despite the fact that each spectrum was continuum normalised before being stacked \autoref{fig:stack}. These broad flux variations result mainly from the smoothly varying average contribution of uncorrelated absorption. The artefacts of the stacking procedure are unwanted in a composite spectrum of the selected systems but vary smoothly enough that one can distinguish them from absorption features of interest. Since they are analogous to  quasar continua, 
\citetalias{Pieri2010Stacking} gave these artefacts in the stacked spectra the name `pseudo-continua'.
They argued that the effect of this contamination in the mean stacked spectrum can be best approximated by an additive factor in flux decrement. This is because quasar absorption lines are  narrower than the SDSS resolution element and hence would typically be separable lines in perfect spectral resolution. These uncorrelated absorbers are present on either side of the feature of interest and it is reasonable to assume that they will continue through the feature of interest contributing to the absorption in every pixel of the feature on average without typically occupying the same true, resolved redshift range. In this regime each contributing absorber makes an additive contributions to the flux decrement in a pixel. The alternative regime where absorption is additive in opacity, leads to a multiplicative correction, but weak absorption features (such as those we measure here ) are insensitive to the choice of a multiplicative or additive correction. In light of these two factors we continue under the approximation of additive contaminating absorption.

We therefore arrive at a composite spectrum of SBLAs by correcting the stacked spectrum using
\begin{equation}
F_C (\lambda_r) = F_S + (1 - P),
\end{equation}
where (again) $F_S$ represents the mean stacked flux and $(1-P)$ represents the flux decrement of the `pseudo-continuum' and can be estimated by fitting a spline through flux nodes representing this pseudo-continuum. 

\begin{figure*}
\begin{center}
\includegraphics[angle=0, width=.99\textwidth]{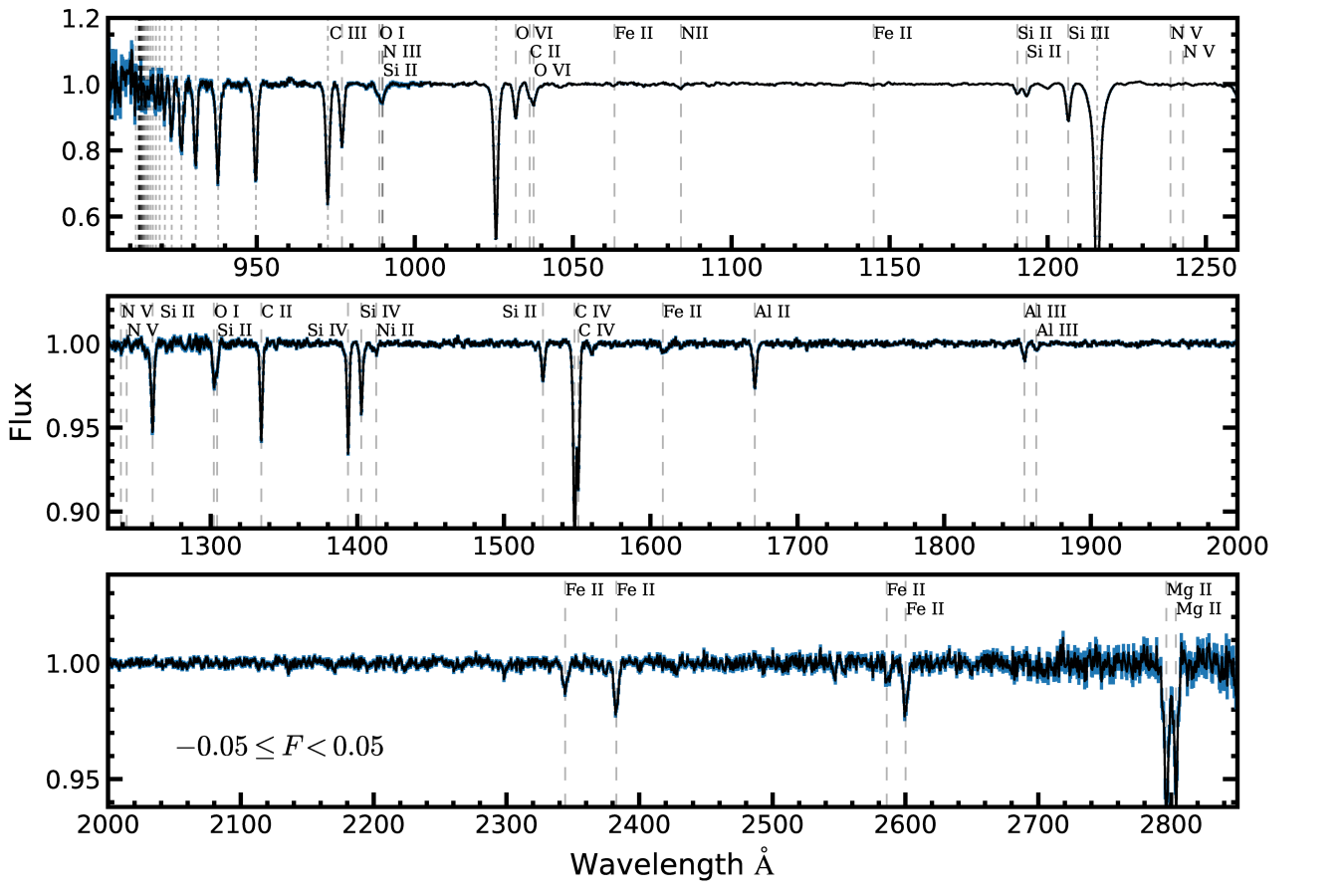}
\end{center}
\caption{Composite spectrum of the SBLA system sample FS0 (systems selected with flux between $-0.05 \le F<0.05$) produced using the arithmetic mean statistic. Error bars are shown in blue. Vertical dashed lines indicate metal lines identified \citep{Cashman2017,Morton2003} and dotted vertical lines denote the locations of the Lyman series. Note the scale of the y-axis in each panel: this is our lowest S/N composite spectrum and yet we measure absorption features with depth as small as 0.0005.}
\label{fig:composite}
\end{figure*}

To calculate these nodes we first manually select regions of the stacked spectrum in areas where signal from correlated absorption is not seen and/or expected. Then for each such `pseudo-continuum patch', we define the corresponding node using the mean of flux and wavelength values of all stacked pixels within this patch. In estimating the pseudo-continuum we typically use $\sim$~10~\AA\ wide "patches" of spectrum. However, smaller continuum patches were used in regions crowded by correlated absorption features, while much wider segments were selected for relatively flat regions of the stacked spectrum. \autoref{fig:stack} shows the pseudo-continuum along with the regions used to estimate it for the mean stacked spectrum corresponding to FS0. The corresponding composite spectrum is shown in \autoref{fig:composite}.

\section{Improved estimations of measurement uncertainty}
\label{sec:errore2e}

In this work, we explore a more inclusive treatment of measurement uncertainty than \citetalias{Pieri2010Stacking} and \citetalias{Pieri2014} allowing more reliable fits and more quantitative model comparison. We will initially summarise the previous method in order to expand on our more precise error estimations.

\subsection{Quick bootstrap method}
\label{subsec:quickboot}

In \citetalias{Pieri2010Stacking} and \citetalias{Pieri2014} the errors were estimated for the stacked spectrum alone, i.e. prior to the pseudo-continuum normalisation step above. In taking this approach, they did not formally include the potential contribution to the uncertainty of the pseudo-continuum normalisation. Instead they took the conservative choice to scale the errors generated by the bootstrap method by a factor of root-2 assuming that pseudo-continuum fitting introduced an equal contribution to the uncertainty of the final composite spectrum. 

Errors in the stacked spectrum were estimated by bootstrapping the stack of spectra.
At every wavelength bin in the stack, 100 bootstrap realisations were produced and the error was calculated as the standard deviation of the means calculated from those random realisations. This was performed independently for each bin.  In the process of exploring improved estimates of uncertainty in the composite spectrum of \lya\ forest systems, we have learned that 100 realisations is not a sufficient number for precision error estimates. Based on these convergence tests we advocate generating 1,000 realisations to have high confidence of accuracy. See Appendix~\ref{extra:errore2e} for more detail on this choice.

\subsection{End-to-end bootstrap method}
\label{subsec:End2Endboot}

In this work we wish to relax the assumption of \citetalias{Pieri2014} that pseudo-continuum fitting introduces an uncertainty to the composite spectrum equal to, but uncorrelated with, the combination of other sources of error. 
In order to do this, we seek to estimate the errors from the telescope all the way to final data analysis step of producing a composite spectrum.
In order to build an end-to-end error estimation framework we begin by bootstrapping the sample of SBLAs and their accompanying spectra. For each random realisation of the sample, we construct a realisation of the stacked spectrum following the same approach as that in the quick bootstrap method. The key difference is that we do not simply calculate an uncertainty in the stacked spectrum and propagate it forward analytically through the pseudo-continuum normalisation to the composite spectrum. Instead we include this process in the bootstrap analysis by performing the pseudo-continuum fit and normalisation upon each realisation.

\begin{figure*}
\begin{center}
\includegraphics[angle=0, width=.99\textwidth]{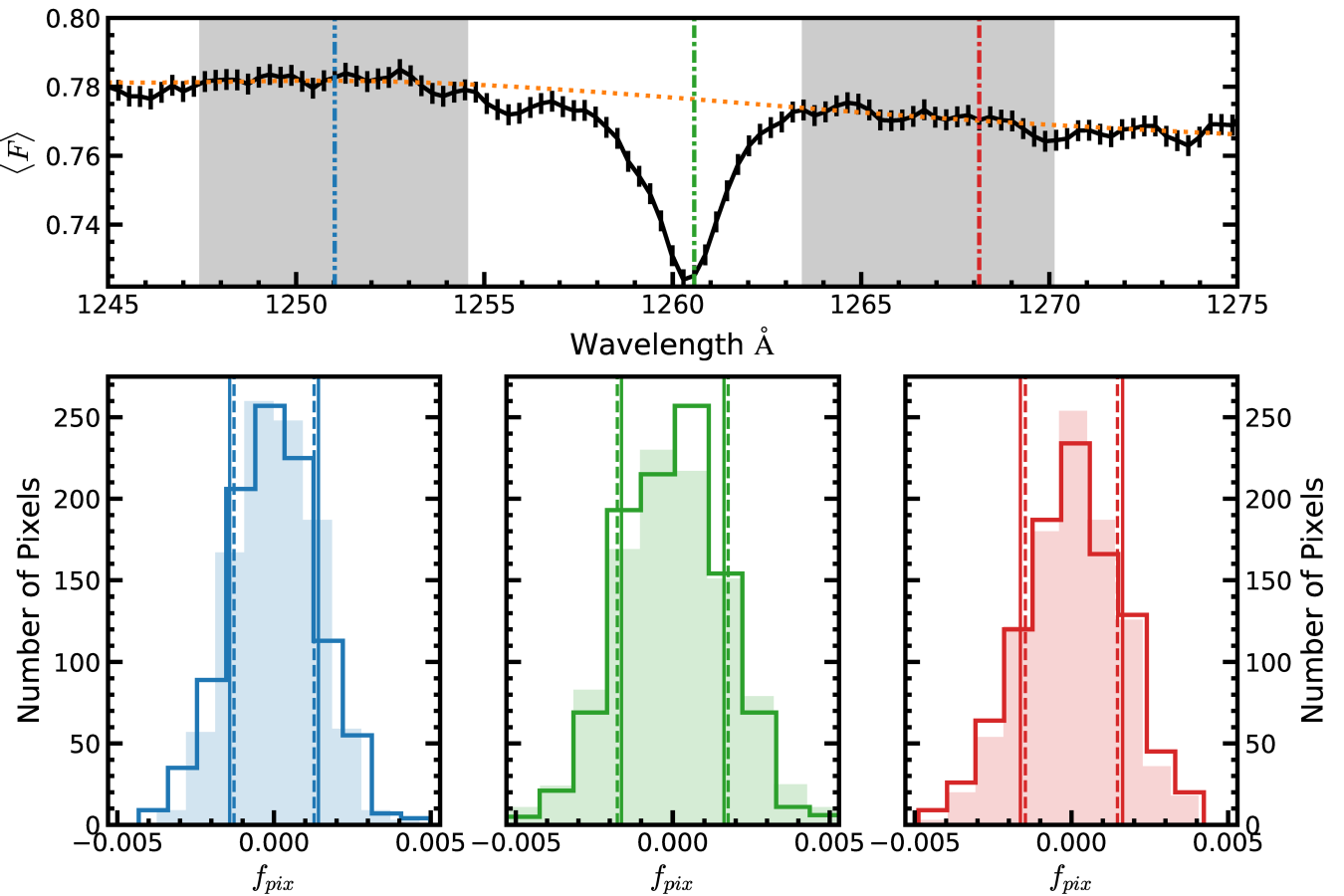}
\end{center}
\caption{Illustration of the end-to-end error estimation mechanism using a regions of the FS0 stack around the \SiII\ $\lambda 1260$ absorption feature. {\bf Top row:} The stacked spectrum around the absorption feature centred at 1260\AA\ is shown using a black curve. The shaded grey regions represent a pair of continuum patches on either sides of the feature. The pseudo-continuum is also shown using the orange dashed curve. The green, blue and red vertical lines mark the locations of three pixels chosen for illustration: the pixel at the centre of the absorption feature and the pixels at the midpoints of the continuum patches located on the left and right of the feature, respectively. {\bf Bottom row:} Each panel shows the distributions of the stacked and composite flux across all the realisations at one of the pixels marked in the upper panel. The wavelength of each distribution is indicated by their colour and the colour of the dot-dash line in the top panel.
The distributions are shown on a linearly shifted flux scale so that the mean of each distribution corresponds to $f_{pix}$ = 0. The stacked flux distribution is shown using a open histogram while the composite flux distribution is shown using a shaded histogram and their corresponding standard deviations are shown using vertical solid and dashed lines, respectively.}
\label{fig:e2eError}
\end{figure*}

\begin{figure}
\begin{center}
\includegraphics[angle=0, width=.99\linewidth]{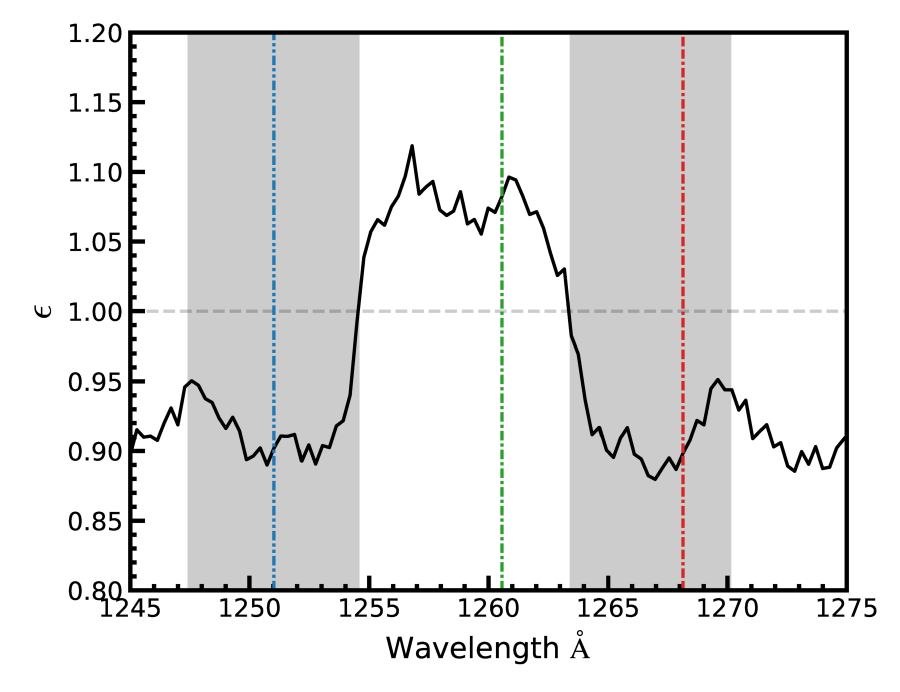}
\end{center}
\caption{The ratio, $\epsilon$,  between 1$\sigma$ error in the composite flux ($\sigma_{F_C}$) to that of the stacked flux ($\sigma_{F_S}$) for the FS0 sample is plotted over the region around the \SiII\ $\lambda 1260$ feature. The shaded grey regions represent a pair of continuum patches on either side of the feature. The vertical lines correspond to the locations of the pixels marked in \autoref{fig:e2eError}.
}
\label{fig:errrat}
\end{figure}

The patches used to fit the pseudo-continuum of our observed stacked spectrum (as described in \autoref{sec:stackprocess}) were applied to each of the bootstrap realisations to obtain spline nodes for a unique pseudo-continuum per realisation. This created an ensemble of 1,000 bootstrapped realisations of the (pseudo-continuum normalised) composite spectrum, $(\tilde{F_{C}})_i$, where $i$ denotes the $i$th bootstrap realisation at every wavelength. Finally, the error in the composite flux $\sigma_{F_C}$ is estimated to be the standard deviation of the  ensemble  $(\tilde{F_{C}})_i$ at every wavelength. The resulting uncertainties in the composite flux derived using the end-to-end error estimation method are shown in \autoref{fig:composite} using blue error bars.

\autoref{fig:e2eError} illustrates the end-to-end error estimation mechanism taking a region of the stack around the \SiII\ $\lambda 1260$ absorption signal. The stack is shown in the top panel of the figure along with a pair of continuum patches on either sides of the absorption feature as well as the pseudo-continuum estimate. This panel also marks the locations of three pixels chosen as example to illustrate the method: the pixel at the centre of the absorption feature and the pixels at the middle of the continuum patches on the `blue' and `red' side of the feature. The panels in the bottom row of \autoref{fig:e2eError} show the distribution of realisations for the stacked spectrum (open histogram) and composite spectrum  (filled histogram). For convenience of comparison, each distribution is plotted with respect to that distribution's mean (i.e $f_{pix, i} = (\tilde{F_C})_i - \langle \tilde{F_C} \rangle$ or $f_{pix, i} = (\tilde{F_S})_i - \langle \tilde{F_S} \rangle$). The wavelength for each distribution is indicated by the vertical dot-dash line of matching colour in the top panel. 
 The interval described by the standard deviation of each distribution is indicated using vertical solid lines for the stacked spectrum ($\pm \sigma_{F_S}$) and vertical dashed lines for the composite spectrum ($\pm \sigma_{F_S}$).

We can further compare the uncertainty derived for the composite spectrum and the stacked spectrum through the ratio $\epsilon$ = $\sigma_{F_C} / \sigma_{F_S}$) as a function of wavelength. An $\epsilon > 1$ indicates that uncertainty is increased by the pseudo-continuum fitting, whereas $\epsilon < 1$ indicates that pseudo-continuum fitting is suppressing variance. We again take the example of \SiII\ $\lambda 1260$ and show $\epsilon$ as a function of wavelength in \autoref{fig:errrat}. As illustrated for \SiII\ $\lambda 1260$, line absorption features show an additional uncertainty and the regions between them show variance suppression. The latter is to be expected because the pseudo-continuum fitting suppresses large-scale deviations in uncorrelated absorption by erasing low order modes in the spectra. On the other hand the absorption features themselves are free to deviate and show the increased uncertainty of interest. 
The value of $\epsilon$ for every measured metal line measurement bin is provided, as we shall see in \autoref{tab:metalmeasmainFS0}.
The pseudo-continuum normalisation process does increase the uncertainty at the absorption locations, but the increase is smaller than the  41\% increase implied by the root-2 assumption of \citetalias{Pieri2014}. Only \CIII\ $\lambda 977$ and \SiII\ $\lambda 1190$ show a greater than 10\% increase in errors and so overall a more accurate (but less conservative) error estimate would have been to neglect the contribution of pseudo-continuum fitting.
We note, however, that the degree of noise suppression in feature free regions and the degree of noise inflation at absorption feature centres are both dependent on the placement and width of patches are used to generate spline nodes (shown in \autoref{fig:stack}). Therefore we advise caution if using quick bootstraps with these $\epsilon$ measurements as correction factors, if precise error estimates are needed. The placement of these patches may change if absorption features are broader/narrower than the results presented here, leading to changes in $\epsilon$.

\section{Measurement of the SBLA halo mass}
\label{sec:sblamass}

We cross-correlate the main FS0 sample of SBLAs with the \lya\ forest in order to measure large-scale structure bias, and constrain SBLA halo mass. The \lya\ forest is prepared in a distinct way for this analysis using the standard method developed for correlation function analyses, as outlined in our companion paper 
(\citealt{Perez-Rafols2022}, hereafter \citetalias{Perez-Rafols2022}).
We summarise the data preparation briefly in Appendix~\ref{extra:corrmethod} and refer the reader to that paper for a detailed discussion. 

\begin{figure}
   \begin{center}
   \includegraphics[width=0.9\linewidth]{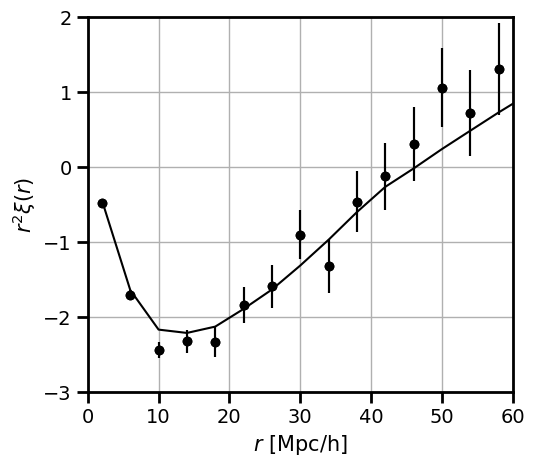}
   \caption{Cross-correlation function averaged over the full angular range $0<|\mu|<1$ for the fitting range $10<r<80~\hMpc$. The solid line shows the best-fit model.}
   \label{fig:slalya}
   \end{center}
\end{figure}

\autoref{fig:slalya} shows the measured cross-correlation and the best-fit model. The best fit has $\chi^{2}=5060.602$ for 4904 degrees of freedom (probability $p=0.058$). The best-fit value of the SBLA bias parameter is
\begin{equation}
b_{\rm SBLA} = 2.34\pm0.06 ,
\end{equation}
where the quoted uncertainty only includes the stochastic errors. The recovered  $b_{\rm SBLA}$ value is consistent with that found by \citetalias{Perez-Rafols2022}. 
If all SBLAs were sited on halos of a single mass, this mass would be $\sim7.8 \times 10^{11}{\rm h^{-1}M_{\sun}}$.
However, SBLAs are likely found in halos with a range of masses. Following what \cite{Perez-Rafols2018} proposed for DLAs (see their equations 15 and 16 and their figure 8), a plausible distribution of the SBLA cross-section, $\Sigma\left(M_{h}\right)$,  is a power law in halo mass, starting with some minimal halo mass:
\begin{equation}
    \Sigma\left(M_{h}\right) = \Sigma_{0}\left(\frac{M_{h}}{M_{\rm min}}\right)^{-\alpha}\,\left(M_{h} > M_{\rm min}\right) ~.
\end{equation}
Using this cross-section, the mean halo mass is computed as
\begin{equation}
    \overline{M_{h}} = \frac{\int_{M_{\rm min}}^{\infty} n(M)\Sigma(M)M{\rm d}M}{\int_{M_{\rm min}}^{\infty} n(M)\Sigma(M){\rm d}M} ~,
\end{equation}
where $n\left(M\right)$ is the number density of halos for a given mass. For plausible values of $\alpha=0.50$, $0.75$ and $1.00$ this yields a mean mass of $1.3\times10^{12}{\rm h^{-1}M_{\sun}}$, $9.4\times10^{11}{\rm h^{-1}M_{\sun}}$, and $7.6\times10^{11}{\rm h^{-1}M_{\sun}}$ respectively. 
We note that a detailed study of this cross-section using simulations is necessary 
to make more accurate mass estimates, but our finding indicate that SBLAs reside in halos of mass $\approx 10^{12}{\rm h^{-1}M_{\sun}}$.

It is informative to compare this with order of magnitude estimates of the halo mass derived by assuming that the width of the SBLA line blend is driven by the circular velocity of virialised halo gas undergoing collapse. This connection between halo circular velocity, halo virial mass, and galaxy populations has been well-explored (e.g. \citealt{ThoulWeinberg1996}). Specifically we apply the relationship between maximal circular velocity and halo mass modelled by \citealt{Zehavi2019}. Using these relations, we infer that a circular velocity of $138\kms$ at $z\sim2.4$ leads to halo mass estimate of $M_h \sim 3 \times 10^{11}{\rm h^{-1}M_{\sun}}$. This value is broadly consistent with our findings from SBLA clustering, supporting our assumption that blending scale is associated with halo circular velocity and so halo mass. This may shed some light on the reason why SBLAs are CGM regions.

\section{Average SBLA absorption properties}
\label{sec:interpret_comp}

As one can see in \autoref{fig:composite}, absorption signal is measurable in the composite spectrum from a a wide range of transitions: Lyman-series lines (\lya ~- {Ly$\theta$}) and metal lines (\OI, \OVI, \CII, \CIII, \CIV, \SiII, \SiIII, \SiIV, \NV, \FeII, \AlII, \AlIII, and \MgII), but care must be taken to measure them in a way that is self-consistent and without bias. Although these features appear to be absorption lines, they are in fact a complex mix of effects that precludes the naive application of standard absorption line analysis methods appropriate for individual spectrum studies.

\citetalias{Pieri2014} demonstrated that the main difference in interpretation of the 3 potentially CGM dependent samples (which we have named) FS0, FS1 and FS2 was the purity of CGM selection in light of spectral noise given the large excess of pixels with higher transmission that might pollute the sample. Since FS0 has the lowest transmission, it is the purest of these samples. Hence, in this work directed at understanding CGM properties, we focus on interpreting FS0 sample properties.

Throughout this work we only present lines measured with 5$\sigma$ significance or greater. \NV, for example, fails to meet this requirement and is not included in the measurements presented below.

\subsection{Line-of-sight integration scale}
\label{subsec:integration-scale}
There are two approaches to the measurement of absorption features seen in the composite spectra (as identified in \citetalias{Pieri2014}); the measurement of the full profile of the feature and the measurement of the central pixel (or more accurately resolution element). In order to understand this choice, it is necessary to reflect, briefly, on the elements that give rise to the shape and strength of the features.

The signal present for every absorption feature is a combination of 
\begin{itemize}
\item the absorption signal directly associated with the selected \lya\ absorption,
\item possible associated absorption complexes extending over larger velocities  (typically associated with gas flows, often with many components), and
\item sensitivity to large-scale structure (including redshift-space distortions) reflected in the well-documented (e.g \citealt{Chabanier2019}) fact that \lya\ forest absorption is clustered, leading to potential clustering in associated absorbers also (e.g \citealt{Blomqvist2018}).
\end{itemize}
In large-scale structure terminology the first two points are `one-halo' terms and the last one is a `two-halo' term. This two-halo effect is clearly visible in the wide wings of the \lya\ absorption feature extending over several thousand $\kms$. Since the metal features seen are associated with \lya\ every one must present an analogous (albeit weak) signal due to the clustering of SBLA. Although this large-scale structure signal is present in the composite, our stacking analysis is poorly adapted to the measurement of large-scale structure since the signal is degenerate with the pseudo-continuum fitting used, and the preferred measurement framework for this signal is the \lya\ forest power spectrum \citep{McDonald2006}.

As outlined in \autoref{sec:sampleselection}, the selection of SBLAs to be stacked includes clustering and therefore both complexes and large-scale structure. Therefore even the central pixel includes all the above effects to some extent but limiting ourselves to the measurement of the central pixel sets a common velocity integration scale for absorption measurement. In fact, since the  resolution of SDSS is 2.4 pixels, the appropriate common velocity scale is two native SDSS pixels. We therefore take the average of the two native pixels with wavelengths closest to the rest frame wavelength of the transition in question as our analysis pixel. This sets the integration scale fixed to 138$\kms$. This mirrors the \lya\ selection function bin scale which is also a 2-pixel average (see \autoref{sec:sampleselection}). The error estimate for the flux transmission of this double width wavelength bin is taken as the quadrature sum of the uncertainty for the two pixels in question (a conservative approximation that neglects the fact that errors in neighbouring pixels are correlated due to pipeline and analysis steps such as pseudo-continuum fitting). Here after we will use `central bin' to refer to this 2-pixel average centred around the rest frame wavelength of the transition of interest.

In contrast \citetalias{Pieri2014} showed that measuring the full profile of the features leads to a different velocity width for every feature indicating either varying sensitivity to these effects or tracing different extended complexes. Critically this means that some absorption must be coming from physically different gas. Since the objective of this work is the formal measurement and interpretation of the systems selected,
we limit ourselves to the central analysis pixels at the standard rest frame wavelength of each transition. We note, however, that information is present in the composite spectra on the velocity scale of metal complexes and this demands further study if it can be disentangled from large-scale structure.

\subsection{\texorpdfstring{Measuring the \HI\ Column density}{Modelling H I absorption}}
\label{subsec:fith1}

Here we compare Lyman series line measurements in the composite spectrum with a variety of models in order to constrain the column density and Doppler parameter. As we have stressed throughout this work, our SBLA samples are a blend of unresolved lines contributing to a 138$\kms$ central bin. As a result a range of \HI\ column densities are present in each SBLA. While the full range of \HI\ columns contribute to the selection, it is reasonable to presume that a high column density subset dominate the signal in the composite. It is, therefore, natural that the further we climb up the Lyman series, the more we converge on a signal driven by this dominant high-column subset. Here we exploit this expected convergence to jointly constrain the integrated dominant \HI\ column density ($\mathrm{N_\HI}$) of lines in the blend and their typical Doppler parameter ($b$).

In the following, the results are presented as equivalent widths to follow standard practise, but the measurements are in fact central bin flux decrements ($1-F_C$) multiplied by the wavelength interval corresponding to the 138$\kms$ central bin interval. In effect, the equivalent widths presented are the integrated equivalent widths of all lines contributing to that central bin measurement.

We build a grid of model\footnote{Produced using VPFIT 10.0 \citep{VPFIT}} equivalent widths for the eight strongest Lyman transitions over the range $13.0\leq \log{\mathrm{N_\HI}} (\mathrm{cm}^{-2}) \leq 21.0$ with interval $\delta\log{\mathrm{N_\HI}}(\mathrm{cm}^{-2})=0.01$, and $5.0 \leq b (\kms) \leq 50.0$ with interval $\delta b=0.1 \kms$. These models are built for the composite spectrum wavelength solution and include instrumental broadening of $167 \kms$. 

In order to measure the dominant \HI\ contribution, we must determine which of the Lyman series lines should be treated as upper limits progressively, starting with \Lya\ and moving up the series until a converged  single line solution of satisfactory probability is reached.
For each line considered as upper limit, if the model prediction lies 1$\sigma$ equivalent width error above the measured equivalent width, the line contributes to the total $\chi^2$ for the model and one degree of freedom gets added to the number of degrees of freedom for the model. If the model prediction lies below this threshold, it does not contribute to the total $\chi^2$ and the number of degrees of freedom for the model remain unchanged. This process `punishes' the overproducing models instead of rejecting them.

The probability 
for each model is calculated based on the total $\chi^2$ and the updated number of degrees of freedom. The best-fit model for a given upper-limit assignment scheme is determined by maximising the probability. The best-fit probabilities, $N$ and $b$-values corresponding to the different upper-limit assignment schemes are compared to determine the number of lowest order Lyman lines assigned to upper limits ($N_{ul}$) necessary to achieve a converged probability.

\begin{figure}
\begin{center}
\includegraphics[angle=0, width=1.\linewidth]{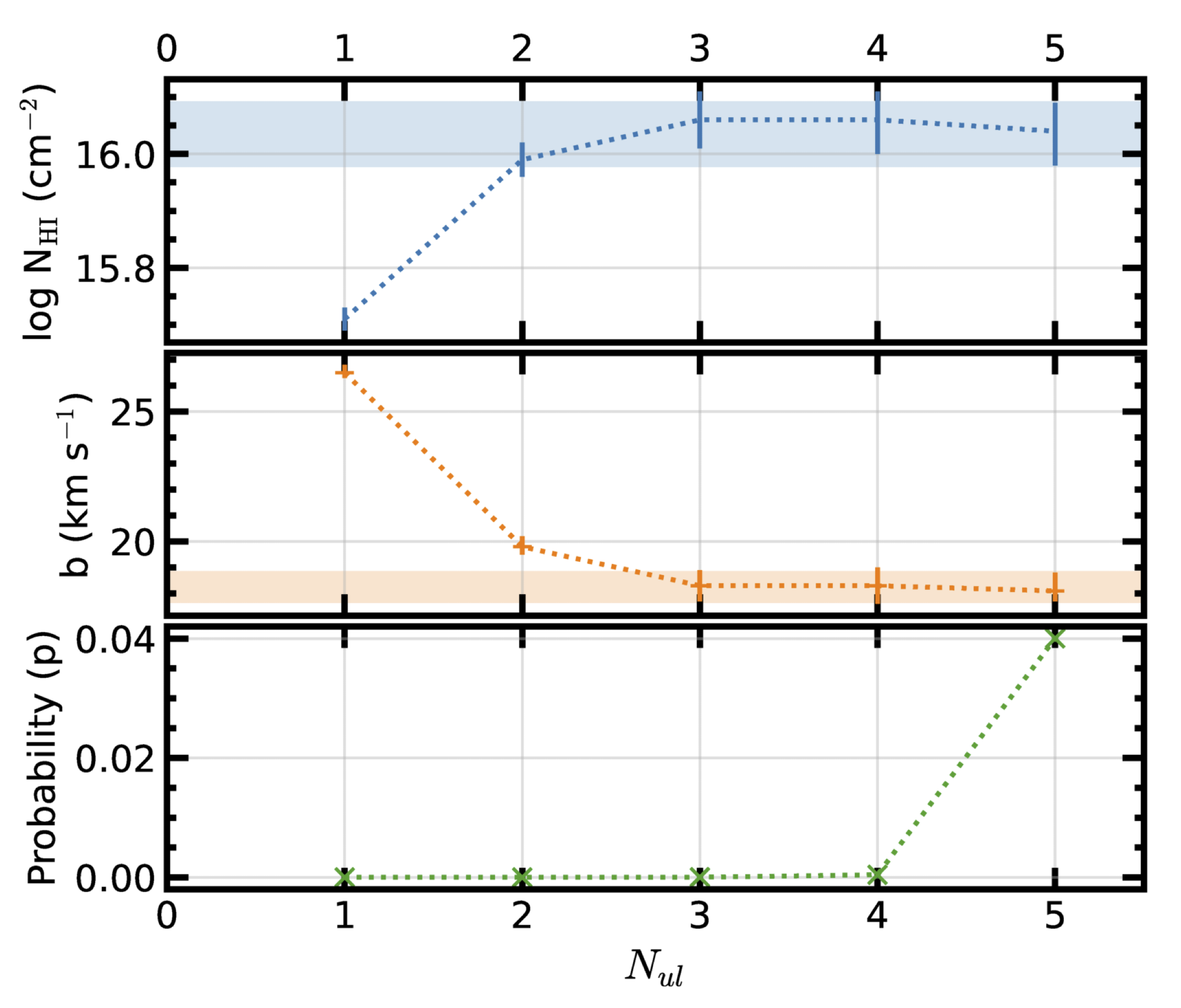}
\end{center}
\caption{Test of \HI\ Lyman series upper limits (starting with \lya\ as an upper limit and progressively adding higher order Lyman lines) for convergence to determine best fit model parameters for the FS0 composite. The shaded bands represent the final best fit parameters for $\log{\mathrm{N_\HI}}$ (top, blue) and $b$ (middle, red). The probability (of a higher $\chi^2$) for each best-fit model, as a function of the number of upper limits, is given in the bottom panel (green).}
\label{fig:h1analysis_FS0}
\end{figure}

\begin{figure}
\begin{center}
\includegraphics[width=0.99\linewidth]{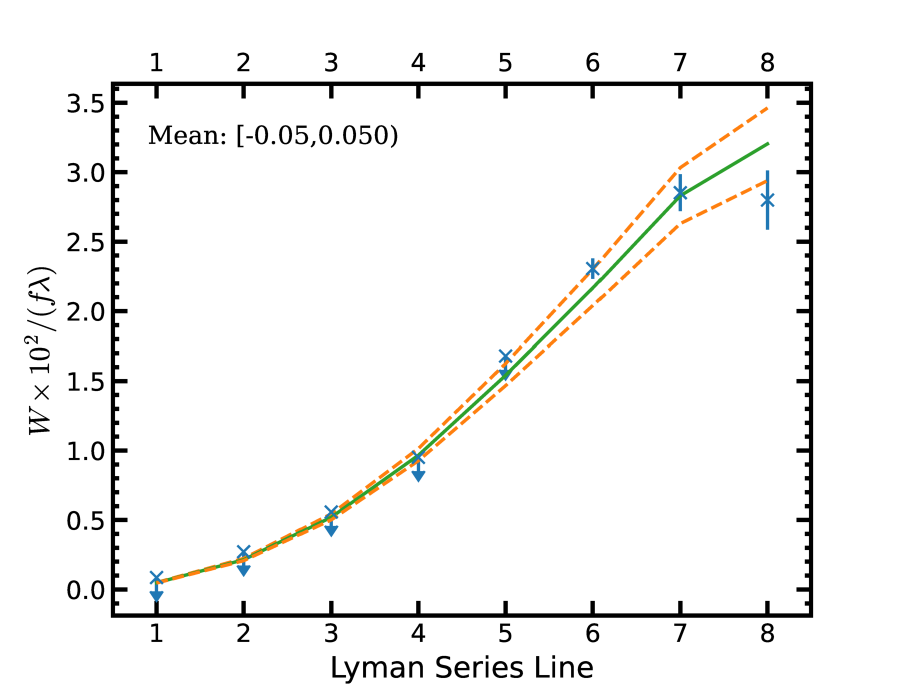}
\end{center}
\caption{The best fit \HI\ model ({\it green solid} line) and the limiting $\pm 1\sigma$ allowed models ({\it orange dashed} line) compared to Lyman series equivalent width measurements for the FS0 sample. The upper limits reflect the convergence described in the text and illustrated in \autoref{fig:h1analysis_FS0}. 
}
\label{fig:h1fit_FS0}
\end{figure}

The convergence for the FS0 sample is shown in \autoref{fig:h1analysis_FS0}. The model that corresponds to the convergence is chosen as the best-fit model for the \HI\ column density and Doppler parameter for that sample. \autoref{fig:h1fit_FS0} shows the measured equivalent widths ($W$) normalised by the oscillator strength ($f$) and rest frame wavelength ($\lambda$) for each transition for the FS0 sample. Also shown is the best-fit model,
along with models for the $1\sigma$ upper and lower confidence intervals on the dominant \HI\ column density. Note that when plotted this way, unsaturated lines would produce a constant $W/(F\lambda)$, and so the dominant \HI\ population is only beginning to show unsaturated properties for the highest Lyman series transitions measured.

\autoref{tab:hicolumns} shows the fit results for this procedure. The differences in measured column densities between FS0, FS1, and FS2 demonstrate that, along with decreasing purity of noiseless $F_{\textrm{\lya}} < 0.25$, higher transmission bands also select lower column densities. The P90, P75 and P30 samples show a similar trend but show a weaker variation in \HI\ column density along with a weaker decline in mean purity. This combined with the large numbers of systems selected indicates that these purity cuts do indeed provide more optimal SBLA samples. While we chose to focus on FS0 in order to preserve sample continuity for comparison with previous work, we recommend a transition to such optimised selection in future work. This supports the choice taken in \citet{Perez-Rafols2022} to use the P30 sample.

\begin{table}
 \caption{Inferred \HI\ column densities from Lyman series measurements.}
 \label{tab:hicolumns}
 \begin{center}
 \begin{tabular}{@{}lcccc}
  \hline
  Sample & $\log{\mathrm{N_\HI}} (\mathrm{cm}^{-2})$  & $b (\mathrm{km\,s}^{-1})$ & $N_{ul}$ & Prob \\
  \hline
  \hline
    FS0 &   $16.04$\ullim{0.06}{0.05} & $18.1$\ullim{0.4}{0.7} & 5 & 0.04 \\
    FS1 &   $15.64$\ullim{0.04}{0.04} & $12.3$\ullim{0.4}{0.6} & 3 & 0.6 \\
    FS2 &   $15.11$\ullim{0.07}{0.06} & $8.5$\ullim{0.5}{1.10} & 5 & 0.13   \\
    P30 &   $15.49$\ullim{0.03}{0.06} & $10.8$\ullim{1.5}{0.4} & 5 & 0.4\\
    P75 &   $15.67$\ullim{0.03}{0.04} & $13.5$\ullim{0.8}{0.3} & 5 & 0.27 \\
    P90 &   $15.79$\ullim{0.04}{0.05} & $14.6$\ullim{1.2}{0.6} & 5 & 0.37 \\
\hline
 \end{tabular}
 \end{center}
 \end{table}
 
\subsection{Average Metal Column Densities}
\label{subsec:metalcolumn}

Unlike the \HI\ measurement above, metal features in the composite are sufficiently weak that  several metal transitions are not necessary to establish a reliable column density. However, the combination of line strength and measurement precision means that the small opacity approximation (that the relationship is linear between equivalent width and column density) is inadequate for our needs. Again given that we lack a large number of metal transitions with a wide dynamic range of transition strengths for each metal species, a suite of model lines (as performed for \HI) is not necessary. We instead fit them directly with column density the only free parameter, treating each feature as fully independent or one another. We assume a Doppler parameter value taken from the \HI\ measurement (see below). 
We fit the mean of the pair of pixels nearest to the transition wavelength with  instrumental broadening set to $167 \kms$ using VPFIT. Since VPFIT was not designed to reliably assess the uncertainty in the column density from a single pixel at time, we pass the upper and lower 1$\sigma$ error envelope through VPFIT for every line to obtain $N_{min}$ and $N_{max}$ respectively. The measurements for our main sample (FS0) are given \autoref{tab:metalmeasmainFS0}. 

\begin{table*}
 \caption{Mean metal columns for the main sample, FS0.}
 \label{tab:metalmeasmainFS0}
 \begin{tabular}{@{}lcccccccc}
  \hline
  Ion &   Wavelength (\AA) & Ionization Potential (eV) & $F_C$ & $\sigma_{F_C}$ & $\epsilon$ & $\log\mathrm{N}(\mathrm{cm}^{-2})$ & $\log \mathrm{N_{max}}(\mathrm{cm}^{-2})$ &  $\log \mathrm{N_{min}} (\mathrm{cm}^{-2})$  \\
  \hline
      \OI\ & 1302.17 &   13.6 &  0.9743 & 0.0011 & 1.084 & 13.470 & 13.449 & 13.489 \\
    \MgII\ & 2796.35 &   15.0 &  0.9376 & 0.0031 & 1.034 & 12.450 & 12.424 & 12.474 \\
    \MgII\ & 2803.53 &   15.0 &  0.9404 & 0.0031 & 1.043 & 12.729 & 12.703 & 12.754 \\
    \FeII\ & 1608.45 &   16.2 &  0.9956 & 0.0007 & 1.020 & 12.509 & 12.433 & 12.573 \\
    \FeII\ & 2344.21 &   16.2 &  0.9878 & 0.0009 & 1.042 & 12.499 & 12.467 & 12.530 \\
    \FeII\ & 2382.76 &   16.2 &  0.9807 & 0.0009 & 1.032 & 12.252 & 12.231 & 12.272 \\
    \FeII\ & 2586.65 &   16.2 &  0.9932 & 0.0013 & 1.041 & 12.415 & 12.321 & 12.493 \\
    \FeII\ & 2600.17 &   16.2 &  0.9798 & 0.0014 & 1.031 & 12.361 & 12.329 & 12.390 \\
    \SiII\ & 1190.42 &   16.3 &  0.9709 & 0.0010 & 1.147 & 12.780 & 12.765 & 12.795 \\
    \SiII\ & 1193.29 &   16.3 &  0.9643 & 0.0010 & 1.165 & 12.574 & 12.561 & 12.586 \\
    \SiII\ & 1260.42 &   16.3 &  0.9481 & 0.0012 & 1.082 & 12.422 & 12.411 & 12.433 \\
    \SiII\ & 1304.37 &   16.3 &  0.9823 & 0.0010 & 1.076 & 13.044 & 13.017 & 13.069 \\
    \SiII\ & 1526.71 &   16.3 &  0.9780 & 0.0006 & 1.032 & 12.886 & 12.872 & 12.899 \\
    \AlII\ & 1670.79 &   18.8 &  0.9740 & 0.0007 & 1.020 & 11.806 & 11.795 & 11.817 \\
    \CII\ & 1334.53 &   24.4 &  0.9428 & 0.0010 & 1.019 & 13.410 & 13.401 & 13.418 \\
   \AlIII\ & 1854.72 &   28.4 &  0.9904 & 0.0005 & 1.031 & 11.805 & 11.780 & 11.828 \\
   \AlIII\ & 1862.79 &   28.4 &  0.9965 & 0.0005 & 1.035 & 11.661 & 11.590 & 11.722 \\
   \SiIII\ & 1206.50 &   33.5 &  0.8904 & 0.0010 & 1.057 & 12.690 & 12.685 & 12.696 \\
    \SiIV\ & 1393.76 &   45.1 &  0.9367 & 0.0007 & 1.016 & 12.838 & 12.832 & 12.844 \\
    \CIII\ &  977.02 &   47.9 &  0.8180 & 0.0025 & 1.259 & 13.444 & 13.434 & 13.455 \\
     \CIV\ & 1548.20 &   64.5 &  0.8764 & 0.0008 & 1.029 & 13.586 & 13.582 & 13.590 \\
     \OVI\ & 1031.93 &  138.1 &  0.8994 & 0.0014 & 1.084 & 13.799 & 13.792 & 13.807 \\

\hline
 \end{tabular}
 \end{table*}

\begin{figure}
    \centering  
    \includegraphics[width=0.9\linewidth] {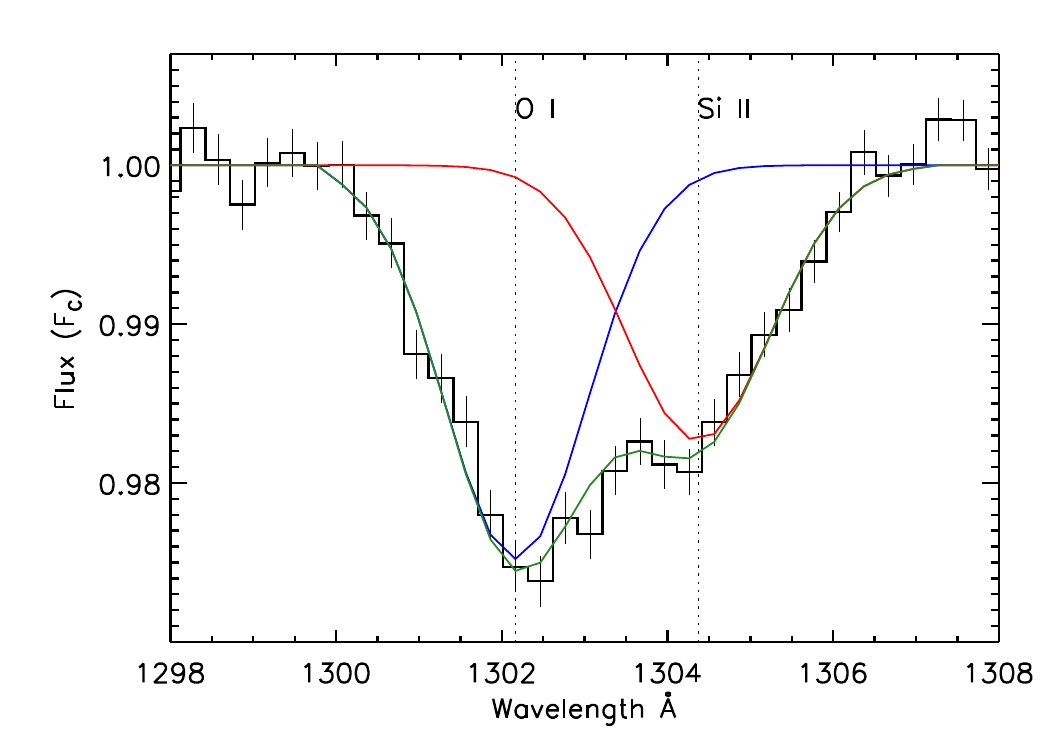}
    \caption{The contribution of \SiII \ $\lambda$1304 to the central bin measurement of  \OI\ $\lambda$ 1302 and vice versa. The {\it blue} curve is the fit to the portion of the \OI\ feature that is \SiII-free (the blue-side of the profile). The {\it red} curve is the fit to the  portion of the \SiII\ feature that is \OI-free (the red-side of the profile). The green curve is the joint fit of the full profiles of both features. The full profile fit is only used to measure the contribution to the measurement bin of the neighbouring line. As discussed in \autoref{subsec:integration-scale}, we do not use the full profile measurement of features in this work. }
    \label{fig:blendoi}
\end{figure}

We exclude from our analysis all transitions where there is a significant contribution to the central 138$\kms$ by the broad wing of a neighbouring feature. In principal, it is possible to fit the superposed features, correct for the profile of the unwanted feature and measure the 138$\kms$ central core of the desired line, but these blended features are incompatible with the population modelling procedure that follows and so are of limited value. Examples of cases where a broad feature wing contaminates the desired feature centre (and are hence discarded) are  \OI\ $\lambda 989 $, \NIII\ $\lambda 990$ and \SiII\ $\lambda$990, and \CII\ $\lambda$1036 and \OVI\ $\lambda$1037. On the other hand  \OI\ $\lambda$1302 and  \SiII\ $\lambda$1304 are retained in our analysis despite being partially blended in our composite spectrum. The contribution of the \SiII\ $\lambda$1304 feature wing to the central \OI\ analysis bin is 3\% of the observed flux decrement.  The \OI\ feature wing contributes 6\% to the observed flux decrement to the \SiII\ $\lambda$1304 measurement. This is illustrated in \autoref{fig:blendoi}. In each case spectral error estimate is similar to the size of the contamination. As we shall see in \autoref{sec:interpret_comp} the error estimates of the composite are too small for any true model fit and instead the limiting factor is the much larger uncertainty in the population model fits of \autoref{sec:abspop}.

Another consequence of our inability to resolve the individual lines that give rise to our metal features (and our lack of a dynamic range of transition strengths) is that we lack the ability to constrain the Doppler broadening parameter. However, we do have a statistical measurement of the Doppler parameter of systems that dominate the blend selected. This is the value of the Doppler parameter obtained from the \HI\ measurement. While the measurement of narrow lines in wide spectral bins is often insensitive to the choice of Doppler parameter, in our measurements it does matter. The theoretical over-sampled line profile is a convolution of the the narrow line and the line spread function. Our choice of 2 spectral bins is much larger than the former but does include the entire line spread function. This means that the choice of Doppler parameter in the model does have an impact. For example, changing the Doppler parameter by 5$\kms$ generates a change of $\Delta (log N) \lesssim 0.1$ (the strongest features are closest to this limit, e.g. \CIII). Normally this degree of sensitivity would be considered small but in the context of the extremely high precision of the average column density statistic, the choice of using the \HI\ Doppler is a significant assumption. Again we shall see in \autoref{sec:abspop} that the population analysis implies larger column density errors.

\subsection{Modelling average metal column densities}
\label{subsec:modmet}

In order to interpret our measurements of SBLA sample FS0 (both for the ensemble SBLA mean and the population properties in \autoref{sec:abspop}) we follow the simple framework in \citetalias{Pieri2010Stacking} and \citetalias{Pieri2014}. We will review this analytic framework here, and for further details see \citetalias{Pieri2014}. 

A key supporting assumption for what follows is that the gas studied follows the optically thin (to ionizing photons) approximation. This assumption is supported by various arguments. First of all, as stated in \autoref{sec:sampleselection} Damped Lyman-$\alpha$ systems in the \citetalias{Lyke2020} 
sample are masked. Secondly, the mean \HI\ column density found (see \autoref{subsec:fith1}) is that of optically thin gas. 
Thirdly, the population analysis (see \autoref{sec:abspop}) indicates that \lye\ is homogeneous indicating that the \HI\ population does not deviate significantly from this mean. Finally DLAs and Lyman limit systems are not sufficiently numerous to significantly modify our mean results (as discussed in \autoref{sec:sampleselection}). However, as we shall see in \autoref{sec:abspop} when one delves further into the metal population behind the mean one finds that such small populations can have an important contribution if the absorption is sufficiently strong. Metal lines consistent with such small populations are identified in \autoref{sec:abspop} and omitted from further analysis in this work.

In order to model the column density of each metal line from the measured the \HI\ column density we need a simple sequence of conversion factors: the neutral hydrogen fraction is needed to obtain the hydrogen column density, the metallicity (and an abundance pattern baseline) is needed to obtain the metal element column density, and finally the metal ionization fraction is needed to obtain the required metal column density. The ionization fractions are provided under the optically thin approximation by runs of the CLOUDY (C23; {\footnote{https://gitlab.nublado.org/cloudy/cloudy/-/wikis/NewC23} \citealt{Ferland1998}) with varying gas density and temperature using a quasar+galaxy UV background model \citep{Khaire2019}.
For relative elemental abundances, we assume a solar abundance and take the solar abundance pattern of \cite{AndersGrevesse1989}. The UV background and abundance patterns used are significant simplifying assumptions (see \autoref{subsec:disc_gasprop} for further discussion).

In this work, we focus on the constraining density and temperature from these ionization models with metallicity as an additional free parameter (acting as a overall normalisation of projected metal column densities). We give the gas density in terms of hydrogen atom number density, but this can be converted to gas overdensity by scaling up by 5.04 dex for a standard cosmology at $z=2.7$. In the process of interpreting these metal features, we take into account that all these features should build up a coherent picture either as a multi-phase medium, or multiple populations of systems or both. By `multiphase' we mean that an individual SBLA in our sample may be associated with multiple phases of gas that are unresolved in our data.
Interpreting our average metal properties in a purely multiphase way presumes that all SBLAs stacked are the same. We will initially explore this straw-man model before going on to explore the underlying population, and combined multi-population and multi-phase fits in \autoref{sec:abspop}.

\begin{figure}
    \centering
    \includegraphics[width=1.\linewidth]{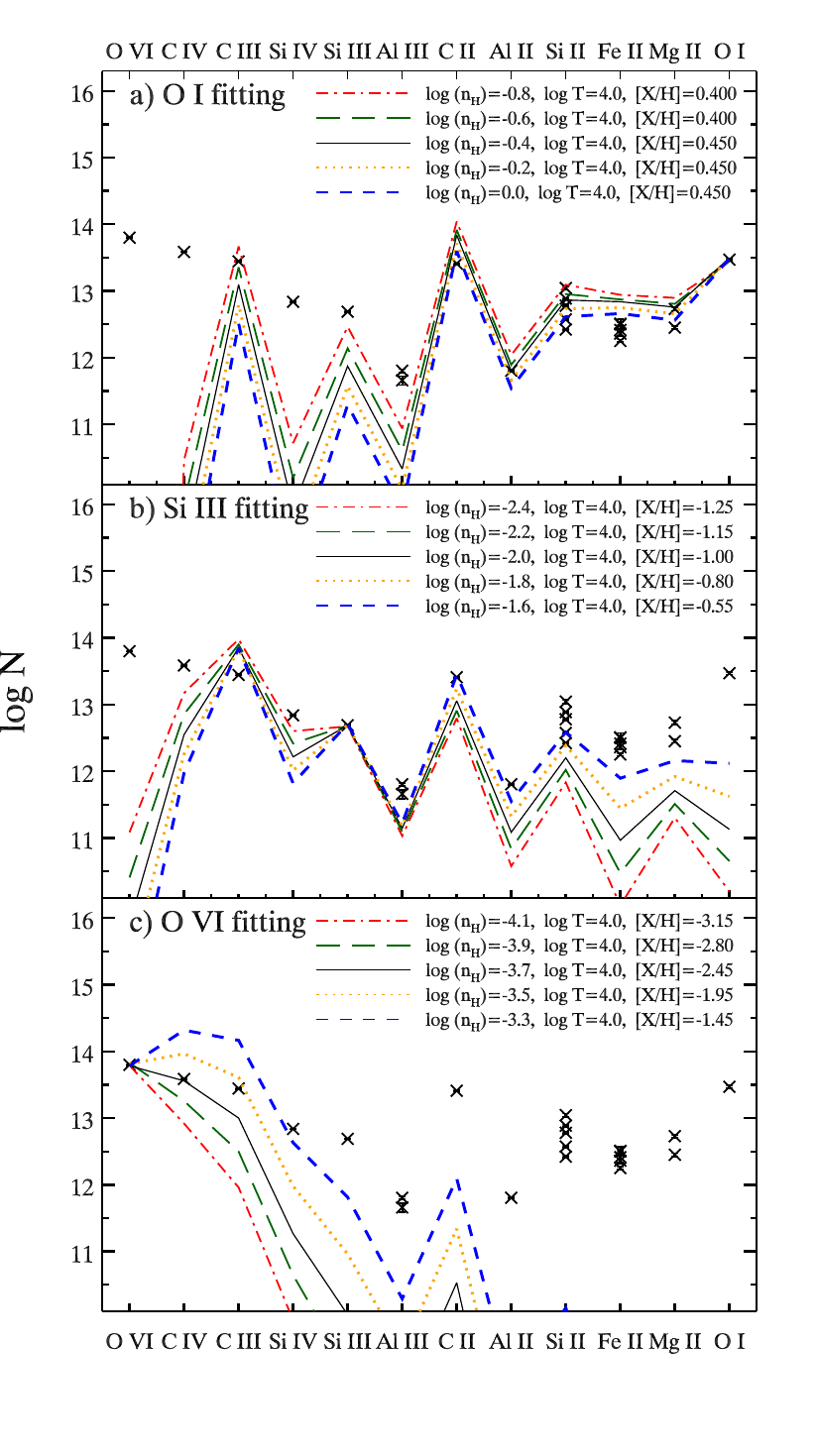}
    \caption{Metal column densities models for the FS0 sample.  Model curves are displayed assuming gas is optically thin, H I columns shown in \autoref{tab:hicolumns}, and a solar abundance pattern.  Models to fit the column densities of \OI\ ({\it top panel}), \SiIII\ ({\it middle panel}), and \OVI\ ({\it bottom panel}) are shown with varying density. Metallicities are tuned in order to fit to the chosen species for a given density and temperature. A preferred value of density for a fixed temperature ($10^4$K) attempting to avoid overproducing any species and avoiding unjustified extremes of density. Density is varied around this preferred value ({\it black line}) in the the middle and bottom panels. In the top panel, we are not able to do this since the maximum density is the favoured one ({\it blue dashed line}) and we are only able to vary the density downwards.}
    \label{fig:model_vary_density}
\end{figure}

\begin{figure}
    \centering
    \includegraphics[width=1.\linewidth]{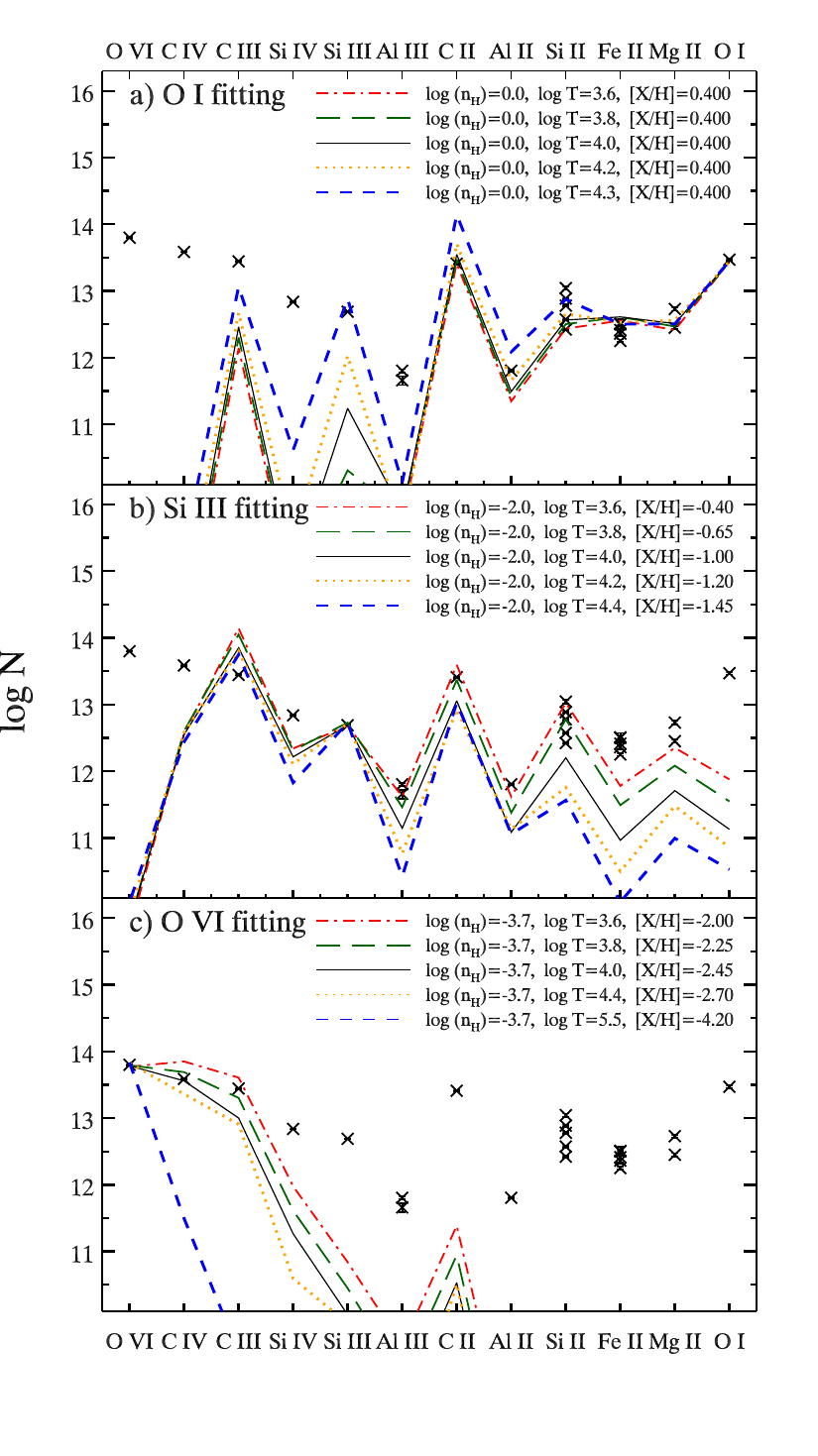}
    \caption{As in \autoref{fig:model_vary_density} metal column densities models are shown for the FS0 sample. Models to fit the column densities of \OI\ ({\it top panel}), \SiIII\ ({\it middle panel}), and \OVI\ ({\it bottom panel}) are shown with varying temperature around the value $10^4$K corresponding to the preferred values of \autoref{fig:model_vary_density}. Metallicities are again varied to provide the best fit to the chosen species for a given density and temperature. }
    \label{fig:model_vary_temperature}
\end{figure}

One cannot fit a model to each of the ionization species in isolation because a fit to one metal column density implies a prediction for another. 
We illustrate this point in figures~\ref{fig:model_vary_density} and \ref{fig:model_vary_temperature}.  In each panel we attempt to fit one of \OVI, \SiIII\ or \OI, varying the metallicity to maintain the fit while exploring density and temperature. In \autoref{fig:model_vary_density} we a take reasonable density for each of the 3 species and a reasonable temperature of $T=10^4$K, and we vary the density around this value.  In \autoref{fig:model_vary_temperature} we vary instead the temperature around these reasonable values.  The temperature, $T=10^4$K, is a standard estimate for a photoionized and photo-heated gas. The central densities are estimates intended to span the range of conditions required without over-production of other species (where possible). Note that the propagated errors associated with the uncertainty in the \HI\ column density are approximately the width of the model lines shown and so can be neglected. In this plot (and all subsequent plots of this section) the measured column densities are those shown in \autoref{tab:metalmeasmainFS0}.

In this plot (and all subsequent plots of this section) the measured metal column densities are those shown in \autoref{tab:metalmeasmainFS0}. Note also that the propagated errors associated with the uncertainty in the \HI\ column density is approximately the width of the model lines shown and so can be neglected. In each panel we attempt to fit one of \OVI, \SiIII\ or \OI. In \autoref{fig:model_vary_density} we a take reasonable density to reproduce each of these 3 species and a reasonable temperature of $T=10^4$K, and we vary the density around this value. In each panel we vary the metallicity to maintain fit to the intended species. In \autoref{fig:model_vary_temperature} we vary instead the temperature around these reasonable values. We include a model of extreme high temperature in the \OVI\ panel to show collisional warm-hot properties (note that at these temperatures the \OVI\ is weakly dependent on density).

Assuming that the gas is multi-phase,  contributions to the column density are additive. In other words, one must not significantly over-produce column densities from any given phase, but under-producing is acceptable so long as the short-fall can be made up by other phases (that do not themselves lead to significant overproduction of other metal column densities).
One can see by eye that two to three phases are sufficient to generate all the ionization species in broad terms. \autoref{fig:model_multiphase_sum} shows the resulting overall model fit from summing these three phases for the reasonable densities and temperatures of figures~\ref{fig:model_vary_density} and \ref{fig:model_vary_temperature}. While not a full parameter search it is clear that this multi-phase model produces the general trend required by the data but only with extremely high density and metallicity for the CGM.

However, it completely fails to offer acceptable statistical agreement required by the very small measured uncertainties. While one might attempt to generate instead four, five, six or more phases (indeed a plausible physical model would not be discrete at all), but each of our current three phases makes strong productions for multiple species and the model lacks the freedom to meet the statistical requirements of the data. For instance, producing more \AlIII\ without overproducing \SiIII, \CII\ and \AlII\ seems implausible. Similarly producing more \SiIV\ without overproducing \SiIII\ or further overproducing \CIII\ seems implausible.

Indeed the data is also not self-consistent in this purely multi-phase picture. For example the five \SiII\ features measured are statistically divergent from one other. A natural solution to this puzzle presents itself; not all SBLAs are alike and treating the composite spectrum as a measurement of a uniform population of lines with multi-phase properties is unrealistic. 

\begin{figure}
    \centering
    \includegraphics[width=0.99\linewidth]{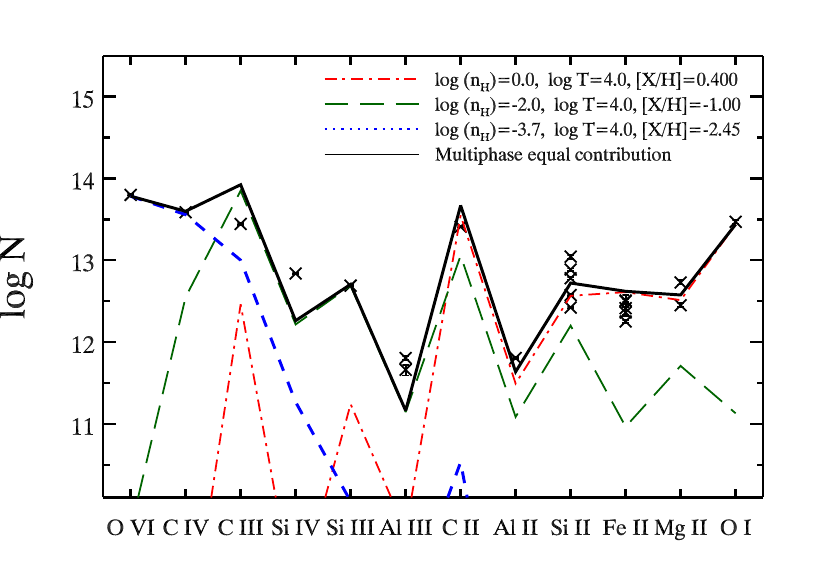}
    \caption{The column densities of metal ionization species in order of decreasing ionization potential the FS0 sample as in \autoref{fig:model_vary_density}. The best three models to fit the column densities of \OI, \SiIII, and \OVI\ are shown. A combined model is showing reflecting the multiphase scenario where each system stacked has same properties and three phases of associated gas. By summing the  columns from the three models without correction we are assuming that the \HI\ is distributed equally in each phase. Each sample receives a third the \HI\ column and therefore the metallicity is a three times larger than values shown in the legend for the model.}
    \label{fig:model_multiphase_sum}
\end{figure}

\subsection{The covariance between SBLA metal features}
\label{subsec:covar}

In order to explore the absorbing population beyond the mean we can study the properties of the stack of spectra used to calculate the mean composite spectrum. Naturally there is variance in the metal population giving rise to any given feature. In order to exploit these metal populations, we must develop an understanding of whether line strengths vary together. For example, it is expected that  \SiII\ $\lambda$1260 will be covariant with \CII\ given the predictions of models shown in figures~\ref{fig:model_vary_density} and \ref{fig:model_vary_temperature}. On the other hand, it is far from clear if \SiII\ $\lambda$1260 will be similarly covariant with  \OI,  \SiIII\ or even \OVI. Insignificant covariance would imply that population variance is negligible.  Similar and significant covariance between all species irrespective of ionization potential would indicate that metallicity variation is the main driver for population variation.
On the other hand significant differences in covariance of  low, medium and high ions with themselves and each other are a sign of more complex multi-population properties.

In order to explore this we calculate the covariance of the transmitted flux between our metal features normalised by their line strengths. The procedure used is set out in Appendix~\ref{extra:covariance}. \autoref{fig:line_covariance} shows the covariance between pairs of lines measured at line centre normalised to the product of the associated mean absorption signal for each line (corresponding to the flux decrement in the composite spectrum at line centre). This normalisation is performed in order to allow meaningful comparisons of the covariance between lines of different intrinsic strengths. In general covariance is approximately as large as the absorption strength or up to $4\times$ larger.

In top panel of \autoref{fig:line_covariance} we focus once again on transitions of our 3 indicative species: \OI, \SiIII, and \OVI, for low, medium and high ionization species respectively. We show the trend of covariance with the best-measured carbon lines, the best-measured silicon lines and remaining low ionization species in subsequent panels. 
We find that high ions are covariant with other ions with little or no signs of ionization potential dependence. Medium ions (\SiIV, \SiIII, \AlIII, and to an extent \CIII\ and \CII) also show an increased (albeit weaker) covariance with low ions and no signs of raised covariance with each other. We can conclude that SBLAs are not all alike with respect to their mix of detected metal lines. High ions appear to be relatively homogeneous, low ions appear to be inhomogenous. Medium ions lie between and their inhomogeneity seems to be linked to the inhomogeneity of low ions. Low ions generally show high levels of covariance with each other aside from the peculiar low covariance between \MgII\ and \OI\ despite their closely related ionization properties. However \autoref{sec:abspop} shows that the \MgII\ population is poorly constrained and is (marginally) consistent with a separate small self-shielded population. 

Overall it seems evident from the line covariance alone that more than one population exists in the ensemble of SBLA properties, and that metallicity variation alone is not sufficient to explain it. Overall covariance with low ionization species is high. It is at least as high as the covariance between high ions, between medium ions and between high ions with medium ions. Hence we conclude that the strong population(s) of low ions is also accompanied by strong populations of all species.

\section{SBLA absorption population}
\label{sec:abspop}

The standard stacking approach of calculating a mean or a median in order to understand the ensemble properties of the sample neglects variation in the ensemble. In this section we seek to explore the underlying properties of the population probed in our fiducial composite spectrum by using the full distribution in the stack at metal line centres. This is a non-trivial matter since the flux transmission distribution provided by this stack of spectra is a mix of different effects. In addition to the metal strength distribution we seek to probe, one can expect contributions from the observing noise (mostly read noise and photon shot noise),
contaminating absorption fluctuations (absorption in the spectra not associated with the selected system but nevertheless coincident with them), any smooth residual continuum normalisation errors in the individual quasar spectra, and finally any errors in the subtraction of the overall mean level of uncorrelated absorption (i.e. the pseudo-continuum). It is not possible to pick apart these various effects from the data alone but we may forward-model potential metal populations and compare them with the observed distribution. One could seek to study each effect in detail and generate synthetic spectra, but a much simpler and more robust method presents itself; we use the data itself as the testbed for our population modelling by adding model signal to null data.

\subsection{The null sample}
\label{subsec:nullsample}

The stack of spectra itself provides the ideal signal-free null sample: the close blueward and redward portions of the stack of spectra beyond the full profile of the feature of interest.  These proximate portions of the spectral stack represent a close approximation of the effects present at line centre excluding the metal signal of interest.  Potential linear variation as a function of wavelength in these effects are dealt with by attempting to mirror as much as possible the null pixels selected on both the blueward and redward sides. 

These nulls wavelength bins are drawn from the sample used in pseudo-continuum fitting as shaded in green in \autoref{fig:stack}. We take 8 wavelength bins on the red-side and 8 wavelength bins on the blue-side for all metal lines except \SiIII\ (where the close proximity of the broad \Lya\ absorption feature limits us to 4 bins on each side). We then average together the flux transmission in red-blue pairs from closest to furthest to the metal transition in order to generate the usual 138 $\kms$ integration scale of the central bin and to cancel out linear evolution with wavelength between red and blue. This leaves us with 8 null bins (or 4 nulls for \SiIII) for every metal feature central bin. In all cases the sampling of the null distribution is sufficient to allow the errors in the true measurement to dominate.

Finally, before assembling our null pixels we rescale them by any residual offset from the pseudo-continuum in the mean spectrum. As a result the nulls show only dispersion and no zero-point offset from the pseudo-continuum before mock signal is added.

\subsection{The population model}

 We model the populations underlying the average metal absorption signal of each feature independently with two main fitted parameters and two further marginalised parameters. These main parameters generate bimodality in the metal populations constrained by a prior that the population mean arrived as is that given by the unweighted arithmetic mean composite spectrum. In effect this unweighted arithmetic mean provides a flux decrement ($D_m=1-F_C$) `metal absorption budget' to be allocated in a way such that the ensemble mean is preserved. 
 Specifically our main parameters are:
\begin{itemize}
\item  $f_{pop}$, the fraction of systems with strong metal absorption, and
\item $f_{move}$, the proportion of the flux decrement by which to reduce the weak metal absorption population and reallocate to the strong population. 
\end{itemize}
The two parameters combined define the degree of asymmetry between the two populations. 

We initially attempted to fit with only $f_{pop}$ and $f_{move}$ as free parameters but found that two forms of random scatter were required and must be marginalised over.  The first is a Gaussian scatter in the strong absorption flux decrements with a standard deviation,  $\sigma_p$. The second is a Gaussian random noise added to the entire sample (both strong and weak components) with a standard deviation,  $\sigma_n$. This additional noise term is typically small (see \autoref{tab:poplnfit}) but appears to be necessary in some cases for an acceptable fit. The addition is a logical one since the pseudo-continuum fitting leads to an asymmetry in the noise properties between the metal measurements and nulls. 
The null pixels are part of the pseudo-continuum fitting and therefore the mean of the noise distribution is suppressed. This suppression is reinforced by our choice to rescale the zero-point of the nulls. In this way, we chose to generate a random noise in the nulls rather than carry-forward a potential different noise deviation already present in the nulls.

Overall, these two normally distributed random variables are sufficiently flexible to account for any scatter in the weak population also, since the sum of two independent normal random variables is also normal. The resulting model represents the simplest that provides an acceptable fit to our data.

More explicitly,
a mock absorption sample is built by taking every null pixel in the ensemble of nulls and applying the model as follows. For strong absorbers the flux decrement applied is
\begin{equation}
D^\prime_{strong} = D_m + \frac{D_m f_{move} (1-f_{pop})}{f_{pop}} \mathcal{G}(0,\sigma_p) + \mathcal{G}(0,\sigma_n)
\label{eq:strong}
\end{equation}
whereas the weak absorbers flux decrement is modelled as
\begin{equation}
D^\prime_{weak} = D_m (1-f_{move}) +\mathcal{G}(0,\sigma_n)
\label{eq:weak}
\end{equation}
where 
$\mathcal{G}(0,\sigma)$ denotes a Gaussian random number with zero mean and a standard deviation $\sigma$. The Gaussian random number that represents scatter in the strong population is bounded such that $\mathcal{G}(0,\sigma_p) < D_m$ in order to ensure that the strong sample never shows unphysical negative absorption. In principal this could lead to an asymmetry in the generated Gaussian numbers, a non-conservation of the `metal budget' and therefore an incorrect mean metal strength for the ensemble. In practise, however, favoured values of  $\sigma_p$ are sufficiently small that this regime is not reached.
The mock absorption sample combines together every null pixel from every member of the stack of spectra.

We randomly assign weak or strong absorber status to each pixel (using a uniform random number generator) in line with the trial $f_{pop}$ and proceed following \autoref{eq:strong} or \ref{eq:weak} as necessary.  

For every model (specified by our 4 parameters) we compare the flux transmission distribution of the mock sample with the measured flux transmission distribution function for the feature of interest. Despite our large number of null pixels, our model distribution functions can be unstable. Hence we make at least 100 random realisations of these mocks and the model distribution function carried forward is the average of these random realisations.
More realisations are produced when it is clear that the flux distribution is more complex or if the favoured models are those with small $f_{pop}$, which therefore require additional statistics to offset the intrinsically smaller sample size. In each case we compare the averaged simulation histogram with the measured true one, by performing a  $\chi^2$ test.  An example of this is shown in \autoref{fig:pophisto}, which compares the distribution function of the flux transmission for the central bin of \SiII\ $\lambda$ 1260 with the distribution of the preferred  model. 

In the development of these mocks and their comparison to data, it became apparent that outliers in the noise distribution lead to high $\chi^2$ values. In order to limit the impact of these outliers we sigma-clip by removing the top and bottom 3\% of the distribution from the $\chi^2$ test. This could in principal impair our ability to constrain very small absorbing populations but this is not true in practise. Furthermore, the favoured models are largely unaffected. This suggests that the tails of the distributions are dominated by noise outliers as expected.  The range of flux transmission shown in \autoref{fig:pophisto} shows for example, the range used in the model comparison.

\begin{figure}
\begin{center}
\includegraphics[angle=0, width=.9\linewidth]{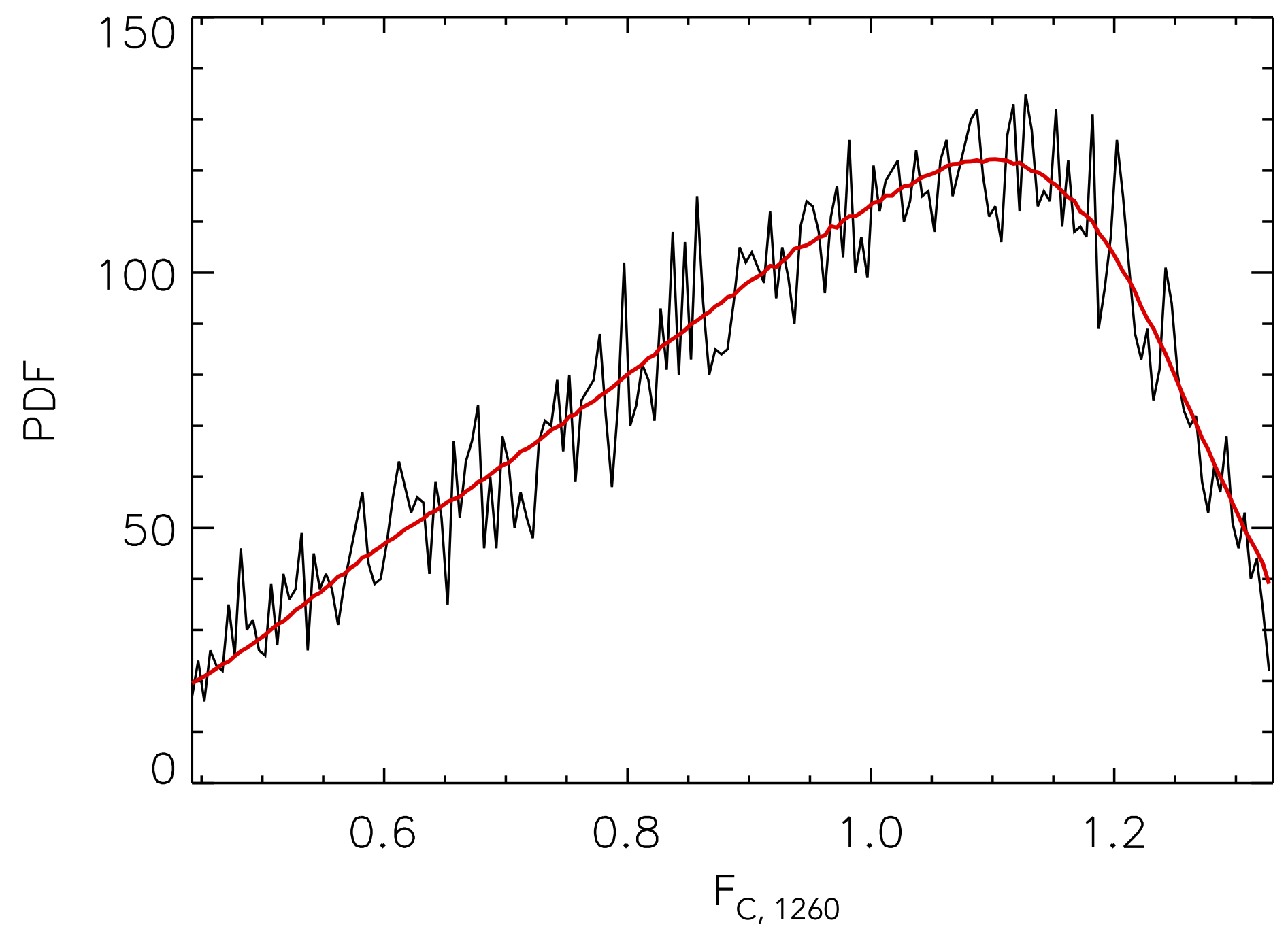}
\end{center}
\caption{An estimate of the probability distribution function of the flux in the stack of spectra corrected for the pseudo-continuum (for consistency with the composite spectrum) at the spectral pixel closest to the rest frame wavelength of \SiII\ $\lambda$ 1260 ({\it black line}). The {\it red line} shows the distribution function of the best fitting model (see \autoref{tab:poplnfit}).}
\label{fig:pophisto}
\end{figure}

We search parameter space from $0.01 \le f_{pop} < 1$ and $0.01 <f_{move}< 0.99$  allowing $\sigma_p$ and $\sigma_p$ to float freely to preferred values in each case following the results of the $\chi^2$ test. We also add grid points in this 4-dimensional parameter space in order to better sample the region with $\Delta \chi^2 < 12$. We then find the minimum $\chi^2$ in this parameter space and calculate the $\Delta$ with respect to this minimum for the entire $\chi^2$ surface. We estimate confidence interval for our two parameters of interest by marginalising over the other 2 in order to produce a $\chi^2$ scan as shown in \autoref{fig:SiIIchisqscan}. Since we are performing a combined fit of the two parameters of interest the standard deviation 68.3\% confidence interval is provided by the region where $\Delta \chi^2 <2.30$. This $1\sigma$ interval is marked in \autoref{fig:SiIIchisqscan}.

\begin{figure*}
\begin{center}
\includegraphics[angle=0, width=0.45\linewidth]{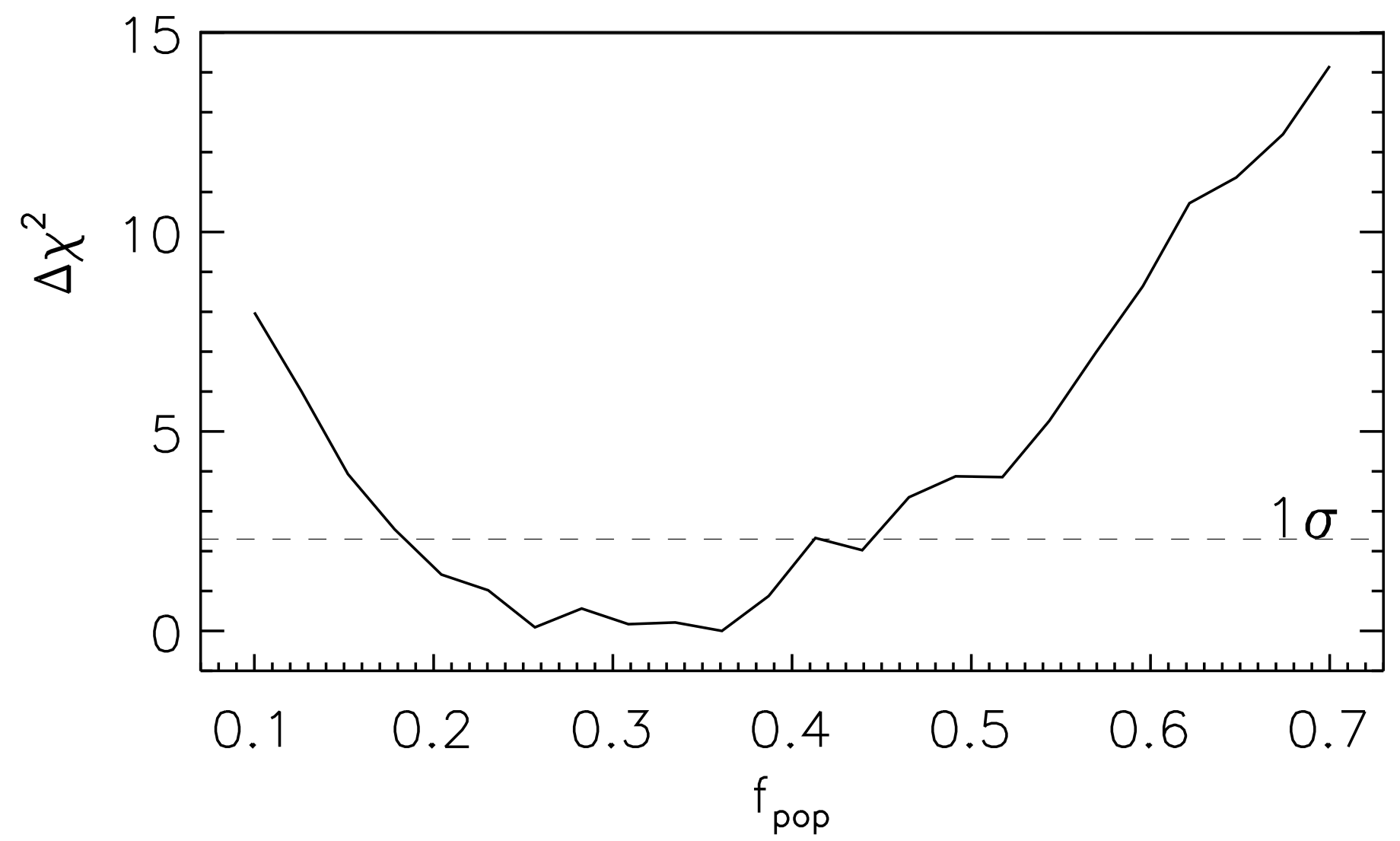}
\includegraphics[angle=0, width=0.45\linewidth]{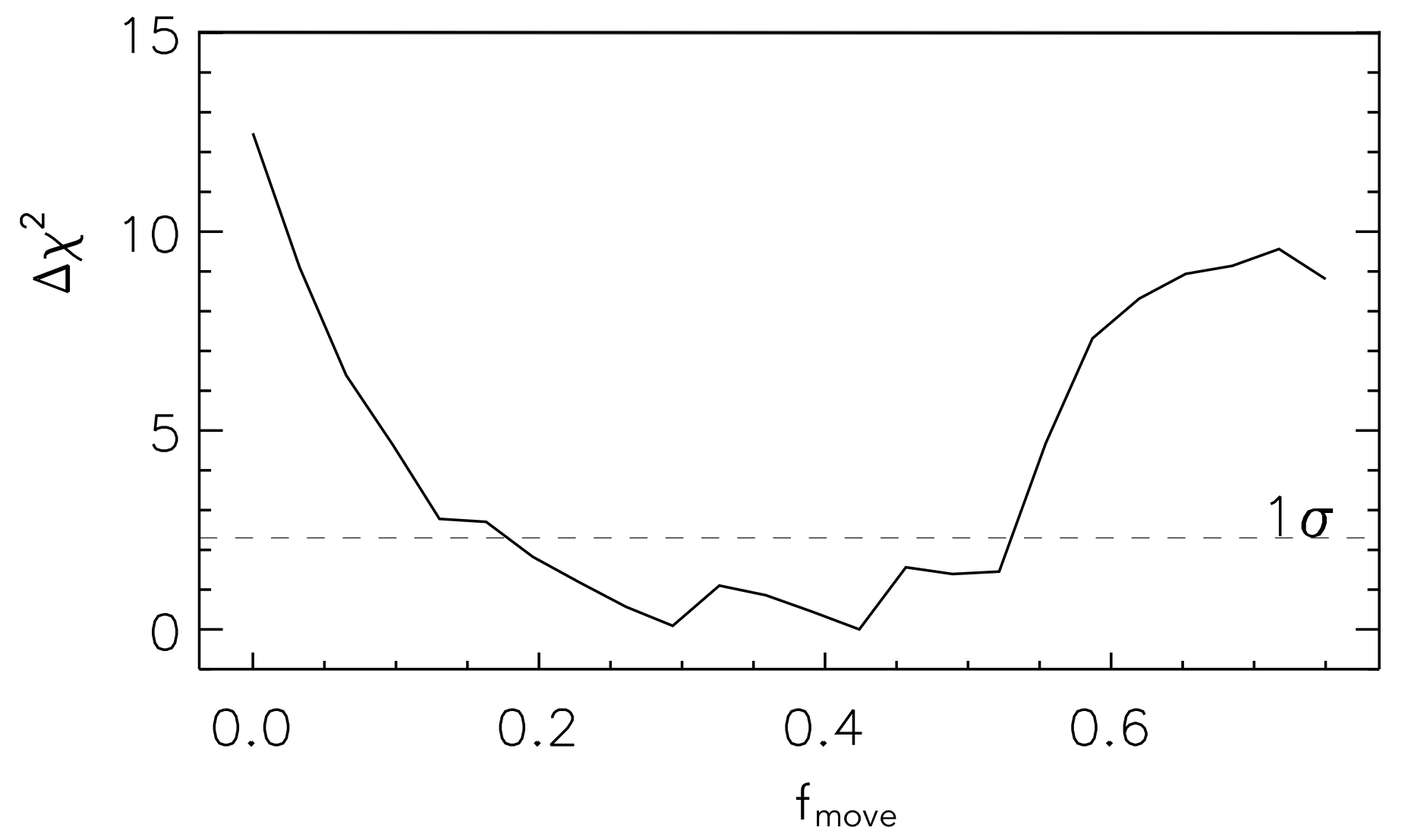}
\end{center}
\caption{The $\chi^2$ scans of both  $f_{pop}$ ({\it left}) and $f_{move}$ ({\it right}) marginalised over all four parameters}
\label{fig:SiIIchisqscan}
\end{figure*}

\subsection{Population analysis results and measuring the strong metal population }

\autoref{tab:poplnfit} shows the resulting favoured model parameters including 1$\sigma$ confidence intervals for our two parameters of interest and the fit probability. Since the constraint is statistically and computationally demanding, we limit ourselves to the most constraining transition for each ionization species. We present only species that have generated statistically meaningful parameter constraints for any feature. 

We study one further parameter, which is a quantity derived from our two parameters of interest. This is the `boost factor'
\begin{equation}
C_{boost}= \frac {f_{move}(1- f_{pop})} {  f_{pop}} +1 ,
\label{eq:cboost}
\end{equation}
which represents for each feature the level of boost in line strength that must be applied to the flux decrement measured in the composite spectrum in order to generate the metal strength of the strong population favoured by the population model search. Note that the best fit $C_{boost}$ is derived from the best fit $f_{pop}$ and $f_{move}$, while the error estimate in $C_{boost}$ is the range given by marginalising over the 1$\sigma$ confidence of $f_{pop}$ and $f_{move}$.

 \begingroup

\renewcommand{\arraystretch}{1.3} 

\begin{table*}
 \caption{Population model fits. We exclude all species where the statistics were insufficient to provide any useful constraint.}
 \label{tab:poplnfit}
 \begin{tabular}{@{}lcccccccccc}
  \hline
  Ion & $\lambda$ (\AA) & $f_{pop}$ & $f_{move}$  & $C_{boost}$ & $\sigma_p$ &  $\sigma_n$ & $\chi ^2$ & DOF & Prob \\
  \hline
  \lye & 937.803  &  0.91 $\substack{+  0.09\\ -  0.05 }$ &  0.050 $\substack{+  0.221\\ -  0.050 }$ &  1.005 $\substack{+  0.087\\ -  0.005 }$ &  0.000  &  0.002  &  81.309  & 76 - 4&  0.212 \\
  \hline
\SiII &1260.422  &  0.36 $\substack{+  0.08\\ -  0.18 }$ &  0.42 $\substack{+  0.11\\ -  0.25 }$ &  1.75 $\substack{+  0.41\\ -  0.25 }$ &  0.21  &  0.009  & 210.7  & 178  - 4 &  0.030 \\
\SiIII &1206.500  &  0.590 $\substack{+ 0.030\\ -  0.037 }$ &  0.298 $\substack{+  0.010\\ -  0.019 }$ &  1.21 $\substack{+  0.03\\ -  0.21 }$ &  0.30  &  0.003  & 214.1  & 183 - 4&   0.038 \\
\SiIV &1393.760  &  0.202 $\substack{+  0.071\\ -  0.068 }$ &  0.526 $\substack{+  0.053\\ -  0.069 }$ &  3.08 $\substack{+  0.93\\ -  0.71 }$ &  0.12  &  0.037  & 117.2  & 94 - 4 & 0.028 \\
 \CII &1334.532  &  0.26 $\substack{+  0.18\\ -  0.16 }$ &  0.67 $\substack{+  0.09\\ -  0.20 }$ &  2.9 $\substack{+  2.7\\ -  1.3 }$ &  0.15  &  0.037  & 127.5  & 98 - 4&  0.012 \\
 \CIII & 977.020  &  0.430 $\substack{+  0.079\\ -  0.068 }$ &  0.870 $\substack{+  0.067\\ -  0.046 }$ &  2.15 $\substack{+  0.29\\ -  0.32 }$ &  0.043  &  0.009  & 253.4  & 130 - 4 &  0.000 \\
 \CIV &1548.205  &  0.373 $\substack{+  0.039\\ -  0.050 }$ &  0.593 $\substack{+  0.024\\ -  0.032 }$ &  2.00 $\substack{+  0.22\\ -  0.11 }$ &  0.15  &  0.043  & 150.5  &  126 - 4 &  0.041 \\
\MgII &2796.354  &  0.05 $\substack{+  0.22\\ -  0.03 }$ &  0.39 $\substack{+  0.49\\ -  0.17 }$ &  8.1 $\substack{+\infty \\ -  6.2 }$ &  0.059  &  0.010  &  18.  & 22 - 4 &  0.444 \\
 \FeII &2382.764  &  0.010 $\substack{+  0.011\\ -  0.010 }$ &  0.38 $\substack{+  0.12\\ -  0.09 }$ & 39. $\substack{+\infty \\ - 32. }$ &  0.000  &  0.028  & 152.6  & 129 - 4 &  0.047 \\
   \OI &1302.168  &  0.19 $\substack{+  0.14\\ -  0.13 }$ &  0.96 $\substack{+  0.04\\ -  0.40 }$ &  5.0 $\substack{+  3.3\\ -  2.0 }$ &  0.043  &  0.004  &  84.1  & 81 - 4 &  0.271 \\
 \OVI &1031.926  &  0.79 $\substack{+  0.09\\ -  0.25 }$ &  0.55 $\substack{+  0.11\\ -  0.03 }$ &  1.15 $\substack{+  0.35\\ -  0.15 }$ &  0.043  &  0.000  & 446.0  & 258 -4 &  0.000 \\
 \AlII &1670.789  &  0.045 $\substack{+  0.043\\ -  0.017 }$ &  0.341 $\substack{+  0.085\\ -  0.068 }$ &  8.3 $\substack{+  1.5\\ -  3.8 }$ &  0.14  &  0.022  & 103.4  & 91 - 4 &  0.111 \\

\hline

 \end{tabular}
 \end{table*}

 \endgroup

\subsection{Inferred column densities for the strong metal population }

\label{subsec:modelpop}

\begin{table*}
 \caption{Strong population column densities. $F_{Corr}$ is  the corrected flux transmission for the strong population of lines derived from the population analysis in \autoref{subsec:modelpop}. $N_{strng}$ is the integrated metal column density associated with SBLAs with strong metals.}
 \label{tab:strongcolumns}
 \begin{tabular}{@{}lccccccc}
  \hline
  Ion  & Wavelength &  $F_{Corr}$ & $F_{Corr,min}$ & $F_{Corr, max}$ & $\log \mathrm{N_{strng}}(\mathrm{cm}^{-2})$  & $\log \mathrm{N_{strng, max}}(\mathrm{cm}^{-2})$  & $\log \mathrm{N_{strng, min}}(\mathrm{cm}^{-2})$ \\
  \hline
      \OI\ & 1302.17  &  0.8708  &  0.7867  &  0.9214 & 14.287 & 14.653 & 14.008 \\
    \MgII\ & 2796.35  &  0.4951  &  0.0000  &  0.8805 & 15.334 &   $\infty$    & 12.798 \\
    \FeII\ & 2382.76  &  0.2484  &  0.0000  &  0.8602 & 18.023 &  $\infty$     & 13.248 \\
    \SiII\ & 1260.42  &  0.9091  &  0.8878  &  0.9218 & 12.707 & 12.825 & 12.628 \\
    \AlII\ & 1670.79  &  0.7844  &  0.7454  &  0.8832 & 12.991 & 13.165 & 12.557 \\
     \CII\ & 1334.53  &  0.8359  &  0.6817  &  0.9097 & 14.005 & 14.748 & 13.645 \\
   \SiIII\ & 1206.50  &  0.8676  &  0.8644  &  0.8904 & 12.802 & 12.817 & 12.690 \\
    \SiIV\ & 1393.76  &  0.8052  &  0.7463  &  0.8499 & 13.513 & 13.770 & 13.322 \\
    \CIII\ &  977.02  &  0.6085  &  0.5550  &  0.6667 & 14.638 & 15.098 & 14.207 \\
     \CIV\ & 1548.20  &  0.7530  &  0.7260  &  0.7664 & 14.125 & 14.257 & 14.065 \\
     \OVI\ & 1031.93  &  0.8845  &  0.8498  &  0.8994 & 13.879 & 14.041 & 13.799 \\

\hline
 \end{tabular}
 \end{table*}

We now have a population analysis fit, as shown in \autoref{tab:poplnfit} and the covariance analysis result in \autoref{subsec:covar}, and so we are able to build up a picture of the dominant strong absorber population with realistic associated measurement errors statistically, even though we make no attempt to recover the sub-population on a case-by-case basis. The population analysis parameter 
$C_{boost}$ allows us infer the typical corrected transmitted flux, $F_{Corr}$, associated with this strong population for each feature (see \autoref{tab:strongcolumns}). Since the uncertainty $C_{boost}$ is much larger than the uncertainty in $F$, the error margin in $C_{boost}$ can be carried forward as the error margin in the flux transmission. This uncertainty is indicated in  \autoref{tab:strongcolumns} as a minimum and maximum transmitted flux, respectively given by $F_{Corr,min}$ and $F_{Corr, max}$. 

The corrected transmitted fluxes shown in \autoref{tab:strongcolumns}                 for the strong population are averaged across a 138$\kms$ velocity window and while this information alone doesn't tell us how many individual components exist, we know there must be at least one component that is strong enough to produce a minimum flux at least this low if resolved. We can conclude therefore that all these lines should be statistically significant in high S/N and high resolution. We don't rule out the possibility that the weak population for any given metal line is detectable, but they are not the focus of this work.

The size of the strong populations (indicated by $f_{pop}$) is not consistent among all features.  Higher ionization lines typically show larger and weaker strong populations. Given the picture, drawn from covariance, that strong higher ions trace a wider range of conditions, this is to be expected. However, it is also true that each feature shows their highest covariance with low ions. The key conclusion of the covariance analysis is that strong low ions appear to be accompanied by medium and high ions. We can therefore treat this sub-population of $\approx$25\% as being traced by all our fitted metal features and fit a multi-phase model to all these features. 

The metal column densities (and their measurement uncertainties) associated with this common strong absorbing population are derived from the corrected transmitted flux (and its error margin), using the same method as set out in \autoref{subsec:metalcolumn}. We recompute each column density value as before using this strong absorber corrected flux transmission. The column densities of the strong population features are given as $N_{strng}$ in \autoref{tab:strongcolumns}, with associated upper low limiting column densities given by $N_{strng, max}$ and $N_{strng, min}$ respectively. Note that while \FeII\ has a very high best fitting column density, the uncertainty is large and the $1\sigma$ lower envelope is a fairly typical integrated column density for strong metal SBLAs.

\subsection{Modelling the column densities for the strong metal population }

Now that we have a series of column densities measurements for a single strong population with multiple phases, we are ready to reassess the model comparisons shown in \autoref{sec:interpret_comp} and thus test the unusually high densities that our comparisons demand. 
 
As explained in the previous section, our metal column density models are dependent on density, temperature, metallicity, the UV background models and abundance pattern. We make standard assumptions for the latter two and explore density and temperature, with metallicity setting the overall normalisation.  A challenge of modelling our measurements lies in the production of the lowest ionization potential species without over-producing others,  driving us towards unusually high minimum densities. 

Thus far we have used this model comparison purely for illustration since no statistical fit to a single mean population was possible. Here we attempt statistical constraints for the dominant strong metal population. We begin with the most conservative choice; we relax the assumption of a solar abundance pattern and  explore the minimum density required by multiple species of a single element. This is possible for both carbon and silicon where we have reliable population analysis results for three ionization species each. Optically thin, photoionized gas is typically heated to $\sim 10^4$K in hydrodynamic simulations \citep{Rahmati2016}, but it is theoretically possible for it to reach $T < 10^{3.7}$K in unresolved gas that is sufficiently metal rich. As a result we consider models with temperatures as low as $10^{3.5}$K.

Figures~\ref{fig:model_strongpop_silicon} and \ref{fig:model_strongpop_carbon} illustrate these limits for silicon and carbon respectively. Only allowed models are shown and in each case the metallicity is tuned such that the low ion is only marginally produced, by treating the lower 1$\sigma$ error bar as the target. The density is then allowed to vary such that it remains below the 1$\sigma$ upper error bar of the high and intermediate ions. The minimum density in each figure is given by the red dot-dash line and the density is free to increase up to {\it and beyond} the density indicated by the blue dashed line. Given that this is a multiphase model, any short-fall in projected column density for the high and intermediate ions can be made up by other phases of gas with a lower density. 

 As one can see from figures~\ref{fig:model_strongpop_silicon} and \ref{fig:model_strongpop_carbon}, silicon provides the more stringent density limited of $\log(n_H / $cm$^{-3}) > -1.65$ assuming $10^4$K gas (equivalent to an overdensity of $\log(\rho /\bar{\rho}) > 3.39$) or $\log(n_H / $cm$^{-3}) > -2.40$ assuming $10^{3.5}$K gas if one allows the temperature to reach the lowest temperature considered here (equivalent to $\log(\rho /\bar{\rho}) > 2.64$). The limit arises from marginally (1$\sigma$) producing enough \SiII\ without marginally (again 1$\sigma$) overproducing \SiIII. Similarly, carbon requires $\log(n_H / $cm$^{-3}) > -2.85$ assuming $T=10^{3.5}$K gas and $\log(n_H / $cm$^{-3}) > -2.50$ assuming $T=10^{4}$K gas.

Since the models imply a hydrogen neutral fraction, the total hydrogen column density can be derived. The characteristic gas clumping scale can be obtained from
\begin{equation}
    l_c= N_H /n_H,
\end{equation}
 where $N_H$ is the {\it total} hydrogen column density and $n_H$ is the hydrogen density (requiring the observed \HI\ column density, the model ionization fraction and the model gas density). For silicon this maximum scale is just $l_c=15$ parsecs assuming a gas temperature of $T=10^4$K and $l_c=210$ parsecs for a gas temperature $T=10^{3.5}$K. Our carbon-only limits produce weaker constraints of 730 parsecs  and 1.6 kpc  respectively. $10^{3.5}$K is rather a low temperature for photoionized gas but as we shall see below, we appear to be forced to allow such low temperatures.

\begin{figure}
    \centering
    \includegraphics[width=0.99\linewidth]{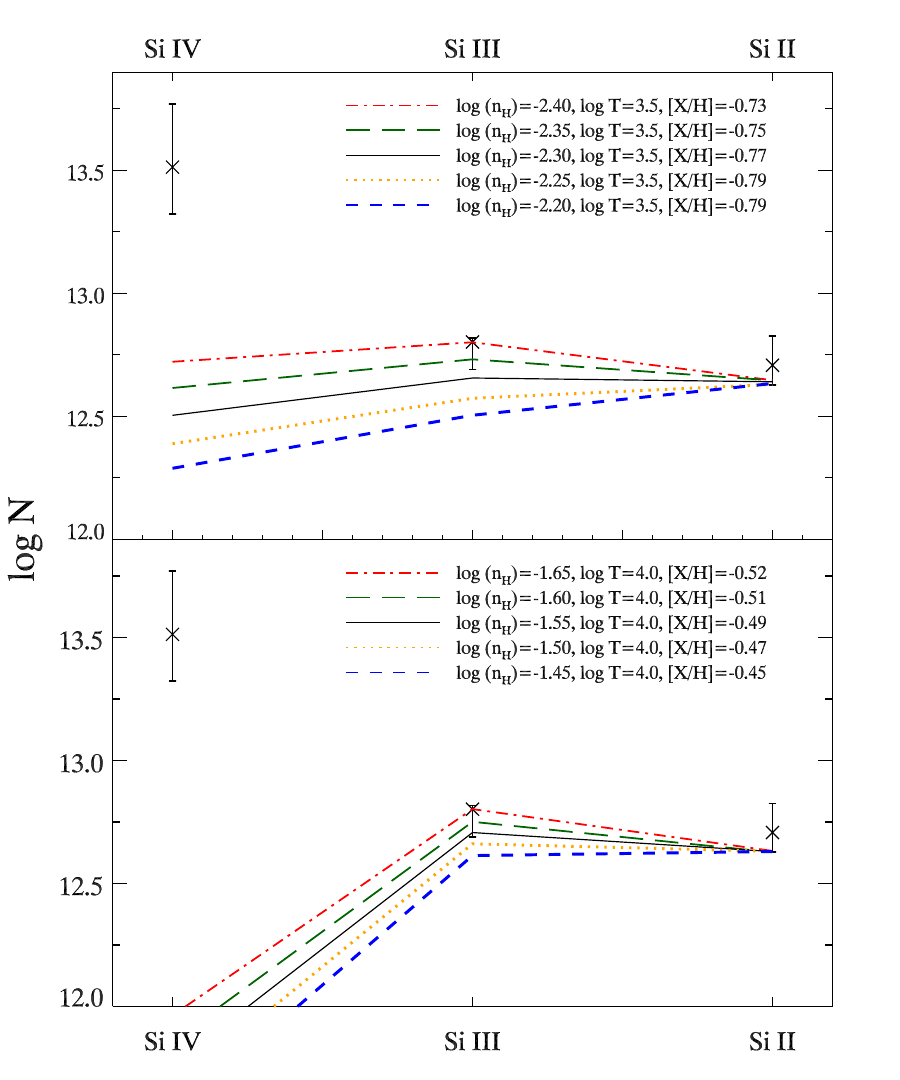}
    \caption{Constraining the minimum density of metal strong SBLAs using silicon species alone. Silicon column densities are modelled as in \autoref{fig:model_vary_density}. The data has been corrected to take into account the column density of the strong metal systems based on the population modelling (including associated model uncertainty). Here we test the minimum density allowed by measurements of \SiII, and \SiIII. \SiIV\ is also shown for completeness but doesn't constrain the analysis since no model produce it in significant amounts and it is evidently produced by gas in a different phase. We conservatively take the 1$\sigma$ lower error bar of our lowest column density measurement of \SiII\ as the target and then tune the density to change the slope while renormalising with the metallicity. The {\it red dash-dot line} shows he lowest density allowed at the 1$\sigma$-level. The {\it top panel} shows the result for the lowest density considered of $10^{3.5}$K and {\it the bottom panel} shows a more standard photoionized temperature of $10^4$K.
     }
    \label{fig:model_strongpop_silicon}
\end{figure}

\begin{figure}
    \centering
        \includegraphics[width=0.99\linewidth]{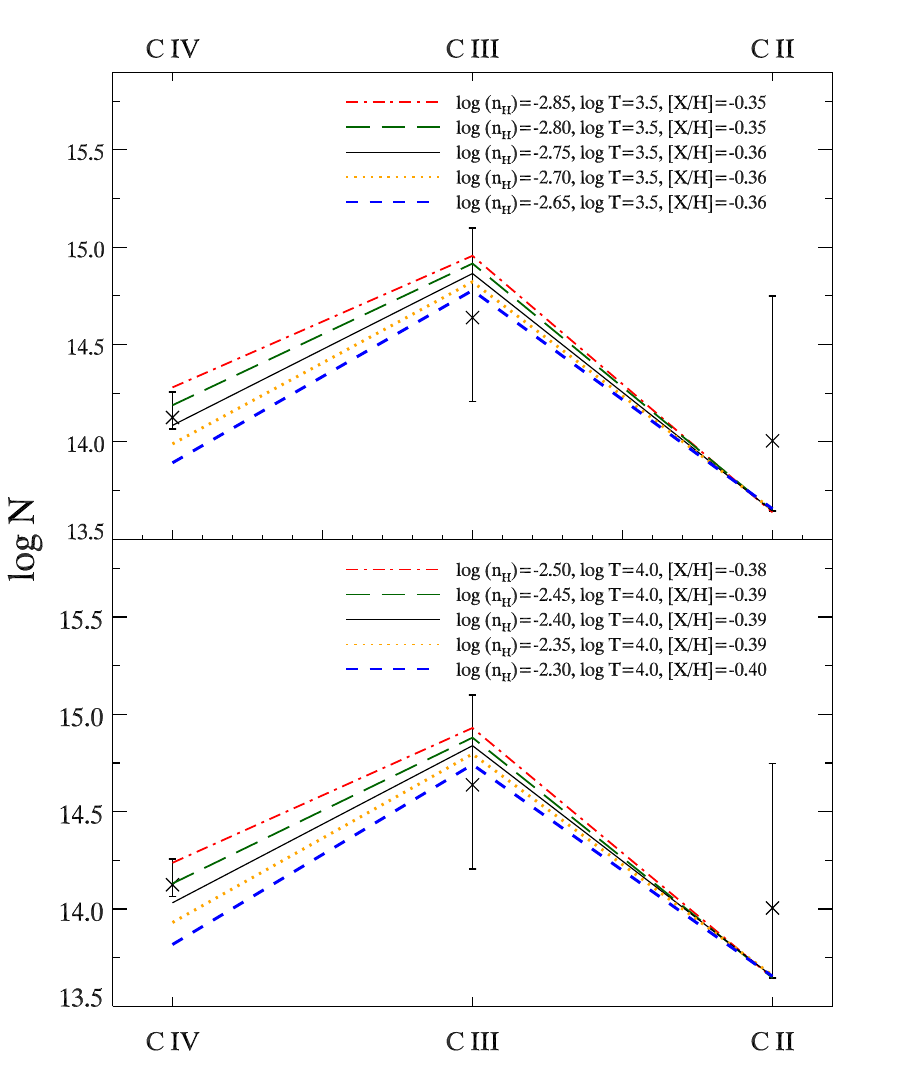}
    \caption{
    Constraining the minimum density of metal strong SBLAs using carbon species alone by following the same procedure used in \autoref{fig:model_strongpop_silicon} for silicon. In this case all of  \CII, \CIII\ and \CIV\ provide useful limits. Again the {\it red dash-dot line} shows he lowest density allowed at the 1$\sigma$-level. The {\it top panel} shows the result for the lowest density considered of $10^{3.5}$K and the {\it bottom panel} shows a more standard photoionized temperature of $10^4$K.
 }
    \label{fig:model_strongpop_carbon}
\end{figure}

Finally, we perform a statistical fit for three gas phases in these dominant strong metal systems by including all species and assuming a solar abundance pattern. We scan through density, temperature and metallicity for two different gas phases: high density and moderate density. As explained below, it was not possible to scan through the third, lower-density phase. Temperature is allowed to vary between $10^{3.5}$K and $10^{4.5}$K in both phases. In the moderate density phase, we searched densities from $\log (n_H / $cm$^{-3}) =-4.8$ to  $\log (n_H / $cm$^{-3}) =-2.8$. In the high density phase we scan through $\log (n_H / $cm$^{-3}) =-0.8$ to  $\log (n_H / $cm$^{-3}) =0$. As usual, metallicity is a free parameter that scales up and down the all projected metal columns simultaneously. 

Extreme small populations may arise due to Lyman limit systems in our sample of SBLAs and require more complex ionization corrections.  \citetalias{Pieri2014} argued that this contamination is likely to be at the level of $\lesssim $1\% and no higher than 3.7\% of our SBLAs. We conservatively require that any strong population that is statistically consistent with 3.7\% contamination should be omitted from our investigation of gas physical conditions.
This leads to the rejection of species \MgII, \AlII\ and \FeII\ from further interpretation using the optically thin to ionizing photons approximation. This is partly a consequence of poor statistical precision and given more data and more refined population modelling these species could be included in future analyses. 

In the process of performing this fit with three phases, it became apparent that only \OVI\ requires the lowest density phase. With one data point and three unknowns (density, temperature and metallicity), this third phases is unconstrained aside from being (by construction) of lower density. As a result, we proceeded with a two phase fit excluding \OVI.

\autoref{fig:model_strongpop_multip} provides the best-fit model based on this parameter scan of the strong metal population.  The fit is of acceptable statistical quality with a $\chi^2=5.5$ for 7 points with 6 degrees of freedom arising from the 6 fitted parameters, equivalent to a probability of statistical consistency between the model and data of 2\%. The favoured conditions for these strong absorbers are $\log (n_H / $cm$^{-3}) =0$, temperature $10^{3.5}$K and super-solar metallicities of [X/H]$=0.77$. In the intermediate density phase we find $\log (n_H / $cm$^{-3}) =-3.05$, again temperature of $10^{3.5}$K and metallicity [X/H]$=-0.81$. Again, a the lowest density phase is required but unconstrained. It should be noted that the favoured density for the dense phase is the limiting value of our parameter scan of $\log n_H =0$. This is driven by the measurement of \OI.  A higher density in the high density phase may provide a better fit to the \OI\ column density, but only to a limited extent since it will lead to a 
a worse fit to \CII\ and \SiII.

Again we can infer a gas clumping scale by dividing the hydrogen column density by the hydrogen density from this final, joint fit of species that reliably probe diffuse, photoionized gas. Our dense gas phase corresponding to $n_H =1$cm$^{-3}$ requires a clumping scale of only $l_c=0.009$ parsecs. If we marginalise the density of this dense component and take the $1\sigma$ minimum value for a 6 parameter fit ($\Delta \chi^2 = 7.04$) we obtain a minimum density of $\log (n_H / $cm$^{-3}) = -0.62$ equivalent to a maximum ($1\sigma$) clumping scales of $l_c=0.08$ parsecs.

The intermediate density gas is expected to have structure predominantly on 4kpc scales. Once again the low density phase traced by \OVI\ is unconstrained.

\begin{figure}
    \centering
    \includegraphics[width=1.\linewidth]{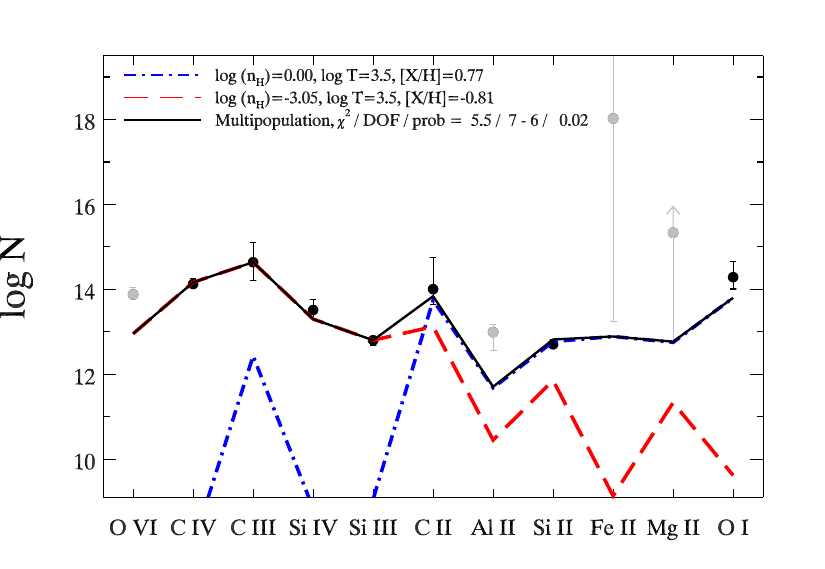}
    \caption{The results of a parameter search for a two phase fit for metal strong SBLAs
limited to species confirmed to arise in optically thin gas
(shown as black data points) assuming that the strong populations overlap. The fit probability is 2\%. Species showing small populations of only $f_{pop} \lesssim 5\%$ are excluded from the fit since they may arise from a self-shielded contaminating population (\FeII, \MgII\ and \AlII).  The measurement of strong \OVI\ is also excluded from the fit since it requires a third (more ionized) phase that is poorly constrained. This is because one can only set a lower limit on density based on the absence of associated \CIV\ (comparing with \autoref{fig:model_vary_density}, one may see that this density is $\log (n_H) \lesssim -4.3$). These four species not included in the fit are shown as grey points for completeness. }
    \label{fig:model_strongpop_multip}
\end{figure}

\section{Discussion}
\label{sec:discussion}

\citetalias{Pieri2010Stacking} and \citetalias{Pieri2014} argued that the presence of high density gas (inferred from the relative strength of low ionization species) indicates the presence of cold dense clumps 10s of parsecs in size, embedded in a more diffuse medium.
We have reviewed and revisited this claim by improving the methodology, challenging several assumptions and interpreting the results more deeply, while quadrupling the amount of data updating from SDSS-BOSS DR9 \citep{Dawson2013, Ahn2012, Lee2013} to SDSS-eBOSS DR16. Specifically we have, 
\begin{enumerate}
\item explored the statistical robustness of the mean composite spectrum error estimation,
\item made robust statistical measurements of \HI\ column density and verified its homogeneity,
\item improved the robustness of metal column densities,
\item explored metal line dependence on density and temperature,
\item measured the covariance and populations of metal species,
\item inferred the properties of the dominant strong metal population,
\item placed limits on the density derived from a single element  (carbon and silicon) for the strong metal population,
\item performed a fit to models of density and temperature for the strong metal population.
\end{enumerate}

From silicon alone we find that gas clumping on scales of at most 36 parsecs is required assuming temperatures of at least $10^{3.5}$K. However, when we include \CIV, \CII, \SiIV, \SiIII, \CII, \SiII\ and \OI\ we find that a clumping scale of 0.009 parsecs is favoured (with a $1\sigma$ upper limit of 0.38 parsecs) and super-solar metallicities are required.

We discuss this chain of reasoning its weak points further below.

\subsection{Metal populations and the nature of SBLAs}

 Our covariance measurements and population models carry wider implications for the nature of SBLAs than simply gas property measurements. Perhaps some SBLAs probe the CGM (with low ions and medium/high ions) and others probe the metal enriched IGM (showing only medium/high ions). Alternatively perhaps all SBLAs probe the CGM with medium/high ions, and when the line of sight happens to pass through a dense clump, low ions are also seen.

 The former implies a high impact cross-section to at least one dense clump with covariance being driven by CGM/IGM separation. The latter, implies a lower impact cross-section to dense clumps and covariance driven by the lines of sight passing through the CGM with or without intersecting a dense clump.

 Naturally these two scenarios are not mutually exclusive. This is self-evident since we cannot exclude the possibility that a metal rich IGM surrounding the CGM plays a significant role. Nor can we argue that there is a perfect association between our SBLA samples and CGM regions. This is likely to be a factor in why the high ion covariance is non-zero, but we cannot rule out the possibility that some CGM is relatively diffuse or metal poor (e.g. inflows).
 In practise the variation in ion strengths must arise due to some combination of SBLA purity,  CGM selection purity of SBLAs and the impact cross-section to various phases. The first term is known since we have measured the FS0 sample purity to be 89\%. Neglecting this minor correction, the fractional size of the low ion strong population, $\approx 30$\%, provides the cross-section to high density phases
 modulated by the CGM purity. We make this assertion because these low ionization species are not expected to occur in significant quantities outside of the CGM.

\subsection{Inferring gas properties from SBLA metals}
\label{subsec:disc_gasprop}

Following on from \citetalias{Pieri2010Stacking} and \citetalias{Pieri2014}, we focus on the surprising appearance of low ionization metal species in forest absorbers that are optically thin to ionizing photons. All the metal line measurements are of interest but the low ionization species drive our interpretation towards a quite narrow set of physical conditions. Specifically, the need for high densities and therefore small-scale clumping. 

Our goal in this work has been to update the measurements of \citetalias{Pieri2014} with the final BOSS/eBOSS dataset, to make error estimates more robust and to perform a thorough multi-phase and multi-population analysis of our measurements in order to generate statistically robust constraints.

Despite our inclusive error analysis, the error estimates on the metal column densities remain so tight that no single population, multi-phase model is satisfactory. This in combination with an analysis of the metal line covariance has led us to go beyond the study of mean properties in the composite spectrum and explore the full properties of the stack. Hence we forward model the metal absorbing population for each of our metal species using the full stack.

The quality of fit provided by the population is largely acceptable, with more complex models unjustified by current data. Exceptions are \CIII\ $\lambda$977 and \OVI\ $\lambda$1032, both of which offer 0\% quality of fit. This is not surprising since these are two of our four strongest metal features. It seems likely that this is a sign that more sophisticated population models are required by larger samples of SBLAs and/or higher signal-to-noise spectra. It is also possible that the metal populations are an exceptionally poor fit in these two cases, however, neither species' strong line fits are critically important for the main results presented in this article.

For each of the metal species we obtain (among other quantities) a constraint on the absorbing population size. All species with a population modelling constraint are included in the fit except \AlII, \FeII\ and \MgII\ since their strong populations are sufficiently small that they could plausibly arise in self-shielded gas (although it is notable that \FeII\ and \MgII\ column density constraints are statistically consistent with preferred models). 

Given the measured column density of $\log (N_{HI}/$cm$^{-2}) =16.04$\ullim{0.06}{0.05} for the FS0 sample, the lack of any significant inhomogeneity in the \HI\ population, the small potential interloper incidence rate, and our efforts to exclude metal species that show populations consistent with the interloper rate, we robustly conclude that our SBLA analysis is not sensitive to complex self-shielding effects expected for Lyman limit systems, or indeed partial Lyman limit systems. The inferred column density is at the limit where these effects are considered to be negligible and therefore the sample under study can be treated as strong, blended groupings of optically thin \lya\ forest absorbers.

The measurements of covariance indicate that strong low ion absorption is also associated with strong medium and high ion absorption, so we proceeded with measurements of the properties of these strong metal SBLA systems in various forms. Measurements of carbon-only and silicon-only were made independent of assumptions about abundance patterns providing lower limits in gas density and so upper limits on gas clumping on sub-kpc scales, but full fits become possible where all elements are included. 
These fits require 3 phases. 2 phases to provide both low and medium ions (defined broadly to include \CIV) and one additional unconstrained phased providing only \OVI\ absorption. The derived density of $n_H =1$cm$^{-3}$ for the dense phase is notably high even for the CGM (corresponding to an overdensity of $10^5$). Leading to a measurement of 0.009 parsecs cold dense clumps. Even if one considers the $1\sigma$ lower limit on density allowed in this fit, the analysis requires sub-parsec scale clumping (0.38 parsecs).

Parsec-scales are required by silicon alone but the sub-parsec scales are driven by \OI\ absorption. We cannot dismiss the measurement of \OI\ absorption since no other metal lines contribute significantly to the measurement spectra bin. \SiII\ $\lambda$1304 is closest but when one fits the full \SiII\ line profile one sees that the contribution to the \OI\ line centre is negligible (as shown in \autoref{fig:blendoi}).
 Note that charge-exchange driving the \OI\ ionization fraction to that of \HI\ \citep{Draine2011} does not apply in this case. This effect occurs at the boundaries of \HI\ and \HII\ regions and as we have discussed, SBLAs are optically thin to \HI\ ionizing photons and no boundary region is present. We must, therefore, conclude that we are probing clumps on scales as low as 1\% of a parsec due to our measurement of \OI\ absorption. 

Small increases in the favoured density above $n_H =1$cm$^{-3}$ are possible since the favoured density is the one selected as a prior. Also lower temperatures than our prior of $10^{3.5}$K are also possible but would stretch the limits of plausibility for a photoionized gas. The relationship between density and temperature warrants further investigation in simulations.

 It should be noted that this work assumes a solar pattern of elemental abundances (taken from \citealt{AndersGrevesse1989}) for the final results in \autoref{fig:model_strongpop_multip}. If the relative abundances of oxygen, carbon and silicon differ significantly from solar in SBLAs then our results would require modification. Our carbon and silicon only measurements are, of course, unaffected.

 Furthermore we assume photoionization reflecting a "quasar + galaxy" UV background following \citet{HaardtMadau2001}. \citet{Morrison2019} and \citet{Morrison2021} demonstrated that large-scale inhomogeneities exist in the UV background at these redshifts on scales or 10s or even 100s of comoving Mpc. \citet{Morrison2021} in particular explored the spatial variation in metals species through large-scale 3D quasar proximity in eBOSS DR16. There we used a mixed CGM sample including the superset of FS0+FS1+FS2 and found 10-20\% variations in \OVI\ and \CIV\ absorption on 100 comoving Mpc h${^{-1}}$ scales with similar variations in \SiIV\ and \SiIII\ also possible but unconstrained. It seems clear that the high ionization species studied here are susceptible to large-scale variation while the low ionization species have not yet been explored. Questions remain about the potential impact of the local galaxy (or galaxies) of these CGM systems.

\subsection{Comparison with simulations}

Wind tunnel simulations indicate that cold clumps of gas should survive  entrainment  by a hot galactic wind despite concerns that they might be destroyed before they can be accelerated by Kelvin-Helmholtz instabilities \citep{McCourt2015, GronkeOh2018, Tan2023}. These simulations are broadly consistent with our findings that such high-densities, low-temperatures (for a photoionized medium) and small-scales are plausible. Indeed many physical effects with characteristic scales of order a parsec are key for the ejection, propagation, entrainment and subsequent accretion of gas in the CGM with important consequences for further galaxy evolution (\citealt{Hummels2019,Faucher-Giguere2023} and citations therein).

For detailed observational predictions, high resolution cosmological simulations are required and cosmological simulations do not resolve  below 10-pc-scales even with zoom-in regions or adaptive refinement \citep{Lochhaas2023,Rey2023}.  CGM scales as small as 18 pc have been studied by \citet{Rey2023} for a single isolated dwarf galaxy although this is currently computationally demanding. They found that increasing resolution does indeed reveal higher densities ($n_H\approx0.5$cm$^{-3}$) and more extreme temperatures in the CGM (both $10^{3.6}$K and $10^{6.5}$K). It is notable that temperatures below our minimum prior of $10^{3.5}$K or our high density prior of  $n_H = 1$cm$^{-3}$ were not required in this simulation. However, we cannot rule out that that more extreme temperatures will be required by yet higher resolutions needed to probe the 0.01pc scales inferred by our multiphase, strong population, multi-element fit.

Although it seems that no simulations currently exist that reproduce the full range of conditions we infer for SBLAs, they can validate our findings that extreme small-scales are a requirement. This can be achieved by simply passing lines of sight through CGM zoom-in simulations and selecting those which meet our with \HI\ properties (an HI column of $\approx 10^{16}$cm$^{-2}$ and distributed in components over $138\kms$ to generate flux transmission <25\%) and comparing with with the metal populations we infer. 

Cosmological simulations can also address the potentially less demanding task of helping us understand the relationship between our selection of strong, blended \lya\ absorption and the galaxies and dark matter halos identified by it. In particular, to learn whether it can be optimised to better recover these systems or be modified to identify others. Such tests would greatly enhance our understanding of how the \lya\ forest traces IGM and CGM properties.

\subsection{Individual systems and SBLA analogues}

As explained in \autoref{subsec:integration-scale}, we advise caution in the interpretation of column densities measured in this work. The features measured here are integrated and averaged quantities. Our population analysis seeks to correct for the impact of averaging SBLAs showing weaker metal absorption with SBLAs showing stronger metal absorption, but the integrated nature of our measurements per SBLA is unavoidable. SBLAs themselves arise due to the blending of \lya\ lines over 138 $\kms$ and we cannot rule out that they correspond to multiple close CGM regions of multiple close galaxies (`close' here referring to both impact parameter and redshift). 
Furthermore within one CGM region we cannot resolve individual metal lines. We don't measure metals over the full observed feature profile as explained in \autoref{subsec:integration-scale}, but even within the narrower $138 \kms$ velocity window the measurements are integrated quantities. They cannot be trivially compared to individual metal line components that one might fit in an individual spectrum.

If one interpreted the measured signal as arising from single lines the metals would be strong and quite evident in high-resolution and high signal-to-noise studies of individual quasar absorption spectra. Those systems drawn from the strong population we have inferred would be even more evident once one takes into account the associated line strength boost, leading to quite high column densities ( $F_{strng}$ in \autoref{tab:strongcolumns}) but once again we stress that these are integrated column densities. 

We illustrate this argument with Appendix~\ref{extra:HR_SBLAs} in which we identify SBLAs at $2.4<z_{abs}<3.1$ in 15 high resolution and high signal-to-noise KODIAK spectra by taking 138$\kms$ bins and the noiseless definition of SBLAs ($-0.05\leq F_{\textrm{\lya}} <0.25$; where in this work we limit ourselves to $-0.05\leq F_{\textrm{\lya}} <0.05$ to prioritise SBLA purity in light of the SDSS noise). 

\autoref{fig:HR_SiII1260} shows the distribution of flux transmissions in native Keck HIRES wavelength bins at the position of \SiII\ $\lambda$1260 in the SBLA rest frame. Distributions are also shown for pixels on both the red and blue side of the \SiII\ feature (selected as usual to be at wavelengths away from lines and on the pseudo-continuum). Error bars show the 75\% spread of these null distributions. At the level of what one can discern by eye the \SiII\ $\lambda$1260 pixel distribution could have been drawn from the null distributions.  
Based on our analysis, around a third of SBLAs should show `strong' \SiII\ absorption with an integrated column density of $N_{strng}= 10^{12.7}$cm$^{-2}$.  Assuming that this signal is present in association with this KODIAK SBLA sample, it must be weak enough to not be clearly detected here. In other words, the \SiII\ absorption signal must be weak and distributed among the native pixels in the 138$\kms$ SBLA window and not a single narrow \SiII\ line with $N= 10^{12.7}$cm$^{-2}$.

One might reasonably ask, then,  what SBLAs should look like in individual spectra of high quality. The inferred column densities may be integrated column densities but the strong metal population should nevertheless be individually significant.

However, high confidence individual line identification isn't simply a matter of observing a significant absorption line. They must also be unambiguously assigned an absorption transition and  redshift. This may be complex task when lines are weak and there are no lines from the same species with which to confirm. It is made more difficult at high redshift where the line density in quasar spectra is generically higher, particularly in the \lya\ forest.
\OI\ is particularly challenging since \SiII\ absorption is expected to be nearby and could be caused by the same galaxy or galaxy group.
Our measurement of statistical excess here is robust and unambiguous because all sources of contaminating absorption are included in our error analysis both in the mean composite and the multi-population decomposition.

We are aware of what appears to be one strong metal SBLA analogue at $z>2$ in the literature, published in \cite{Nielsen2022}. Following up on systems in their catalogue of \MgII\ absorbers they discovered an associated compact group of galaxies and DLA absorption.  Among many interesting structures seen, there is an group of seven \HI\ absorbers with velocities offset blueward from the central velocity by between 350 and 450 $\kms$. The \HI\ column density of these lines is between $\approx10^{13.5}$ and $\approx 10^{15.8}$cm$^{-2}$, with a group total of approximately  $10^{16}$cm$^{-2}$. The velocity range of this structure and the resulting integrated column density are consistent with our SBLA sample. In \cite{Nielsen2022} this SBLA seems to have been found because of its association with this wider clustering of strong \HI\ and strong \MgII. It should be noted that this system would not have been selected by our methods because the SBLA \lya\ absorption is masked by the wide damping wing of the close DLA in the spectrum. Of course SBLAs in groups with DLAs will be missing from our sample in general, but the loss will be minimual because, as mentioned elsewhere (e.g. \autoref{sec:sampleselection}), SBLAs are much more numerous than DLAs.
\cite{Nielsen2022} measure the \HI\ column densities of these individual lines using higher order Lyman lines. 

The average metal absorption strengths over a $138\kms$ window is similar to our strong metal population in all the lines which are measured by both studies: \SiII, \SiIII, \CIII, \SiIV, and \CIV. Their intermediate metal ion models are also broadly similar to what we find. For low ionization species \cite{Nielsen2022} infer that components are present with solar or super-solar metallicities, high densities (between $-2< log n_H <-1 $), low temperatures ($3< log T (K)<4.5 $) and sub-parsec gas clouds. They do not infer densities as high as here nor gas clouds as small but they do not present detailed \OI\ measurements, which are the main driving factor our extreme inferences. They point out that the observed \OI\ column density of the DLA portion of the group is high compared to their model, but they are not able to measure \OI\ for the SBLA (private communication).

The analysis of KODIAQ data presented in \cite{Lehner2022} presumably include SBLAs among their sample but when they define their sample of strong \lymana\ absorption systems (or `SLFS' as they call them) they do not include the blending requirement critical for SBLAs selection and CGM properties that we, \citetalias{Pieri2010Stacking}, \citetalias{Pieri2014} and \citet{Yang2022} have seen. Instead their SLFS appear better characterised as IGM systems. However, they do show an example which superficially seems to quality for SBLA selection, and it appears to be an example of a weak metal system in contrast to the strong metal system case discussed above.

Studies of individual low ionization systems in photoionized ($N_{HI} \approx 10^{16}$cm$^{-2}$) are more common at low redshift. Examples of such works are \cite{Lehner2013}, \cite{Sameer2021} and \cite{Qu2022}. These works also produce a similar picture of multiphase gas showing small clumps (or clouds or shells) on parsec scales with temperatures of around $10^4$K.

Studies such as these (that focus on the detailed properties of individual absorbers) have particular virtues compared to our work, including probing detailed velocity structure and temperature from line widths. However, they cannot (yet) study the statistical properties of well-defined and unbiased large samples of CGM systems with our wide range of metal species. 
Our work demonstrates that the \cite{Nielsen2022} SBLA with super-solar metallicity and high densities is not simply an isolated oddity 
but a member of a population of around 125,000 in current surveys (taking a $\sim$25\% strong population 0.5 million SBLAs expected in eBOSS).

Simulators aiming to reproduce the results of these studies can seek to generate gas clouds that reproduce these properties among the clouds in their simulations. Whereas simulators can aim to compare the global properties of their CGM systems by simply reproducing our simple selection function. In this sense our statistical work complements the detail gas properties derived form those observations.

\subsection{Comparison with other observations based on stacking}

We have referred \citetalias{Pieri2010Stacking} and \citetalias{Pieri2014} throughout this work. They showed evidence of dense, parsec-scale, photoionized gas, and the goal has been to build upon their stacking methods, improve on the exploitation of their composite spectra, and verify their conclusions. There is a another study, \cite{Yang2022}, that has been inspired to apply these methods to SDSS-IV/eBOSS DR16 data. 

Our work is different in many respects from that publication. Referring back to the list at the beginning of this section, only point (iv) regarding investigating the density and temperature of gas proved by the composite spectrum is in common between the two papers. In a sense \cite{Yang2022} follows on directly from \citetalias{Pieri2014} in that they take a range of composite spectra for different \lya\ absorption strengths and explore more sophisticated ionization models to interpret them. \citetalias{Pieri2014} measured both the full profile of the metal features and the core of the absorption profile associated with the 138$\kms$ velocity window `central pixel' matched to the \lya\ selection. The former is a more inclusive integration and therefore generates a higher column density for both metals and \HI\ (see for example the comparison between their table A1 and table A3). 
\cite{Yang2022} take the full profile approach only, while we take the central pixel approach only. The motivation for our choice is set out in \autoref{subsec:integration-scale}. \cite{Yang2022} will, therefore, naturally present higher metal column densities than us derived from the composite spectrum.  This difference makes direct comparison difficult. 

There are further complications from differences in analysis choices. We select and stack \lya\ absorbers and their associated spectra in precisely the same way as \citetalias{Pieri2014} in bins of flux transmission (and so take advantage \citetalias{Pieri2014} progress on understanding SBLAs with tests on hydrodynamic simulations and comparison with Lyman break galaxy samples). On the other hand \cite{Yang2022} selects \lya\ samples in windows of flux transmission contrast (see Appendix~\ref{extra:corrmethod}), have a different S/N requirement for selection, apply no strong redshift cut (sacrificing sample homogeneity for statistics in the process) and weight their stack of spectra to compute the composite. On this final point regarding weighting, we do not weight the spectra by S/N because we wish to preserve the equal contribution of every system stacked, which simplifies our population analysis. We are also conscious of the fact that weighting the stacking by S/N would bias us towards stronger \lya\ absorption in complex in difficult to control ways.\footnote{Higher S/N for  \lya\ selection provides a purer selection of strong \lya. This higher S/N is typically associated with a higher S/N spectrum as a whole (quasar brightness varies and the S/N is highly covariant across each spectrum), therefore the weighting applied at the metal line is a complex mix of weighting towards stronger \lya\ systems modulated by any quasar shape change between 1216\AA\ and the metal line placement in the absorber rest frame.}

With all these caveats in mind, the results of \cite{Yang2022} and our measurements of the mean composite spectrum present broadly the same picture of multiple gas phases in the CGM exhibiting low ionization species tracing at least one high density phase, high ionization species tracing at least one low density phase, and intermediate ionization species probing intermediate densities. They do not go into a detailed error analysis to understand what is allowed statistically, and so did not conclude (as we do) that column densities and their small error estimates force us to go beyond fits to the composite spectrum and study the underlying population behind the mean. When we do this, we appear to disagree with some of the findings of \cite{Yang2022}. Our population analysis leads us to rule out a significant higher column density \HI\ sub-population, forces us to higher densities, sub-parsec clumping and lower temperatures for agreement with low ionization species. We are also forced to similarly low temperatures for intermediate/high ionization species (excluding \OVI) along with elevated densities and metallicities. 

In this work we explored a more precise and demanding error analysis method compared to \citetalias{Pieri2014} and included not just the statistical errors in the stacking but also absorbed uncertainty in the pseudo-continuum fitting to generate the final composite spectrum. \citetalias{Pieri2014} conservatively assumed that the errors in the final step were equal to the rest of the errors combined and scaled their error estimates of the stacked spectra by $\sqrt{2}$ for the composite spectra. Our end-to-end bootstrap error analysis shows that the pseudo-continuum fitting step contributes weakly to the errors. This is quantified by $\epsilon$ as shown in \autoref{tab:metalmeasmainFS0}. Assuming that the pseudo-continuum fitting is performed with similar care to this work, this contribution can typically be neglected and the step of pseudo-continuum fitting an entire suite of bootstrapped realisations of the stack can be foregone. This is assuming that the error estimate need only be known to around 10\% precision. A notable exception is \CIII, for which the error contribution is estimated at 26\%  due to the challenge of separating it from absorption by Lyman series lines. Overall, we advocate moving beyond  studies of the mean (or median) composite spectra alone and in doing so make the need for precise error estimates redundant. Instead we advocate a focus on forward modelling the underlying population, and measuring covariance between the metal features in order to obtain a deeper understanding of the SBLA population studied.

\subsection{Future surveys}

Despite the extreme high signal-to-noise in the composite spectrum presented here, our work demonstrates that more data is needed. Our population analysis requires not only high S/N in the composite spectrum but excellent sampling over the entire SBLA ensemble to build high a S/N measurement of the distribution function of the flux for every metal line studied. Only the metal transitions presented here were sufficiently well-sampled to obtain a population estimate. On the other hand, the distributions functions of some metal transitions are sufficiently well-measured that our 5 parameter fit does not appear to capture the characteristics of the population and a more complex parametrisation is  required. 

More quasar absorption spectra are required to both widen the range of transitions (and species) measurable and help define improved metal populations for a more extensive round of forward modelling. The DESI survey \citep{DESI2016} began in 2021 and is expected to grow to produce around 700,000 $z>2.1$ quasar spectra. The WEAVE-QSO survey (\citealt{Jin2023,Pieri2016}, Pieri et al. in prep.) is expected to begin imminently and will observe around 400,000 $z>2.1$ quasar spectra. 4MOST \citep{deJong2019} is also in preparation and looks set to include $z>2.1$ quasars among its spectroscopic sample. These surveys will also  generate greater numbers of the moderate-to-high signal-to-noise spectra (S/N$\gtrsim 3$) spectra required to identify SBLAs. 

These next generation surveys will also provide spectral resolution that is twice (DESI and 4MOST), three-times (WEAVE-QSO LR) or even ten-times (WEAVE-QSO HR) the resolution BOSS spectra. This will allow us the freedom to treat the velocity scale of the selection blend as a free parameter. In this work, we noted the striking similarity between the inferred halo mass derived from the large-scale 3D clustering of the \lya\ forest with SBLAs and the virial mass inferred by treating the velocity-scale of the blend as the halo circular velocity. This may be a coincidence but if there is some connection it raises the attractive possibility of identifying specific galaxy populations or halo populations from \lya\ absorption blends/groups alone. This warrants further study using next generation surveys and simulations with accurate small-scale IGM and CGM \lya\ clustering. 

The diversity of environmental properties for IGM/CGM gas studied in the \Lya\ forest is also expected to grow substantially in the coming years. Maps of the cosmic web are expected using IGM tomography applied to data from WEAVE-QSO \citep{Kraljic2022}, DESI, PFS \citep{Takada2014, Greene2022} and further to the future MOSAIC \citep{Japelj2019} and a potential DESI-II survey, allowing us to study SBLA properties in filaments, sheets and voids of structure. Furthermore large $z>2$ galaxy surveys are expected over the coming years associated with
these surveys allowing us to study gas properties near confirmed galaxies of known impact parameter with galaxy properties. These surveys promise to shed new light on the formative epoch of galaxy formation in the build-up towards cosmic noon.

\section{Conclusions}
\label{sec:conslusions}

In this work we have sought to establish the potential of Strong, Blended \Lymana, or SBLA, absorption systems for the study of the CGM. In this work we define "strong" as a flux transmission less than 25\% and "blended" as average absorption in bins of $138\kms$. We build on the work of \citetalias{Pieri2014} in various ways such that we conclude a new widespread class of circumgalactic system must be defined and we explore the properties of these CGM systems.

Specifically we find,
\begin{enumerate}
    \item SBLA samples can be defined various ways to prioritise sample size of sample purity, though we focus on the main sample of \citetalias{Pieri2014} for continuity, we which label FS0.
    \item We make the first statistical constraint of the \HI\ column density of the FS0 SBLA sample and find it to be $\log (N_{HI}/$cm$^{-2}) =16.04$\ullim{0.06}{0.05} with a Doppler parameter of $b=18.1$\ullim{0.4}{0.7}$\kms$. This is not an individual line measurement but a constraint of the dominant \HI\ column density in the $138\kms$ spectra window driven by a convergence to a solution ascending the Lyman series.
    \item By studying the mean composite of the FS0 sample we find that at least 3 phases of gas are present in SBLAs but that no single multiphase solution can be found that would agree with the tight error bars and so a multiphase {\it and} multi-population model is needed.
    \item We explore the SBLA population by forward-modelling trial populations using portions of the stack of spectra without correlated absorption as a null test-bed. In doing this we find good agreement with a bi-modal population, and we exclude from further study metal transitions which are consistent with populations small enough to plausibly arise from rare Lyman limit system interlopers.
    \item  We find that low ionization metals (traced by optically thin gas) are present in a 1/4 of SBLAs while higher metal ionization species are typically more common in SBLAs (present in 40--80\% cases). We also find that \HI\ shows a high degree of homogeneity as measured from the \lye\ population.
    \item We study the covariance between our metal features and find that metals species are significantly covariant with one another spanning all ionization potentials. In general low ions show a high excess covariance with one another, moderately excess covariance with intermediate ions and a mild excess covariance with high ions. This is consistent with the picture presented by the population analysis where low ions appear 25\% of the time and tend to appear together, while other ions are more common in SBLAs. It also indicates that when SBLAs are strong low ions, they are strong in all metal ions and so defines a sub-class of metal strong SBLAs.
    \item By conservatively focusing only silicon species \SiIV, \SiIII, and \SiII\ we find densities in metal strong SBLAs of at least $\log(n_H / $cm$^{-3}) > -2.45$ are required assuming $>10^{3.5}$K. This corresponds to gas clumping on $< 255$ parsecs scales.
    \item Focusing conservatively only carbon species \CIV, \CIII, and \CII\ we find that densities in metal strong SBLAs of at least $\log(n_H / $cm$^{-3}) > -2.95$  are required assuming $>10^{3.5}$K. This corresponds to gas clumping on $< 2.5$~kpc scales.
    \item We fit a mixture of three gas phases to
    all metal lines associated with the metal strong SBLA sub-population (excluding species that could arise due to self-shielding). The highest ionization phase is required by \OVI\ but it unconstrained. The intermediate ionization and low ionization phases both require our minimum temperature of $T=10^{3.5}$K. The intermediate ionization model shows a density of $\log(n_H / $cm$^{-3}) > -3.35$ (equivalent to 15~kpc clumping) withr metallicity $[X/H]=-1.1$. The favoured low ionization phase model has a density of $n_H=1$cm$^{-3}$ corresponding to scales of only 0.009 parsecs and metallicity $[X/H]=0.8$. The minimum allowed density for this phase is $\log n_H > -0.93$ (at $1\sigma)$ corresponding to a clumping of 0.38 parsecs.  These extreme and yet common CGM conditions required further study in simulations.

\end{enumerate}

\section*{Acknowledgements}

We thank KG Lee for his continuum fitting code that was used in a modified form to produce the continua used in this work. We  thank Ben Oppenheimer supplying the ionization tables and providing helpful discussions. We also thank Nikki Nielsen for her useful comments about this work.

This work was supported by the A*MIDEX project (ANR-11-IDEX-0001-02) funded by the ``Investissements d'Avenir'' French Government program, managed by the French National Research Agency (ANR), and by ANR under contract ANR-14-ACHN-0021.

Some the data presented in this work were obtained from the Keck Observatory Database of Ionized Absorbers toward QSOs (KODIAQ), which was funded through NASA ADAP grant NNX10AE84G.

This research has made use of the Keck Observatory Archive (KOA), which is operated by the W. M. Keck Observatory and the NASA Exoplanet Science Institute (NExScI), under contract with the National Aeronautics and Space Administration.

Funding for the Sloan Digital Sky Survey IV has been provided by the Alfred P. Sloan Foundation, the U.S. Department of Energy Office of Science, and the Participating Institutions. 

SDSS-IV acknowledges support and resources from the Center for High Performance Computing  at the University of Utah. The SDSS website is www.sdss4.org.

SDSS-IV is managed by the Astrophysical Research Consortium for the Participating Institutions of the SDSS Collaboration including the Brazilian Participation Group, the Carnegie Institution for Science, Carnegie Mellon University, Center for Astrophysics | Harvard \& Smithsonian, the Chilean Participation Group, the French Participation Group, Instituto de Astrof\'isica de Canarias, The Johns Hopkins University, Kavli Institute for the Physics and Mathematics of the Universe (IPMU) / University of Tokyo, the Korean Participation Group, Lawrence Berkeley National Laboratory, Leibniz Institut f\"ur Astrophysik Potsdam (AIP),  Max-Planck-Institut f\"ur Astronomie (MPIA Heidelberg), Max-Planck-Institut f\"ur Astrophysik (MPA Garching), Max-Planck-Institut f\"ur Extraterrestrische Physik (MPE), National Astronomical Observatories of China, New Mexico State University, New York University, University of Notre Dame, Observat\'ario Nacional / MCTI, The Ohio State University, Pennsylvania State University, Shanghai Astronomical Observatory, United Kingdom Participation Group, Universidad Nacional Aut\'onoma de M\'exico, University of Arizona, University of Colorado Boulder, University of Oxford, University of Portsmouth, University of Utah, University of Virginia, University of Washington, University of Wisconsin, Vanderbilt University, and Yale University.

\section*{Data Availability}
Catalogues and derived data products from this article are available at \url{https://archive.lam.fr/GECO/SBLA-eBOSS} The data underlying this article were accessed from SDSS-IV DR16 (\url{https://www.sdss.org/dr16/}) and Keck Observatory Database of Ionized Absorption toward Quasars (KODIAQ; \url{https://koa.ipac.caltech.edu/applications/KODIAQ}).




\bibliographystyle{mnras}
\bibliography{SBLAs_DR16} 


\appendix

\section{Correlation function methodology}
\label{extra:corrmethod}

We use the same general data quality requirements  as set out in \autoref{sec:data} with two exceptions; the boxcar smoothing S/N cut is not used but no data is used with $\lambda < 3600$\AA, and DLA wings are not masked but corrected for. The sample of Ly$\alpha$ forest quasars is selected from DR16Q in the redshift range $2.05<z<3.5$. For computational convenience, we combine three adjacent spectral pixels into wider analysis pixels while determining the continuum normalisation. Forests containing less than 20 analysis pixels are discarded. Finally, we normalise the continuum to each spectrum using the method described in \cite{Blomqvist2019} and \cite{deSainteAgathe2019} which evolved from ``method 1'' of \cite{Busca2013}. Note that this normalisation is not a fit of the continuum. A mean quasar spectrum is fit to each spectrum and what results is a fit to both the quasar continuum and the mean flux of the forest (or `mean flux decrement' as it is sometimes called). Given that the normalisation is not to the 100\% transmission level of each quasar but to the mean, what is actually obtained is a flux transmission contrast, $\delta$ (often called `flux transmission fluctuations' or simply the `delta field').

Stated briefly, this involves fitting a mean quasar spectrum to the forest with only 2 free parameters for amplitude and slope. The total number of Ly$\alpha$ forests included in the final sample for cross-correlation studies is 335,259.

Our procedure for measuring the Ly$\alpha$ forest flux-transmission field and its cross-correlation with the SBLA distribution, to which a theoretical model is fitted to measure the SBLA bias parameter $b_{\rm SBLA}$, closely follows the methods established by the BOSS collaboration \citep{Slosar2011,Busca2013,Slosar2013,Kirkby2013,Font-Ribera2014,Delubac2015,Bautista2017,duMasdesBourboux2017,duMasdesBourboux2020}. The particularities of using the SBLA sample are described in detail in \citetalias{Perez-Rafols2022}. To measure the correlation functions we use version 4 of the publicly available code package \texttt{picca}\footnote{\url{https://github.com/igmhub/picca/}}. The fits for the SBLA bias are done using the package \texttt{vega}\footnote{Available at \url{https://github.com/andreicuceu/vega/tree/master/vega}}. 

Here we summarise the procedure used but we refer the reader to \citetalias{Perez-Rafols2022} for details. The correlation between the flux transmission contrast $\delta$ in the Ly$\alpha$ forest and the SBLA distribution is estimated as
\begin{equation}
\xi_{A}=\frac{\sum\limits_{(i,k)\in A}w_{i}\delta_{i}}{\sum\limits_{(i,k)\in A}w_{i}}\ ,
\end{equation}
where the sum runs over all pairs of deltas $i$ and SBLAs $k$ in a separation bin $A$. The weights $w_{i}$ are defined as the inverse of the total variance of the delta field and take into account both instrumental noise and the large-scale structure contribution. Our coordinate grid is defined by square bins of size $4~\hMpc$ in comoving separations along the line of sight $r_{\parallel}$ and transverse to the line of sight $r_{\perp}$. Each bin has an associated redshift defined as the weighted mean redshift of the included pixel-SBLA pairs. The ($r_{\perp},r_{\parallel}$) separations are calculated assuming a flat $\Lambda$CDM model with parameter values taken from the Planck 2015 result (using the TT+lowP combination; \citealt{Planck2016}): $\Omega_{\rm c}h^{2}=0.1197$, $\Omega_{\rm b}h^{2}=0.0222$, $\Omega_{\nu}h^{2}=0.0006$, $h=0.6731$, $N_{\nu}=3$, $\sigma_{8}=0.8298$, $n_{s}=0.9655$, $\Omega_{m}=0.3146$. The coordinates are equivalently expressed in terms of ($r,\mu$), where $r=\sqrt{r_{\perp}^2+r_{\parallel}^2}$ and $\mu=r_{\parallel}/r$.

Following \cite{duMasdesBourboux2017,Blomqvist2018,duMasdesBourboux2020}, the covariance matrix is estimated using a sub-sampling technique in which the sky is divided into small regions defined by HEALPix pixels \citep{Gorski2005}.

We fit the measured forest-SBLA cross-correlation to a physical model adapted from the forest cross-correlation with quasars \citep{duMasdesBourboux2017} and DLAs \citep{Perez-Rafols2018}. Besides the standard linear-theory prediction involving the bias and redshift-space distortion parameters \citep{Kaiser1987} for the Ly$\alpha$ forest and the SBLAs, the model also takes into account the effect of absorption by high-column density system and smoothing of the correlation function due to the $4~\hMpc$ binning. The linear matter power spectrum is obtained from CAMB \citep{Lewis2000} for the fiducial cosmology at reference redshift $z_{\rm ref}=2.334$. Distortions of the correlation function due to the continuum fitting procedure are corrected for using the ``distortion matrix'' \citep{Bautista2017,duMasdesBourboux2017,duMasdesBourboux2020}. Correlations arising from absorption by metals in the Ly$\alpha$ forest are modelled using a linear bias parameter for each metal line and mapped onto the Ly$\alpha$ coordinate grid using the ``metal distortion matrix'' \citep{Blomqvist2018}.

We perform a joined fit using the correlations in the Ly$\alpha$ and Ly$\beta$ regions. We limit the fit to separations in the range $10<r<180~\hMpc$, cover the full angular range $-1<\mu<1$ and remove pixels with $r_\perp<30~\hMpc$, corresponding to 4940 data bins included in the fit.

\section{Test of the bootstrap realisations}
\label{extra:errore2e}

As described in \autoref{sec:errore2e}, the error in the stacked flux was calculated as the standard deviation of the flux distribution across 1,000 realisations of the stack created by bootstrapping the \lya\ absorber sample. 
The assumption that for a given wavelength bin, the distribution of flux realisations captures the uncertainty of the stacked flux, also requires the mean of the distribution to equal the stacked flux value. 
The degree to which they are unequal can be characterised by the bootstrap bias,

\begin{equation}
B(\lambda_r)=\left|\frac{F_S-{\langle}\tilde{F_S}{\rangle}}{F_S}\right|,
\label{eq:bootcheck}
\end{equation}
where $F_S$, and ${\langle}\tilde{F_S}{\rangle}$ are the mean stacked flux, and the mean of the bootstrap realisations of the mean stacked flux, respectively.

The bootstrap bias can differ at every rest frame wavelength and to verify that the number of realisations is sufficient for our analysis we must verify that the bias is low at all rest frame wavelengths. To this end, we have calculated the the bias for  900 $< \lambda_r <$ 2900~\AA\ (since this encompasses all the rest frame wavelengths used in our work). We plot the fractional cumulative distribution function of these biases in \autoref{fig:bootstrap}. Over $98\%$ of the stacked pixels have a bootstrap bias smaller than 0.01. 
It is evident from \autoref{fig:bootstrap} that the suite of 1,000 realisations of the stack adequately represents the noise for most of the stacked pixels.
\begin{figure}
\begin{center}
\includegraphics[angle=0, width=.9\linewidth]{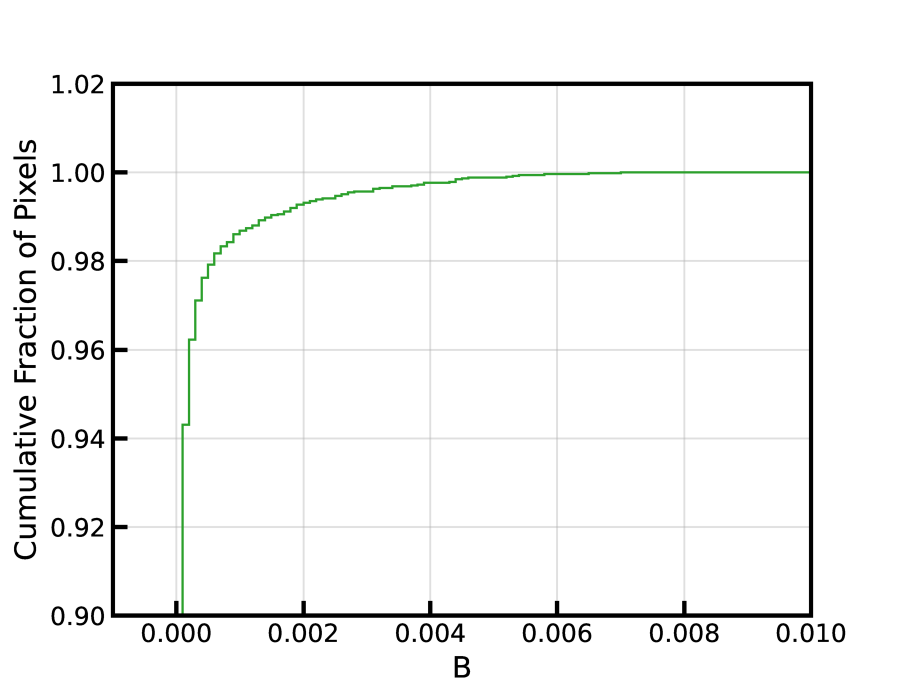}
\end{center}
\caption{Fractional cumulative distribution function of the bootstrap bias for the mean stacked spectrum of SBLA sample FS0. 
}
\label{fig:bootstrap}
\end{figure}


\section{Comparison with Absorbers in High Resolution Spectra}
\label{extra:HR_SBLAs}

Here we explore the the resolved pixel properties of metals associated with SBLAs using 15 high resolution and high S/N Keck HIRES spectra (\autoref{tab:HR}) taken from the Keck Observatory Database of Ionized Absorption toward Quasars (KODIAQ; \citealt{KODIAQ2015,KODIAQ2017,KODIAQ2014}).  KODIAQ provides extracted, continuum-normalised, and combined public HIRES spectra of quasars. The KODIAQ spectra have high S/N but in order to build an SBLA sample of extremely high purity, we impose the flux boundaries of FS0, namely $-0.05\leq F_{\textrm{\lya}} <0.05$. In the process we conservatively focus on the strongest SBLAs in our sample. We follow all the usual requirements of SBLA selection here; we rebin the \lya\ forest to bins of 138 $\kms$ width and for continuity we limit ourselves to $2.4<z_{abs}<3.1$. This yields 108 SBLAs with an incidence rate of $dn/dz=12.4$.  The incidence rate of the SBLA population as a whole ($F_{\textrm{\lya}} <0.25$ assuming that the data is effectively noise-free) at these redshifts is $dn/dz=23.4$.

\begin{table}
    \caption{HIRES Quasars used from KODIAQ}
    \label{tab:HR}
    \begin{center}
    \begin{tabular}{@{}lcc}
        \hline
        Quasar & $z$ & Keck Program \\
        \hline
        \hline
    J134004+281653  & 2.517000	&  U17H, D.Tytler\\
    J220852-194400	& 2.558000	&  C55H, W.Sargent\\
    J094202+042244	& 3.284050	&  U35H, Wolfe\\
    J083933+111207	& 2.696000	&  U11H, A.Wolfe\\ 
    J103456+035859	& 3.388386	&  H13H, A.Cowie\\
    J234451+343348	& 3.053000	&  U46H, Wolfe\\
    J113130+604420	& 2.905892	&  U131H, Wolfe\\
    J124610+303117	& 2.560000	&  U02H, D.Tytler\\
    J155152+191104	& 2.822700	&  C168Hb, Steidel\\
    J110610+640009	& 2.202364	&  U17H, D.Tytler\\
    J121134+090220	& 3.291703	&  U34H, Wolfe\\
    J005700+143737	& 2.648508	&  N033Hb, Bida\\
    J142438+225600	& 3.620000	&  H12H, A.Cowie\\
    J003501-091817	& 2.422646	&  A185Hb, Pettini\\
    J235050+045507	& 2.633003  &  N033Hb, Bida\\
    \hline  
   \end{tabular}
    \end{center}

\end{table}
We then take all the (native resolution) pixels that fall within the central pixel of \SiII\ $\lambda$1260 using our main analysis wavelength solution. In doing this we are exploring what we would find if we were able to resolve the absorption that yields our  \SiII\ $\lambda$1260 central pixel measurement. The black, solid curve in \autoref{fig:HR_SiII1260} shows the resulting flux transmission PDF. The blue (dotted) and red (dashed) PDFs  and represent the null measurements as measured by the mean (and the 75\% spread) of 5 samples each collected over a velocity width 138 $\kms$  ($\pm7,8,9,10,11$\AA) from the central pixel measurements. These results are consistent with the distribution favoured by the population analysis; a strong population of \SiII\ absorption with $\approx 90$\% flux transmission. There is no indication of a small but very strong population that would lead us to conclude that the signal arises from Lyman limit systems. Here equivalent composite spectrum flux transmission is $F_C=0.97$ compared to  our main eBOSS measurement of $F_C=0.9481$ to be explained by the larger errors due to uncorrelated absorption evident in the difference between the redside and blueside nulls in \autoref{fig:HR_SiII1260}.

\begin{figure}
    \centering
    \includegraphics[angle=0, width=0.99\linewidth] {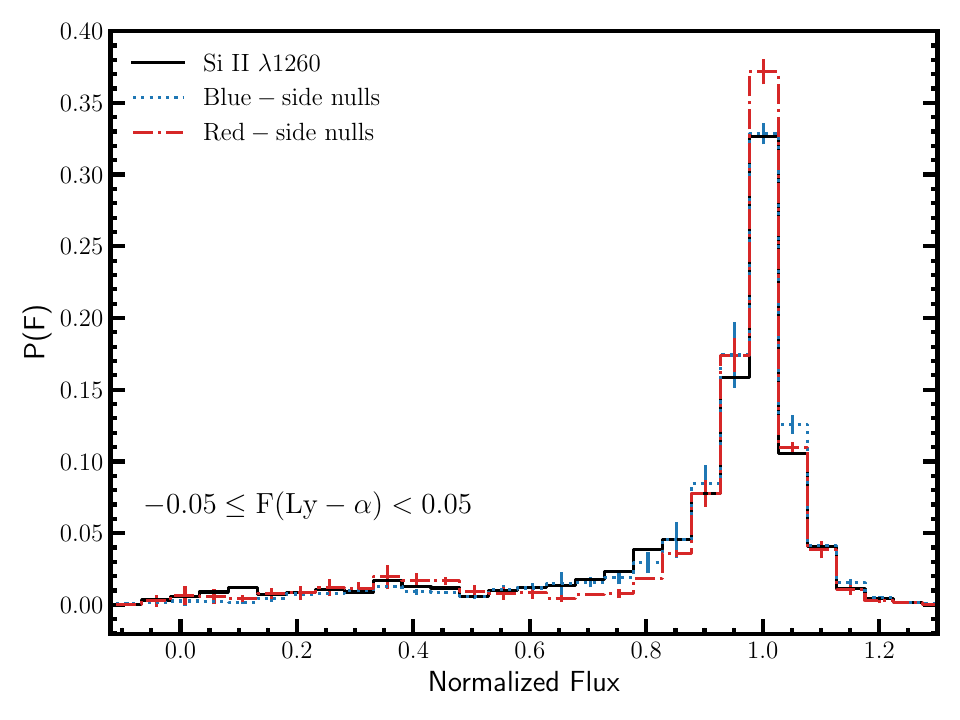}
    \caption{The probability distribution of  resolved flux transmission equivalent to our measured of \SiII\ 1260 absorption recovered in Keck HIRES spectra. The pixels included span the wavelength bin we call the `central pixel' and include the information usually lost due to the moderate resolution of SDSS spectra. For comparison 10 local equivalent null samples centred on $\pm7,8,9,10,11$\AA\ are also analysed. These null pixels sample the same uncorrelated absorber population contaminating the \SiII\ measurement. The \SiII\ sample and associated nulls samples are derived from 108 SBLAs obtained from 15 HIRES spectra in 138$\kms$-wide bins. The error bars on the null distributions correspond to 75\% the spread the nulls.}
    \label{fig:HR_SiII1260}
\end{figure}

\section{Measuring Metal Feature Covariance}
\label{extra:covariance}

In this work we have demonstrated that there is significant variance in the underlying metal populations that give rise to the signal measured for each metal transition measured in the composite spectrum of SBLAs. These populations have been forward modelled independently for each metal transition. Here we explore the degree to which these populations vary together from SBLA to SBLA (accepting that for any given SBLA unresolved phases may be present).

We complete the full covariance matrix of our stack of spectra corrected for the 
\begin{equation}
Cov(F_C(\lambda_1), F_C(\lambda_2))= \frac{1}{n}\sum_{i=1}^{n}\Big(F_i(\lambda_1)- F_S\Big)\Big(F_i(\lambda_2)- F_S\Big),
\end{equation}
noting that the covariance in the composite spectrum is equal to the composite of the stacked spectrum since the outcome is invariant of the fixed contribution of the pseudo-continuum normalisation step. 

The covariance matrix is computed for the entire wavelength range of the composite spectrum but of particular interest are the positions giving the covariance between wavelength transitions of interest. An example is given in \autoref{fig:line_covariance_zoom} for \OI\ $\lambda$1302 and \CII\ $\lambda$1335. Note that no covariance calculation is possible between \SiII\ $\lambda$1260 and \OI\ $\lambda$1302 because we never measure both for the same systems. This is because \SiII\ $\lambda$1260 is only measured in the forest and \OI\ $\lambda$1302 is never measured in the forest and because we suppress pixels within the quasar proximity zone (including the \Lya emission line itself). 

In order to ascertain whether the covariance measurements between every pair of metal transitions considered are statistically significant, we also sample covariance matrix at 64 local positions corresponding to  8 local pixels for both lines (see \autoref{fig:line_covariance_zoom}). These local pixels are determined using the same method as set out in \autoref{subsec:nullsample}. The error on each covariance measurement is derived from the standard deviation of the ensemble of 64 null values.

Finally in order to allow meaningful comparison of the measured covariance from one pair of transitions to another, we normalise our measurements (and the local null measurements) signal in the composite for the two lines
\begin{equation}
    Cov_n(X_1,X_2))= \frac{Cov(X_1,X_2))}{\Big(1-F_C(X_1)\Big)\Big(1-F_C(X_1)\Big)}
\end{equation}
where $X_1$ and $X_2$ correspond to the two metal transitions of interest. Figure~\autoref{fig:line_covariance} shows the normalised covariance for every pair transitions used in the population analysis.

\begin{figure}
    \centering
    \includegraphics[angle=0, width=1.\linewidth]{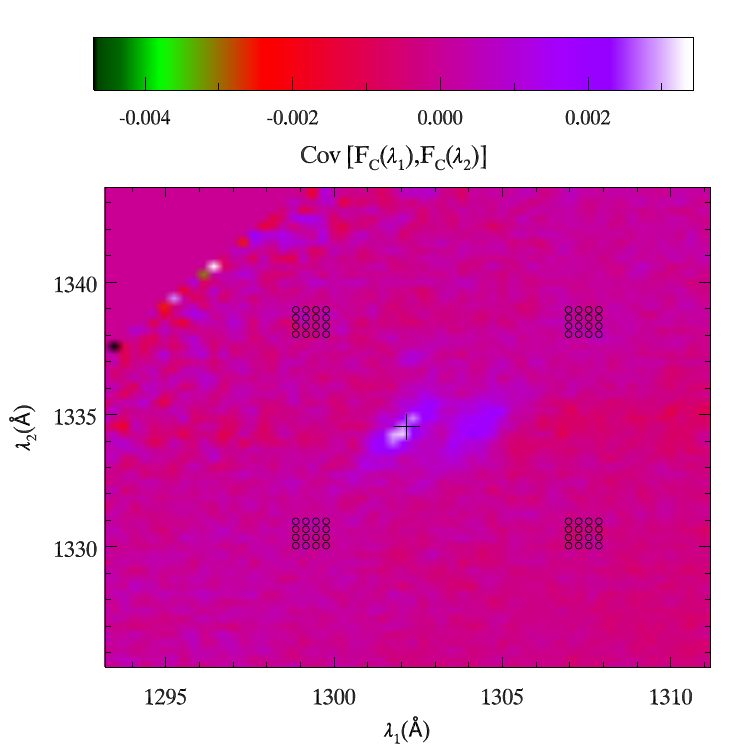}
    \caption{The covariance matrix of the stacked spectrum (and therefore of the composite spectrum) centred on ({\it black cross}) the covariance of flux transmission measurements of \OI\ $\lambda$1302 and \CII\ $\lambda$1335. Also shown as open circles are the local positions in the covariance matrix used as null measurements providing an estimate of measurement error (see the text).}
    \label{fig:line_covariance_zoom}
\end{figure}

\begin{figure}
    \centering
    \includegraphics[angle=0, width=0.99\linewidth]{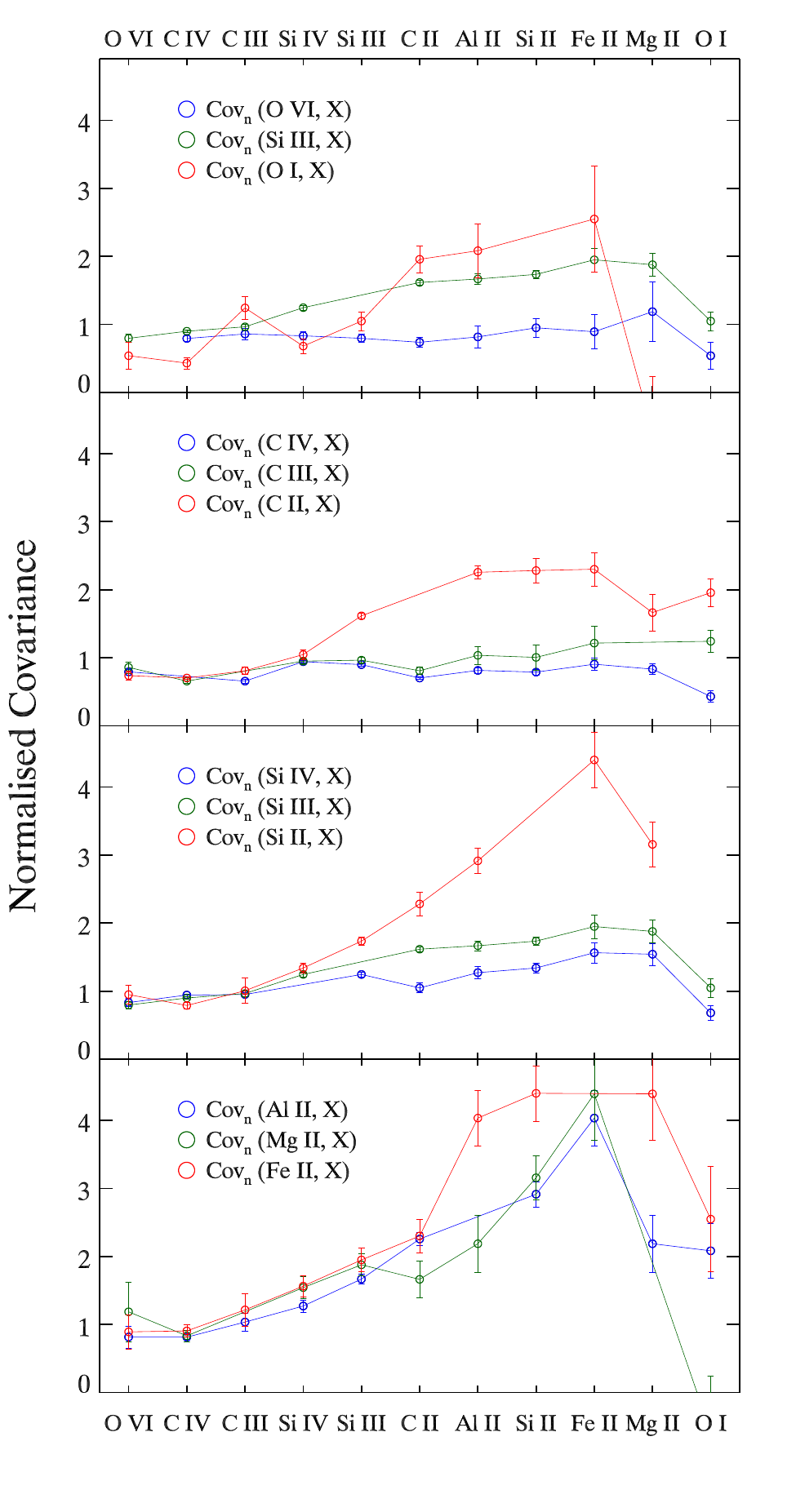}
    \caption{The normalised covariance with error estimates (derived from local null points in the covariance matrix) for every pair of lines used in our population analysis. The {\it top panel} shows the main 3 fiducial lines considered as typical low, intermediate and high ions in this work, the {\it 2nd panel} shows ions of carbon, the {\it third panel} shows ions of silicon, and the {\it bottom panel} shows the low ions that (to varying degrees) presented populations small enough to be plausibly derived from self-shielded gas in \autoref{tab:poplnfit} and associated text.}
    \label{fig:line_covariance}
\end{figure}


\bsp	
\label{lastpage}
\end{document}